\documentclass[lettersize,journal]{IEEEtran}
\usepackage{textcomp}
\usepackage{makecell}
\usepackage{multirow}
\usepackage{xcolor}
\usepackage{bm}
\usepackage{amsmath,amsfonts}
\usepackage{stfloats}
\usepackage{algorithmic}
\usepackage{algorithm}
\usepackage{amsmath,amsfonts,amsthm,bm}
\usepackage{amsthm,amssymb}
\usepackage{mathrsfs}
\usepackage{xcolor}
\usepackage[caption=false,font=tiny,labelfont=sf,textfont=sf]{subfig}
\usepackage{textcomp}
\usepackage{stfloats}
\usepackage{url}
\usepackage{verbatim}
\usepackage{graphicx}
\usepackage{cite}
\usepackage{lipsum}
\usepackage{cuted}
\usepackage{array}
\def\BibTeX{{\rm B\kern-.05em{\sc i\kern-.025em b}\kern-.08em
    T\kern-.1667em\lower.7ex\hbox{E}\kern-.125emX}}
\begin{document}

\title{Sensing for Communication: RIS-Assisted ISAC Coordination Gain Enhancement With \\Imperfect CSI}

\author{Xiaohui Li,~\IEEEmembership{Member,~IEEE}, Qi Zhu, Yunpei Chen, Chadi Assi,~\IEEEmembership{Fellow,~IEEE}, Yifei Yuan,~\IEEEmembership{Fellow,~IEEE}
%	\thanks{Copyright (c) 2015 IEEE. Personal use of this material is permitted. However, permission to use this material for any other purposes must be obtained from the IEEE by sending a request to pubs-permissions@ieee.org.}
%\thanks{This work is supported by the Fundamental Research Program of Shanxi Province (202203021212290) and the Research Project Supported by Shanxi Scholarship Council of China (2023-053). }
%	\thanks{(\textit{Corresponding author}) is with the College of Electronic Information and Optical Engineering, Taiyuan University of Technology, Taiyuan, China. (e-mail: lixiaohui@tyut.edu.cn)}
	\thanks{Xiaohui Li is with the School of Communication and Information Engineering, Nanjing University of Posts and Telecommunications, Nanjing, China, and also with the College of Electronic Information and Optical Engineering, Taiyuan University of Technology, Taiyuan, China (e-mail: lixiaohui@njupt.edu.cn)}
\thanks{Qi Zhu is with the School of Communication and Information Engineering, Nanjing University of Posts and Telecommunications, Nanjing 210003, China. (e-mails: zhuqi@njupt.edu.cn)} 
\thanks{Yunpei Chen is with the School of Marine Information Engineering, Hainan Tropical Ocean University, Sanya 572022, China, with the School of Communication and Information Engineering, Nanjing University of Posts and Telecommunications, Nanjing 210003, China. (e-mail: 18913823955@189.cn)}
\thanks{Chadi Assi are with the Concordia Institute for Information Systems Engineering(CIISE), Concordia University, Montreal, QC H3G 1M8, Canada (email:chadi.assi@concordia.ca)}
\thanks{Yifei Yuan is with Future Mobile Technology Lab, China Mobile Research Institute, Beijing 100053, China (e-mail: yuanyifei@chinamobile.com).}
\thanks{(Corresponding author: Yunpei Chen.) }
%\thanks{Shuran Sheng is with School of Electronic Information Engineering, Shanghai Dianji University, Shanghai, China (email: shengsr@sdju.edu.cn)}
%	\thanks{Tianqi Yu is with the School of Electronics and Information Engineering, Soochow University, Suzhou, China. (e-mail: tqyu@suda.edu.cn) }	
}

\maketitle

\begin{abstract}
Integrated sensing and communication (ISAC) has the potential to facilitate coordination gains from mutual assistance between sensing and communication (S\&C), especially sensing-aided communication enhancement (SACE).
	Reconfigurable intelligent surface (RIS) is another potential technique for achieving resource-efficient communication enhancement. 
	Therefore, this paper proposes an innovative RIS-assisted SACE (R-SACE) mechanism with the goal of improving the systemic communication performance of the ISAC system in practical scenarios where the channel status information (CSI) is imperfectly known.
	In the proposed R-SACE mechanism, a dual-functional base station (BS) provides downlink communication services to both the communication user and the dynamically changing target that is detected using the communication signals.
	RIS assists in both sensing and communications of the BS.
	A typical scenario is investigated in which either or both the direct and RIS-assisted reflected communication links are available depending on sensing results.
	The average systemic throughput (AST) over the entire timeline of the R-SACE mechanism is maximized by jointly optimizing both temporal and spatial resources under the probabilistic constraint and the sensing performance, transmission power, and communication interference constraints.
	The non-convex probabilistic mixed optimization problem is transformed and then solved by the proposed fixed-point iterative (FPI) algorithm.
	Simulation results demonstrate that the proposed FPI algorithm and R-SACE mechanism outperform the baseline algorithms and communication enhancement mechanisms in achieving higher systemic communication performance.

%Therefore, this paper focuses on how to effectively improve the ISAC systemic communication performance with assistance of RIS and the help of sensing. 
% that employs Quadratic transform (QT) and convex optimization methods.
%In particular, the dual-functional ISAC BS in our proposed novel RIS-assisted SACE (R-SACE) model provides downlink communication services to both the communication UE and the sensing target.
%In particular, downlink communication services in our proposed RIS-assisted SACE (R-SACE) model are provided to both the communication UE and the sensing target.

\end{abstract}

\IEEEpeerreviewmaketitle

\section{Introduction}
\IEEEPARstart{I}{ntegrated} sensing and communication (ISAC) has been formally approved as one of key usage scenarios in 6G vision framework.
% \cite{0}.
Different from conventional separate sensing and communication systems, ISAC is capable of unifying sensing and communication (S\&C) functionalities in a resource-efficient way by sharing hardware and wireless resources\cite{01,02}.
 ISAC has the potential to support both integration gain and coordination gain of S\&C\cite{itergation gain,1}.
The integration gain is for the dual purposes of S\&C by sharing the congested wireless resources, while the coordination gain benefits from mutual assistance between S\&C, allowing for communication-aided sensing enhancement (CASE) and sensing-aided communication enhancement (SACE)\cite{sensing as a service}. 
% providing intelligent network services including sensing and communication
% (S\&C)\cite{01}.
This paper focuses on improving the coordination gain, particularly the SACE, of ISAC systems.

Reconfigurable intelligent surface (RIS) is another promising technology in 6G for improving resource-efficient sensing and communication performances.
%\cite{RIS-zhangrui}.
In particular, RIS is typically a planar array with numerous low-cost and hardware-efficient passive reflecting elements installed.
By designing the phase shift intelligently, RIS is capable of reshaping the radio propagation environment, particularly creating useful line-of-sight (LoS) links to support S\&C\cite{RIS-ISAC-zongshu}. 
Currently, there are four main types of RIS: passive RIS, active RIS \cite{ActiveRIS?}, hybrid passive/active RIS \cite{hybrid ris}, and simultaneously transmitting and reflecting RIS (STAR-RIS) \cite{STAR-RIS-Rician}. 
In particular, active RIS deploys signal amplifiers into elements, which effectively overcomes severe channel fading. 
	The hybrid RIS combines passive reflectors and active sensors, allowing it to reflect incident signals while also sensing essential information.
	When the transmitter and receiver are situated on opposite sides of the RIS, the STAR-RIS is more effective because it can simultaneously send and receive incident signals onto both sides of the space around the RIS. 
	Among the four types of RIS designs, passive RIS has the following advantages in assisting the SACE mechanism.
	First, the passive RIS-assisted SACE method has the least hardware complexity in practice because it requires no extra hardware deployments. 
	Second, the passive RIS-assisted SACE mechanism holds promise for facilitating more resource-efficient performance improvements in communication and sensing.
	Third, the passive RIS-assisted SACE mechanism has the potential to improve systemic communication performance because it does not amplify the thermal noise as additional communication interference.
	Lastly, the situation in which the transmitter and receiver are on the same side of the RIS is examined in this work.  
	As a result, the passive RIS benefits SACE more than the STAR-RIS, particularly the energy splitting (ES) based STAR-RIS, because all of the energy can be used to reflect incident signals toward the desired side of the RIS.

The benefits of RIS have attracted lots of research interests in incorporating RIS into ISAC to improve the integration and coordination gains.
%\cite{RIS-ISAC3}.

\subsection{Related Works and Motivations}
Numerous efforts
 \cite{RIS-ISAC2+1,RIS-ISAC5,RIS-ISAC6,RIS-ISAC9,RIS-ISAC10,RIS-ISAC11,RIS-ISAC12,RIS-ISAC13,RIS-ISAC13+1}
%\cite{RIS-ISAC1,RIS-ISAC2,RIS-ISAC2+1,RIS-ISAC3,RIS-ISAC4,RIS-ISAC5,RIS-ISAC6,RIS-ISAC7,RIS-ISAC9,RIS-ISAC10,RIS-ISAC11,RIS-ISAC12,RIS-ISAC13,RIS-ISAC13+1}
 have  studied the RIS-assisted integration gain improvement for ISAC systems. 
%In these studies, the passive RIS\cite{RIS-ISAC2,RIS-ISAC3,RIS-ISAC5,RIS-ISAC6,RIS-ISAC2+1,RIS-ISAC9} and the active RIS\cite{RIS-ISAC10,RIS-ISAC11,RIS-ISAC12,RIS-ISAC13,RIS-ISAC13+1} are utilized to reflect ISAC signals and contribute to improving the
% communication\cite{RIS-ISAC1,RIS-ISAC2,RIS-ISAC3} or S\&C\cite{RIS-ISAC4,RIS-ISAC5,RIS-ISAC6,RIS-ISAC7,RIS-ISAC9,RIS-ISAC10,RIS-ISAC11,RIS-ISAC12,RIS-ISAC13,RIS-ISAC13+1}.
%The phase shift of RIS, as well as the beamforming of the ISAC signal, which includes sensing and communication beamforming, are designed to allow for effective resource sharing between S\&C.
%The 
%communication performance\cite{RIS-ISAC5,RIS-ISAC2+1,RIS-ISAC10,RIS-ISAC11}, sensing performance\cite{RIS-ISAC6,RIS-ISAC9,RIS-ISAC12,RIS-ISAC13,RIS-ISAC13+1}, and S\&C overall performance\cite{RIS-ISAC2} of ISAC systems.
%are respectively maximized under the sensing performance constraint, the communication performance constraint, and the total transmission power constraint of the BS.
%	A denominator of the above integration-gain-oriented ISAC systems is that the sensing and communication are simultaneously but independently performed within one time unit. In more details, the sensing and communication are only tightly coupled in wireless resources, e.g. the power, but loosely coupled in functional assistance. The mutual assistances between S\&C, involving both resource and function couplings,  remain to be further explored for boosting the coordination gain of ISAC systems\cite{coordination gain of ISAC}.
In the integration-gain-oriented ISAC systems, the effective wireless resource sharing between S\&C is investigated for maximizing the communication performance\cite{RIS-ISAC5,RIS-ISAC2+1,RIS-ISAC10,RIS-ISAC11} or the sensing performance\cite{RIS-ISAC6,RIS-ISAC9,RIS-ISAC12,RIS-ISAC13,RIS-ISAC13+1} under constraints of sensing or communication performances. 
%the sensing and communication are performed simultaneously yet independently. 
%In more detail, the dual-functional BS simultaneously performs sensing and communication. 
Nonetheless, mutual assistance between S\&C, such as the aids of sensing results in facilitating communication, is rarely involved in integration-gain-oriented ISAC systems, despite its significance for enhancing coordination gain of ISAC systems.
%Specifically, 
%S\&C in integration-gain-oriented ISAC systems are only coupled in wireless resources but loosely coupled in functional assistance. 
%Conversely, 
Specifically, S\&C in the coordination-gain-oriented ISAC system are not only coupled in wireless resources, as in integration-gain-oriented ISAC systems, but also interdependent in their functions.
As a result, the research on integration-gain-oriented ISAC systems can not be directly applied into the coordination-gain-oriented ISAC systems. 
Especially as sensing begins to rise as a fundamental service in ISAC\cite{sensing as a service}, it's necessary to investigate the SACE mechanism for improving ISAC coordination gains.
Moreover, taking advantage of RIS in establishing controllable supplementary links, the RIS-assisted SACE mechanism could provide extra sensing degrees of freedom while suppressing co-channel communication interference, allowing for enhanced communication performance with the help of desirable sensing results. 
Therefore, the study on RIS-assisted SACE mechanisms, which involves S\&C couplings in both wireless resource and functional assistance, is crucially important and meaningful for ISAC coordination gain enhancement, which motivates the research in our paper.

\begin{table*}[tbp]
	\renewcommand{\arraystretch}{1.2}
	\caption{Comparison of related work about RIS-assisted SACE }
	\label{Table_relatedworks}
	\centering
	\begin{tabular}{|m{0.95cm}|m{1.5cm}|m{2.7cm}|m{2.95cm}|m{4.25cm}|m{3.1cm}|}
%		{|c|c|c|c|c|c|}
		\hline
		\bfseries  Related work & \bfseries RIS structure & \bfseries Communication scenario & \bfseries Optimized Resource &  \bfseries Communication performance & \bfseries CSI condition\\
		\hline
		\cite{RIS-ISAC14}  & Semi-passive & \multirow{7}{*}{Blocked BS-user link} & \multirow{3}{=} {BS Beamforming and RIS phase shift} &  Received signal power at BS & \multirow{3}{=}{Acquiring user location information instead of CSI}\\ \cline{1-2}\cline{5-5}
%		\hline
		 \cite{RIS-ISAC15} & Passive and semi-passive &  &  &  Sum rate of users & \\\cline{1-2}\cline{4-5}\cline{6-6}
%		\hline
		 \cite{RIS-ISAC-SFC2} & \multirow{7}{*}{Passive} &  & RIS phase shift &  \multirow{2}{=}{Achievable data rate of a user} & Acquiring user orientation instead of CSI \\\cline{1-1}\cline{4-4}\cline{6-6}
%		\hline
		\cite{RIS-ISAC-SFC1} &   &   & \multirow{3}{=}{Time and RIS phase shift} &  & \multirow{3}{=}{Perfect CSI} \\\cline{1-1}\cline{3-3}\cline{5-5}
%		\hline
		\cite{ourWCL} &   & Available BS-user link or BS-RIS-user link &  &  Average achievable capacity of a user &  \\\cline{1-1}\cline{3-3}\cline{4-6}
%		\hline
		\bfseries {Proposed} &   & Available BS-RIS-user link and probabilistic available direct link &  Time, BS beamforming, and RIS phase shift &   Average systemic throughput &  Imperfect CSI \\\cline{1-1}\cline{3-6}
		\hline
	\end{tabular}
\end{table*}

To the best of our knowledge, a limited number of works\cite{RIS-ISAC14,RIS-ISAC15,RIS-ISAC-SFC1,ourWCL,RIS-ISAC-SFC2} have begun to investigate the RIS-assisted SACE for improving ISAC coordination gain.
The S\&C in a coordination-gain-oriented ISAC system are mutually constrained by limited wireless resources and mutually helpful, especially in a SACE system where communication signals support sensing and sensing results help facilitate communication further.
RIS in SACE is to assist in sensing or sensing-aided communication.
The goal of RIS-assisted SACE is maximizing the communication performance with aids of RIS and performance-guaranteed sensing results.

%It is critical to maintain well-balanced S\&C performances in the RIS-assisted SACE because the reliability of sensing results has a significant impact on communication performance.
%the time protocol of the RIS-assisted SACE mechanism is divided into two time blocks. At each time block, 

A summary of the related work about RIS-assisted SACE is shown in Table I. Specifically, in \cite{RIS-ISAC14}, a sub-passive RIS helps uplink communication while also estimating user location in each time unit using the communication signal. 
The estimated user location helps with the precoding design of RIS in the next time unit.
In \cite{RIS-ISAC15}, authors extend \cite{RIS-ISAC14} by using passive and sub-passive RIS for location estimations and uplink communications of multiple users.
The sum rate is maximized by jointly optimizing the beamforming and the phase shift of RIS using estimated user locations.
In \cite{RIS-ISAC-SFC2}, the ISAC frame structure consists of a sensing time unit and a communication time unit. 
During the sensing unit, the RIS helps estimate the angle of departure (AoD) of the communication user. 
Then, based on the obtained AoD, the RIS assists in modifying the transmission signal of the base station (BS) during the communication unit. 
The spatial codebook of the RIS is optimized to enhance sensing accuracy and improve the achievable rate.
In summary, the authors in \cite{RIS-ISAC14,RIS-ISAC15,RIS-ISAC-SFC2} use sensed user location information to replace the acquisition of CSI to facilitate communication. 
The spatial resource, including beamforming and phase shift, is optimized in \cite{RIS-ISAC14,RIS-ISAC15,RIS-ISAC-SFC2}. The optimization of temporal resources, such as time allocation among different time units, remains to be investigated further.
In \cite{RIS-ISAC-SFC1} and \cite{ourWCL}, the BS communicates with a legacy receiver and simultaneously detects a target in each time unit. 
Based on the sensing result and RIS assistance, the downlink communication links, which include the direct and RIS-assisted reflected links, are optimally designed in the subsequent time unit.
The average achievable capacity of the receiver is maximized by jointly optimizing the phase shift of RIS and the time allocation between two time units.
The authors in \cite{RIS-ISAC-SFC1} and \cite{ourWCL} assume perfect CSI for evaluating the upper bond of communication performance.

%The BS then uses the AoD data to design the RIS's beamforming at the next communication subframe for maximizing the communication signal-to-noise ratio (SNR).

%\textcolor{red}{Sum-rate maximization is a valuable research topic for improving communication in RIS-assisted ISAC systems, especially . } Specifically, the authors in \cite{5} and \cite{RIS-ISAC3} maximize the achievable sum rate of a RIS-assisted multi-user multi-input single-output (MU-MISO) ISAC system under the sensing performance constraint, transmit power budget, and RIS restriction. The authors in \cite{RIS-ISAC11} maximize the multi-user sum secrecy rate in a RIS-assisted ISAC system. In \cite{RIS-ISAC12} and \cite{RIS-ISAC13}, the authors optimize the sum rate in a RIS-assisted downlink ISAC system and a RIS-assisted integrated sensing and backscatter communication system, respectively. In these systems, the RIS-assisted phase shifts and beamforming of the BS are jointly optimized. In \cite{RIS-ISAC15}, the authors maximize the sum rate of the RIS-assisted ISAC system, considering the interference between sensing and communication functionalities.

The above works lay important groundwork for improving ISAC communication performance with assistance of sensing and RIS.
These works improve the ISAC communication performance based on either the perfect channel status information (CSI) or the estimated location information instead of CSI. 
Nevertheless, due to the fast-varying communication environment, the perfect CSI is challenging to obtain in practice. Meanwhile, the location information is also hard to perfectly estimate especially when the communication user is dynamically changing. 
As a result, evaluating and maximizing the communication performance when CSI is imperfectly known in the RIS-assisted SACE system is critically important for improving the coordination gain for practical ISAC systems, which is the focus of this paper.

Furthermore, current research on RIS-assisted SACE focuses on communication performance in an atypical scenario where the direct link between BS and communication users is blocked. 
While in the typical communication scenario where both the direct link and the RIS-assisted reflected link are available, the communication performance of the RIS-assisted SACE system has yet to be studied.
Therefore, it is worthwhile to investigate the systemic communication performance of the RIS-assisted SACE system when either or both the direct link and the RIS-assisted reflected link are available. 
Moreover, jointly optimizing both temporal and spatial resources is also important for further coordinating S\&C performances and ultimately improving systemic communication performance.
These further motivate our research in this paper. 
%is unavailable or coexists with 

%Motivated by this, the communication performance under both two kinds of scenarios where only the RIS-assisted reflected link is available or the 
%The direct link may still be available in typical scenarios.

%typically considers a separate S\&C service requirement model in which the sensing target (or communicator) is only to be sensed (or communicated) but not communicated (or sensed).
%This model is overly simplistic and unsuitable for practical ISAC systems, where a target, for example, also requires communication services in addition to being detected.
%In practice,  S\&C are tightly integrated in terms of resources, functions, and services within a typical integrated ISAC system.
%As a result, studying a generalized RIS-assisted SACE model applied to typical integrated ISAC systems in practice is important both theoretically and practically.

\subsection{Contributions}

Inspired by the above findings, this paper focuses on the systemic communication performance of an ISAC system where the CSI is imperfectly known.
We aim at improving the systemic communication performance with aids of sensing and RIS. 
A static communication user and a dynamically changing target that also has communication needs,
%further emerges as a potential communication user, 
are considered in the proposed RIS-assisted SACE (R-SACE) mechanism.
Moreover, the proposed R-SACE mechanism takes account of various potential communication scenarios where either or both the direct link and the RIS-assisted reflected link are available.
The allocation of both temporal and spatial resources and the functional assistance between S\&C are investigated for maximizing the systemic communication performance of the proposed R-SACE mechanism.
%Our work builds on previous research by proposing a more generalized and realistic ISAC model in which S\&C are tightly integrated in terms of resources (time and space), functional assistance (namely, sensing facilitates communication), and service requirements (namely, the sensing target also requires communication services).
%Our goal is to improve the systemic communication performance of an ISAC system using RIS and reliable sensing results. To accomplish this goal, the proposed RIS-assisted SACE (R-SACE) mechanism jointly designs the precoding of RIS and the time-space resource allocation between S\&C.
%namely, the S\&C time resource, the phase shift of RIS, and the transmission beamforming of the dual-functional BS 
%both space and time resources are jointly allocated between S\&C to improve the systemic communication performance with the performance-guaranteed sensing results. 
The main contributions of this paper are summarized as follows:
\begin{itemize}
	\item We propose a novel R-SACE mechanism in which the time protocol is designed as an ISAC period followed by a pure communication (PC) period. 
	In the ISAC period, RIS assists BS in sensing the target while simultaneously communicating with the static communication user.
	In the PC period, the BS provides downlink communication services to either or both the static communication user and the target, with the aid of RIS and the sensing results obtained in the ISAC period.
	We derive the closed-form expressions of sensing performance and the systemic communication performance with imperfect CSI.

	\item We evaluate the systemic communication performance as the average systemic throughput (AST) of all potential communication users in various potential communication scenarios over the entire timeline. Then, we investigate the AST maximization problem by jointly optimizing the time allocation between ISAC and PC periods, the transmission beamforming of the BS,
%	 to all potential communication users, 
	 and the phase shift of RIS. 
	Probabilistic constraints on imperfect CSI-based communication performances, as well as constraints on sensing performance, transmission power, and communication interference, are taken into account.
	
	\item  We transform the non-convex probabilistic mixed AST maximization problem into a non-probabilistic one by integrating the probabilistic constraints into the objective function. 
	Then,  we propose an iterative fixed-point iterative (FPI) algorithm to solve the transformed problem. 
	In particular, the optimal time allocation is numerically obtained through the convex theory. The joint optimization of  transmission beamforming and phase shift is solved by converting it into a sequence of convex problems through the quadratic transform (QT) method.
%	Then, we convert the joint optimization of  transmission beamforming and phase shift into a sequence of convex problems through the quadratic transform (QT) method.
	
	\item  Extensive simulations are performed to evaluate the performance of the proposed R-SACE mechanism and FPI algorithm. 
	Simulation results demonstrate the effectiveness of the proposed FPI algorithm in obtaining better solutions for maximizing the AST. 
	Two baseline mechanisms are provided for comparison, with communication aided by sensing or RIS.
	Simulation results indicate that the proposed R-SACE mechanism outperforms baseline mechanisms in achieving higher communication performance especially  from a system perspective.
\end{itemize}
% investigates the assistance and resource-constraint relationships between S\&C during two ISAC time blocks. 
%provides a statistical analysis of the sum rate performance in RIS-assisted ISAC systems. 
%Thirdly, the S\&C time allocation, the transmitting beamforming at the BS, and the reflection precoding at the RIS are jointly optimized under sensing performance, transmission power, and communication interference constraints. 

%\textcolor{red}{In this paper, we first derive the closed-form expressions for the sensing and communication performance metrics of the proposed R-SADCE mechanism.
%Then,  we utilize the alternating algorithm (AO) and quadratic transform (QT) method for decoupling and solving the non-convex optimization problem.
%Finally, simulation results demonstrate the advancement of the proposed R-SADCE mechanism.}
%\subsection{Paper Organization}
This paper is organized as follows. Section II presents the system model.
% of the proposed R-SACE mechanism.
%in which the sensing performance and communication performance with imperfect CSI are derived.
Section III formulates and transforms the AST maximization problem.
% and  it further. 
 Section IV proposes the FPI algorithm for solving the transformed problem. Simulation results are presented in Section V and Section VI concludes this paper. 

%\subsection{Notation}
\textbf{Notations:} 
%In this paper,
The lower-case letter $ a $, the bold lower-case letter $ \mathbf{a} $, and the bold upper-case letter $ \mathbf{A} $ denote a scalar variable, a vector, and a matrix, respectively. 
%For vector $ \mathbf{a} $,
$ |\mathbf{a}| $ denotes the Euclidean norm;
%and $ \mathrm{diag}\left\lbrace \mathbf{a}\right\rbrace  $ denotes the diagonal operation. 
%For matrix $ \mathbf{A} $, 
$ ||\mathbf{A}||$ denotes the Frobenius norm and $ \mathbf{A}^{-1}$ is its inverse. 
$ (\cdot)^\mathrm{H} $ denotes the complex conjugate transpose operation. 
%$ \mathbb{C} $ denotes the set of complex numbers, and 
$ \text{Re}\left\lbrace \cdot\right\rbrace  $ means the real part of a complex number. 
%$ \mathbb{R} $ denotes the set of real numbers.
$ 	\otimes $ denotes the Hadamard product.
$ \mathcal{CN}(\mu,\sigma) $ denotes the distribution of a
circularly symmetric complex Gaussian (CSCG) random variable with
mean $ \mu $ and variance $ \sigma $. 
$ \mathbf{I}_N $, $ \mathbf{0}_N $ and $ \mathbf{1}_N $ denote the $ N\times N $ identity matrix, $ N\times 1 $ all-zero matrix, and $ N\times 1 $ all-one matrix, respectively. 
%Other main notations used in this paper are listed in Table I.

\newcommand{\tabincell}[2]{\begin{tabular}{@{}#1@{}}#2\end{tabular}}  

\begin{figure}[tb]
	\centering
	\includegraphics[width=2.2in]{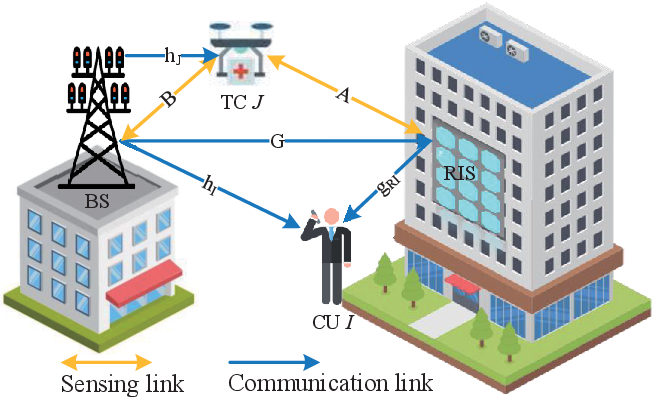}
	\caption{System model of the proposed R-SACE mechanism.}
	\label{model}
\end{figure}

\section{System Model}
This paper considers a single-cell downlink multiple-input single-output network.
As illustrated in Fig. 1, the system model of the proposed R-SACE mechanism consists of a dual-functional ISAC BS with $ K$ transmitting and $ K$ sensing antennas, a static communication user CU $ I $, a dynamic target TC $ J $, and a RIS equipped with $ N_{\mathrm{R}} $ elements.
The CU $ I $ is a pure communication user who receives communication signals from the BS throughout the timeline.
The TC $ J $ is an aperiodically present device-based target that needs to be detected and also a communication terminal that has communication requirements.
The TC $ J $ could be a person, vehicle, or drone that takes communication equipments\cite{sensing as a service} in practice\footnote{As a potential scenario, TC $ J $ can be a manoeuvrable UAV that provides emergency communication supports for users in challenging environments, such as earthquake areas, that cannot be directly served by the BS. 
Under this scenario, the TC $ J $ dynamically supports a variety of emergence areas. Consequently, it is necessary to detect TC $ J $ as it is not a fixed authorized communication user of the BS.
%at that particular location.
When TC $ J $ is present and correctly detected by the BS, it provides communication supports by acting as a communication relay between the BS and the CU $ I $.}. Thus, when TC $ J $ is present and correctly detected by the BS, TC $ J $ also emerges as a potential communication user of the BS.
This is different from \cite{RIS-ISAC2+1,RIS-ISAC11,RIS-ISAC12,RIS-ISAC13,RIS-ISAC15}, where the target has no communication needs but only to be detected.
RIS assists in both sensing and communication in the proposed R-SACE mechanism. 
Specifically, RIS helps BS detect TC $ J $ and reconstruct another feasible communication link that facilitates the downlink communication while suppressing mutual interference\footnote{Note that the proposed R-SACE mechanism is an orthogonal multiple access (OMA) system, which is different from the rate splitting multiple access (RSMA) and non-orthogonal multiple access (NOMA) in terms of interference management. In particular, the main contribution of the proposed R-SACE mechanism is utilizing sensing and the RIS-assisted reconstructed link to coordinate mutual interference. This is different from NOMA and RSMA mechanisms in which power allocation and successive interference cancellation are typically adopted to manage interference.}.

%Furthermore, the proposed R-SACE model considers a practical ultra-dense communication scenario in which UEs are densely distributed and thus not well-separated spatially in desired communication directions.

%As a result, there is harmful mutual interference between the direct BS-UE $ I $ communication link and the direct BS-UE $ J $ communication link (when the BS correctly detects the present UE $ J $).
%To address this issue, the RIS is intelligently deployed to improve the received communication signal-to-interference-and-noise ratio (SINR) while suppressing harmful direct mutual interference.

%\textcolor{red}{Conventional ISAC models consider one communication user and one sensing target exist. Different the conventional model, this paper study a new model where the UE $  J  $ exists not only as a target to be detected but also as a communication user to receive downlink signals from the BS. That is to say, our proposed ISAC model consists one sensing target (i.e., the periodically present UE $ J $) and two communication users (i.e., the UE $ I $ and the present UE $ J $). }

\begin{figure}[tb]
	\centering
	\includegraphics[width=2.3in]{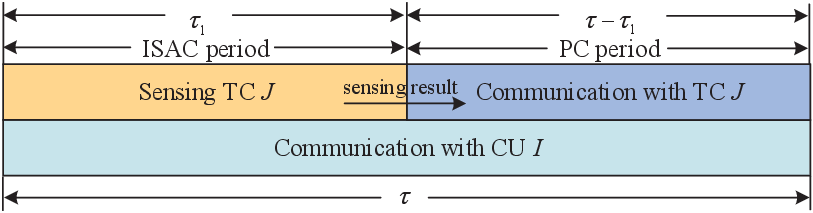}
	\caption{Illustration of the time protocol of the proposed R-SACE mechanism.}
	\label{ISAC period}
\end{figure}

The time protocol
%\cite{RIS-ISAC9} 
of the proposed R-SACE model is illustrated in Fig. \ref{ISAC period}. 
The entire timeline with $ \tau $ time slots includes an ISAC period with $ \tau_1 $ time slots and a pure communication (PC) period with $ \tau-\tau_1 $ time slots.
During the ISAC period, the BS communicates with CU $ I $ through both direct BS-CU $ I $ link and the RIS-assisted BS-RIS-CU $ I $ link and meanwhile detects TC $ J $ using the communication signal.
%\footnote{We use sense and detect interchangeably in this paper.} 
During the PC period, the BS performs downlink communications with assistance of RIS and based on the sensing result obtained at the ISAC period.
Specially, when BS detects TC $ J $ being present, it communicates with CU $ I $ through the RIS-assisted BS-RIS-CU $ I $ link and meanwhile communicates with TC $ J $ through the direct BS-TC $ J $ link.
Additionally, when BS detects TC $ J $ being absent, it communicates with CU $ I $ through both the direct BS-CU $ I $ link and the BS-RIS-CU $ I $ link.

%, and its communication links are designed based on the sensing result about UE $ J $ obtained during the ISAC period. 
%Specifically, when BS detects the UE $ J $ being absent, it will communicate with the UE $ I $ through the direct BS-UE $ I $ communication link. Alternatively, when BS detects the presence of the UE $ J $, it will communicate with the UE $ J $ through the direct BS-UE $ J $ link 
%\footnote{In the R-SACE mechanism, we assume that UE $J $ has a higher priority to be served than UE $I $. As a result, the BS serves UE $ J $ through the direct BS-UE $ J $ link and UE $ I $ through the reflected BS-RIS-UE $ I $ link.} 
%while communicating with the UE $ I $ through the reconstructed BS-RIS-UE $ I $ link with the assistance of RIS.

We use $ a $ and $ d $ to indicate the actual and detected statuses of TC $ J $, respectively. Specially, $ a=1 $ (or $ a=0 $) indicates that the TC $ J$ is actually present (or absent). $ d=1 $ (or $ d=0 $) indicates that the BS detects TC $ J $ being present (or absent). 

%The S\&C models of the proposed R-SACE mechanism, as well as the capacities considering both imperfect and perfect CSI, are illustrated below.

%\footnote{To investigate the upper-bound performance of the R-SADCE mechanism, this paper assumes perfect CSI knowledge of the involved channels.}

\subsection{Sensing Model}
%Let  denote the transmission signal of the BS. 
During ISAC  period, the BS transmits the signal $ s_1 $ to CU $ I $ and simultaneously uses $ s_1 $ to detect TC $ J $ with assistance of RIS\cite{response matrix1}. 
%Note that the self- or mutual interference that was produced in the conventional radar and communication systems no longer exists because the radar sensing and communication tasks are coordinated using a special integrated DRC BS.
As shown in Fig. 1, when TC $ J $ is truly present, the signal $ s_1 $ is reflected back to the BS from both TC $ J $ and RIS. 
As a result, the echo signal received at the BS at time slot  $ \forall t \in\!\left\lbrace 1,\cdots,\tau_{1} \!\right\rbrace$ can be given by\cite{Response Matrix,response matrix1} 
\begin{equation}\label{YB}
\mathbf{z}_B(t)=a(\mathbf{G}^{\mathrm{H}}\bm{	\Lambda}\mathbf{A}\bm{\Lambda}\mathbf{G}+\mathbf{B})\mathbf{w}_0s_1(t)+\mathbf{n}_B,
\end{equation}
where $ \mathbb{E}\left[|s_1|^2 \right]=1  $.  $ \mathbf{G}\!\in\!\mathbb{C}^{N_{\mathrm{R}}\times K} $ represents the channel between BS and RIS. 
$ \bm{\Lambda}\in\mathbb{C}^{N_{\mathrm{R}}\times N_{\mathrm{R}}}  $ represents the phase shift matrix of RIS at the ISAC period. 
$ \mathbf{A}=\alpha_A\mathbf{a}_{\textit{Ar}}(\theta_A)\mathbf{a}_{\textit{At}}^{\mathrm{H}}(\theta_A)
%\in\mathbb{C}^{N_{\mathrm{R}}\times N_{\mathrm{R}}}  
$ and $ \mathbf{B}=\alpha_B\mathbf{a}_{\textit{Br}}(\theta_B)\mathbf{a}_{\textit{Bt}}^{\mathrm{H}}(\theta_B)$ represent the response matrices \cite{response matrix1,Response Matrix} of RIS and BS, respectively.
%\footnote{Note that the response matrices $ \mathbf{A} $ and $ \mathbf{B} $ can be perfectly obtained using efficient estimation techniques as introduced in \cite{response matrix1,response matrix2}.}. 
%$ \alpha_A=\alpha_T\alpha_{RT} $ and $ \alpha_B=\alpha_T\alpha_{BT} $ denote the reflection coefficients incorporating the effects of radar cross section (RCS) $ \alpha_T $ and path losses $ \alpha_{RT} $ and $ \alpha_{BT} $\cite{RIS-ISAC2+1}. 
$ \alpha_A$ and $ \alpha_B$ denote the reflection coefficients incorporating effects of radar cross section (RCS) $ \alpha_T\sim(0,\sigma_T^2) $ and path losses of RIS-TC $ J $ and BS-TC $ J $ links, respectively\cite{RIS-ISAC2+1,Response Matrix}. 
$ \theta\!_A $ and $ \theta\!_B $ represent the direct-of-arrivals (DoAs) of the target with respect to RIS and BS.
$ \mathbf{a}_{\textit{At}}$, $ \mathbf{a}_{\textit{Bt}} $ and $ \mathbf{a}_{\textit{Ar}} $, $ \mathbf{a}_{\textit{Br}} $ represent the transmit and receive steering vectors at RIS and BS, respectively.
In particular, $ \mathbf{a}_{\textit{At}}\!=\!\left[  1,e^{-j\pi\sin\theta\!_A},\cdots,e^{-j(N_{\mathrm{R}}-1)\pi\sin\theta\!_A} \right]\!^{\mathrm{T}}   $ and $ \mathbf{a}_{\textit{Bt}}\!=\!\left[  1,e^{-j\pi\sin\theta\!_B},\cdots,e^{-j(K-1)\pi\sin\theta\!_B} \right]\!^{\mathrm{T}} $.
$ \mathbf{a}_{\textit{Ar}}$ and $ \mathbf{a}_{\textit{Br}}$ can be similarly expressed based on $ \mathbf{a}_{\textit{At}}  $ and $ \mathbf{a}_{\textit{Bt}} $, respectively.
%$ s_0 $, with $ |s_0(t)|^2 = 1 $, is the signal sent by BS to detect TC $ J $.
$ \mathbf{w}_0\in \mathbb{C}^{K\times 1} $ is the beamforming vector of $ s_1 $. 
$\mathbf{n}_B
%=\left[n_1,\cdots,n_K\right]
%\in\mathbb{C}^{K\times 1} 
$
%. $ n_k\sim\mathcal{CN}(0,\sigma_0^2), \forall k $ 
represents the CSCG noise 
%at $ k $-th sensing antenna 
at the BS, with
%Without losing generality, 
%Accordingly, 
%we have 
$ \mathbf{n}_B\sim\mathcal{CN}(\mathbf{0}_{K},\mathbf{1}_{K}\sigma_0^2) $.

%In $ \mathbf{z}_B(t) $, the BS receives the echo signal only when the TC $ J $ is truly present, i.e., when $ a=1 $. Otherwise, $ \mathbf{z}_B(t)=\mathbf{n}_B $ when TC $ J $ is truly absent, i.e., when $ a=0 $.

Then, the BS samples the received $ \mathbf{z}_B(t) $ over the duration of $ \tau_1 $ with a sampling frequency of $ f_s $, thereby conducting a hypothesis test statistic  $ T_e $.
%The test statistic
%\cite{MOMO-radar,Energy-radar,Neyman} 
%%\cite{sensing as a service,MOMO-radar,Neyman} 
%can be given by $  T_e=\frac{1}{\tau_1f_s }\sum_{n_s=1}^{\tau_1f_s }||\mathbf{X}(n_s)||^2$, where $\mathbf{X}(n_s)=a(\mathbf{G}^{\mathrm{H}}\bm{\Lambda}\mathbf{A}\bm{\Lambda}\mathbf{G}+\mathbf{B})\mathbf{w}_1s_1(n_s)+\mathbf{n}_B(n_s)$ is the $ n_s $-th discretely sampled signal at the BS, $  1\le n_s\le N_s $.
Following the chi-square distribution of $ T_e $ and 
%the Neyman-Pearson criterion
central limit theorem (CLT)\cite{sensing as a service,RIS-ISAC10,Neyman,ourWCL}, 
%chi-square distribution of the test statistic
% the central limit theorem and 
% the Neyman-Pearson criterion
%Accordingly, we have $ N_s=\tau_1f_s $ as the total sampling number of the BS. 
%\cite{MOMO-radar,Energy-radar,Neyman}
%\cite{MOMO-radar,Energy-radar,Neyman} 
the sensing performance including detection probability $ P_d $  and false alarm probability $ P_f $ obtained at the BS can be given respectively  by
\begin{equation}\label{P_d}\!
%$ 
\begin{split}
\!\!P_d&=\Pr(T_e>\epsilon|a=1)\\
%=Q\left(\frac{\epsilon\!-\!\mu_1}{\sqrt{\nu_1}}\right)
%=Q\left(\frac{\epsilon\!-\!||\mathbf{B}\mathbf{W}_0||^2-||\mathbf{n}_B||^2}{\sqrt{N_s}(||\mathbf{B}\mathbf{W}_0||^2+||\mathbf{n}_B||^2)}\right), 
&\!\!=\!Q\!\left(\!\! \sqrt{\!\tau_1f_s }\!\left(\! \frac{\epsilon}{||(\mathbf{G}^{\mathrm{H}}\bm{\Lambda}\mathbf{A}\!\bm{\Lambda}\!\mathbf{G}\!+\!\mathbf{B})\mathbf{w}_1||^2\!+\!||\mathbf{n}_B||^2}\!-\!1\!\right)\! \right),
\end{split}
%$
\end{equation}
\begin{equation}\label{P_f}
P_f=\Pr(T_e>\epsilon|a=0)
% =Q\left(\frac{\epsilon-\mu_0}{\sqrt{\nu_0}}\right)
% =Q\left(\frac{\epsilon-||\mathbf{n}_B||^2}{\sqrt{N_s}||\mathbf{n}_B||^2} \right) 
=Q\left( \sqrt{\tau_1f_s }\left( \frac{\epsilon}{||\mathbf{n}_B||^2}-1\right) \right) 
\end{equation}
% $
%\end{align}
%and
%respectively, 
where $ Q(x)\!\!=\!\!\frac{1}{\sqrt{2\pi}}\int_{x}^{\infty}\!\!\exp(\frac{-t^2}{2})dt$, 
%is Q-function, 
$  \epsilon $ is the detection threshold.  To be specific, $ P_f $ represents the probability of that BS wrongly detects the presence of TC $ J $ when TC $ J $ is truly absent. $ P_d $ represents the probability of that BS correctly detects the presence of TC $ J $.
Let $ P_{\mathrm{J_1}} $ and $ P_{\mathrm{J_0}}=1-P_{\mathrm{J_1}} $ denote the probabilities of TC $ J $ being actually present or absent, respectively. 
There exist four potential cases with corresponding probabilities as follows:
\begin{itemize}
	\item \textit{Case 1:} When TC $ J $ is truly present and is correctly detected by the BS, i.e., $ a=1 $ and $ d=1 $,  the corresponding probability is $ P_{11}=\Pr(a=1,d=1) =P_dP_{\mathrm{J_1}}$. 
	\item \textit{Case 2:} When TC $ J $ is truly present but not correctly detected by the BS, i.e., $ a=1 $ and $ d=0 $, the corresponding probability is $ P_{10}=\Pr(a=1,d=0) =(1-P_d)P_{\mathrm{J_1}}$. 
	\item \textit{Case 3:} When TC $ J $ is truly absent and is correctly detected by the BS, i.e., $ a=0 $ and $ d=0 $, the corresponding probability is $ P_{00}=\Pr(a=0,d=0) =(1-P_f)P_{\mathrm{J_0}}$. 
	\item \textit{Case 4:} When TC $ J $ is truly absent but detected as present by the BS, i.e., $ a=0 $ and $ d=1 $, the corresponding probability is $P_{01}=\Pr(a=0,d=1) =P_fP_{\mathrm{J_0}}$. 
\end{itemize}

\subsection{Communication Model}
%In the proposed R-SACE mechanism, the downlink communication from BS to CU $ I $ is performed during the entire timeline. While the downlink communication from BS to the present TC $ J $ is performed during the PC period.
\subsubsection{ISAC period}
During the ISAC period, the BS communicates with CU $ I $ through both the direct BS-CU $ I $ link and the RIS-assisted BS-RIS-CU $ I $ link. 
	Therefore, the signal received at the CU $ I $ at time slot $ t\in \left\lbrace1,\cdots,\tau_1 \right\rbrace $ can be given by
	\begin{equation}\label{S_I_ISAC}
Z_{\textit{I}(1)}(t)=\left(\mathbf{g}_{\textit{RI}}^{\mathrm{H}}\bm{\Lambda}\mathbf{G}+\mathbf{h}_I^{\mathrm{H}} \right)\mathbf{w}_0s_1(t)+n_I,
\end{equation}
where $ \mathbf{g}_{\textit{RI}}\in\mathbb{C}^{N_{\mathrm{R}}\times1} $ represents the channel between RIS and CU $ I $, $ \mathbf{h}_I\!\in\!\mathbb{C}^{K\times1} $ represents the channel between BS and CU $ I $, $ n_I\sim\mathcal{CN}(0,\sigma_I ^2)  $ is the receiver noise at the CU $ I $.

\subsubsection{PC period}
During the PC period, the BS communicates with CU $ I $ and TC $ J $ based on the sensing result and with assistance of RIS.
To be specific, when the BS detects the TC $ J $ being present, i.e., $ d=1 $, it transmits signals $ s_1 $ and $ s_2 $ to CU $ I $ and TC $ J $, respectively.
%will provide downlink communication services to CU $I  $ and TC $ J $ simultaneously. 
The RIS-assisted BS-RIS-CU $ I $ link is used for the communication of CU $ I $ while the direct BS-TC $ J $ link is used for the communication of TC $ J $.
%, aiming at suppressing the harmful direct interference between BS-CU $ I $ link and BS-TC $ J $ link.
On the other side, when BS detects the TC $ J $ being absent, i.e., $ d=0 $, it  transmits $ s_1 $ to CU $ I $ through the direct BS-CU $ I $ link and the RIS-assisted BS-RIS-CU $ I $ link. 

%In the PC period, the BS designs its downlink communication links based on the sensing result about TC $ J $ obtained during the ISAC period, with the assistance of RIS. 

As a result, the signal received at CU $ I $ at time slot $ t\in \left\lbrace1,\cdots,\tau-\tau_1 \right\rbrace $ can be given by
	\begin{equation}\label{S_I0}
	\begin{split}
	%$
	Z_{\textit{I}(2)}^{(d)}\!(t)\!\!=\!\!\left(\mathbf{g}_{\textit{RI}}^{\mathrm{H}}\bm{\Theta}\mathbf{G}\!\!+\!\!(1\!-\!d)\mathbf{h}_I^{\mathrm{H}} \right)\!\mathbf{w}\!_Is_1(t)\!\!+\!d\mathbf{h}_I^{\mathrm{H}} \mathbf{w}\!_Js_2(t)\!\!+\!n_I
	% $.
	\end{split}
	\end{equation}
	%$ d\!\!=\!\!1 $ means that the BS detects the TC $ J $ being active and $ d\!=\!0 $ means that the BS detects the TC $ J $ being inactive. 
	where $ \mathbf{w}_I $ and $ \mathbf{w}_J $ represents the beamforming vectors of $ s_1$ and $ s_2$, respectively; $\bm{\Theta}
%	=\mathrm{diag}([e^{j\theta_1},...,e^{j\theta\!_{N\!_{\mathrm{R}}}}]^\mathrm{T})
	\in\mathbb{C}^{N_{\mathrm{R}}\times N_{\mathrm{R}}} $ is the phase shift matrix of RIS\footnote{Note that $ \mathbf{w}_I $ and $\bm{\Theta}$ in the PC period are different from $  \mathbf{w}_0 $ and $ \bm{\Lambda} $ in the ISAC period, since BS communicates with both CU $ I $ and TC $ J $ in the PC period. 
%	 That is to say, there exist mutual interference between communications of CU $ I $ and the present TC $ J $. 
	 Therefore, $ \mathbf{w}_I $, $ \mathbf{w}_J $, and $\bm{\Theta}$ need to be optimally designed in the PC period for restricting mutual interference and ultimately facilitating the systemic communication enhancement.}.
	The first term in $ Z_{I(2)}^{(d)}(t) $ represents the desired signal received at CU $ I $ from the BS-RIS-CU $ I $ link and the direct BS-CU $ I $ link (i.e., when $ d=0 $).
	The second term in $ Z_{I(2)}^{(d)}(t)$ represents the interference experienced by  CU $ I $ from $ s_2 $ when BS  detects the TC $ J $ being present (i.e., when $ d=1 $). 

%Let $ s_2 $ represent the communication signal from BS to TC $ J $. 
Additionally,  the signal received at the truly present TC $ J $ at time slot $ t\in \left\lbrace1,\cdots,\tau-\tau_1 \right\rbrace $ can be given by
\begin{equation}\label{Yj}
\begin{split}
Z_{J}^{(d)}(t)=d\mathbf{h}_J^{\mathrm{H}}\mathbf{w}_Js_2(t)+\mathbf{g}_{\textit{\textit{RJ}}}^{\mathrm{H}}\bm{\Theta}\mathbf{G}\mathbf{w}_Is_1(t)+n_J,
\end{split}
\end{equation}
where $\mathbf{h}_J\in\mathbb{C}^{K\times1} $ represents the channel between BS and TC $ J $, $ \mathbf{g}_{\textit{RJ}}\in\mathbb{C}^{N_{\mathrm{R}}\times1}  $ represents the channel between RIS and TC $ J $.
%$ A_IJ$ and $ \mathbf{f}_I $ represent path loss coefficient and Rayleigh fading coefficient, respectively.
The first term in $ Z_{J}^{(d)}(t) $ is the desired signal received at the present TC $ J $ when it is correctly detected by BS (i.e., when $ d=1 $). 
The second term in $ Z_{J}^{(d)}(t) $ is the interference experienced by the present TC $ J $.  $ n_J\sim\mathcal{CN}(0,\sigma_J ^2)  $ represents the CSCG noise at the TC $ J $.

%Based on the the central limit theorem (CLT)\cite{Liang_quanwei}, for a large $ N $, the test statistic $ E $ subjects to a Gaussian distribution with mean $  \mu_{(a)}=\frac{1}{K}\sum_{k=1}^{K}(a\sigma_{S_k}^2+\sigma_{k}^2) $ 
%%	$  \mu_{(a)}=\frac{1}{K}\sum_{k=1}^{K}(a\sigma_{S_k}^2+\sigma_{Rk}^2) $ 
%	and variance	$ \nu_{(a)}=\frac{1}{K^2N}\sum_{k=1}^{K}(a\sigma_{S_k}^2+\sigma_{k}^2)^2 $. 
%%	$ \nu_{(a)}=\frac{1}{K^2N}\sum_{k=1}^{K}(a\sigma_{S_k}^2+\sigma_{Rk}^2)^2 $. 
%As a result, the false alarm probability $ P_f $ and the detection probability $ P_d $  obtained at the BS can be given by
%%\begin{align}\label{P_f,k}
%$ P_f\!=\!Pr(E\!>\!\epsilon|a\!=\!0)\!=\!Q\!\left(\frac{\epsilon\!-\!\mu_0}{\sqrt{\!\nu_0}}\right) $
%%\end{align}
%and
%%\begin{equation}\label{P_d,k}
%$ P_d\!=\!Pr(E\!>\!\epsilon|a\!=\!1)\!=\!Q\!\left(\frac{\epsilon\!-\!\mu_1}{\sqrt{\nu_1}}\right), $
%%\end{equation}
%where $ Q(x)\!\!=\!\!\frac{1}{\sqrt{2\pi}}\!\!\int_{x}^{\infty}\!\exp(\!\frac{-t^2}{2}\!)dt$ is the Q-function\cite{RIS-ISAC8}, $  \epsilon $ is the detection threshold. 
%Below we assume $ \sigma_k^2\!=\!\sigma_0^2, \forall k $ to simplify the equations.

\subsection{Systemic Communication Performance With Imperfect CSI}

In practice, it is hard to obtain the CSI of $ \mathbf{h}_I $ and $ \mathbf{h}_J $ perfectly\footnote{
This paper primarily studies the imperfect CSI of direct links $ \mathbf{h}_I $ and $ \mathbf{h}_J $.
The estimation of cascaded channels $ \mathbf{g}_{\textit{RI}}^{\mathrm{H}}\bm{\Lambda}\mathbf{G} $, $ \mathbf{g}_{\textit{RI}}^{\mathrm{H}}\bm{\Theta}\mathbf{G} $ and $ \mathbf{g}_{\textit{RJ}}^{\mathrm{H}}\bm{\Theta}\mathbf{G} $ in RIS-assisted systems is a more challenging task that can be addressed by effective methods, such as the joint bilinear factorization and matrix completion method\cite{cascaded channel estimation1} and 
the progressive channel estimation method\cite{cascaded channel estimation2}.
%Due to the limited space, here we only present the estimation model of the cascaded channel, taking $ \mathbf{g}_{\textit{RI}}^{\mathrm{H}}\bm{\Theta}\mathbf{G} $ as an example. 
For instance,
% the least-square (LS) estimation model\cite{cascaded channel estimation2} of $ \mathbf{g}_{\textit{RI}}^{\mathrm{H}}\bm{\Theta}\mathbf{G} $ can be given as follows. 
let $ \mathbf{g}_{\textit{RI}}^{'}\!\!=\!\!\mathrm{diag}(\mathbf{g}_{\textit{RI}}^{\mathrm{H}})\mathbf{G} $ 
%denotes the cascaded channel vector without the phase shift 
and $ Z_{I\!(\!2\!)}^{'}\!(t)\!=\! \mathbf{g}_{\textit{RI}}^{\mathrm{H}}\bm{\Theta}\mathbf{G}\mathbf{w}\!_I\!s_1\!(t) $ in Eq. \eqref{S_I0}. 
Then, the least-square (LS) estimation model\cite{cascaded channel estimation2} of $ \mathbf{g}_{\textit{RI}}^{'} $ can be given as $ \hat{\mathbf{g}}_{\textit{RI}}^{'}\!=\!(\!\mathbf{w}\!_Is_1\!(t)\!)^{-\!1}\bm{\theta}^{-\!1}Z_{I\!(2)}^{'-1}\!(t)\!=\!\mathbf{g}_{\textit{RI}}^{'}\!+\mathbf{e}_{\textit{RI}} $, where $ \bm{\theta} $ is the diagonal vector of $ \bm{\Theta} $; $ \mathbf{e}_{\textit{RI}}\!=\!(\mathbf{w}\!_I\!s_1\!(t))^{-\!1}\bm{\theta}^{-\!1}\!\Delta $ is the channel estimation error, and $ \Delta\!\!=\!\!Z_{\textit{I}(2)}^{(d)}\!(t)\!-\!Z_{I\!(2)}^{'}\!(t) $.
Then, the mean-square error (MSE) $ \mathbb{E}\left[ \|\mathbf{g}_{\textit{RI,e}}\|^2 \right]   $ can be minimized by designing $ \bm{\Theta} $ as referred to \cite{cascaded channel estimation2}.}. Therefore, the systemic communication performance with imperfect CSI is worthwhile to explore in this paper. 

We assume $ \mathbf{h}_I=L_I\mathbf{f}_I $ and $ \mathbf{h}_J=L_J\mathbf{f}_J$, where $ \mathbf{f}_I\sim\mathcal{CN}(\mathbf{0}_K,\mathbf{1}_K)$ and $ \mathbf{f}_J\sim\mathcal{CN}(\mathbf{0}_K,\mathbf{1}_K)$.
Generally, the small-scale Rayleigh fading coefficients $ \mathbf{f}_I $ and $ \mathbf{f}_J $ have server effects on the channel estimation accuracy than the large-scale pass loss coefficients $ L_I $ and $ L_J $. 
%Nevertheless, the perfect CSI of $ \mathbf{h}_I $ or $ \mathbf{h}_J $ is hard to obtain in practice. The BS only has the imperfectly estimated small-scale fading coefficients.
%\footnote{The large-scale pass loss coefficients, i.e., $ A_I $ and $ A_J $, vary slowly and can be estimated perfectly by the BS\cite{largescale-perfect1,largescale-perfect2}}. 
Therefore, according to the minimum mean square error channel estimation error model\cite{estimation-error-model1,estimation-error-model2,estimation-error-model3}, $ \mathbf{h}_I $ and $ \mathbf{h}_J $ can be rewritten as
%realistic Rayleigh fading coefficients $ \mathbf{f}_I $ and $ \mathbf{f}_J $ of $ \mathbf{h}_I $ and $ \mathbf{h}_J $ can be modeled as
\begin{equation}\label{f_I}
\mathbf{h}_I=L_I\mathbf{f}_I=L_I\left( \hat{\mathbf{f}}_I+\mathbf{e}_I\right) 
\end{equation}
%and 
\begin{equation}\label{f_j}
\mathbf{h}_J=L_J\mathbf{f}_J=L_J\left( \hat{\mathbf{f}}_J+\mathbf{e}_J\right) 
\end{equation}
respectively. 
$ \mathbf{e}_I $ and $  \mathbf{e}_J$ denote the estimation error vectors in which each element follows complex Gaussian distribution with mean zero and variance $ \sigma_e^2 $.
$ \hat{\mathbf{f}}_I  $ and $ \hat{\mathbf{f}}_J $ denote the estimated Rayleigh fading coefficients that follow $ \mathcal{CN}\left(\mathbf{0}_K,\mathbf{1}_K(1-\sigma_e^2)
\right) $.

Then, according to Eqs. \eqref{S_I_ISAC}-\eqref{f_j}, 
%Given the estimated fading channel $ \hat{\mathbf{f}}_I  $ and $ \hat{\mathbf{f}}_J $, 
the data rates of CU $ I $ and TC $ J$ obtained with imperfect CSI during the ISAC period and the PC period can be given as
\begin{equation}\label{hat_R}
\hat{R}_l=\!\begin{cases}
%\hat{R}_0=
\hat{R}_{I(1)}=\log_2(1+\hat{\gamma}_{I(1)})=\log_2(1+\hat{\gamma}_0), &  l=0,\\
%\hat{R}_1=
\hat{R}_{I(2)}^{(1)}=\log_2(1+\hat{\gamma}_{I(2)}^{(1)})=\log_2(1+\hat{\gamma}_1), & l=1,\\
\hat{R}_{I(2)}^{(0)}=\text{log}_2(1+\hat{\gamma}_{I(2)}^{(0)})=\log_2(1+\hat{\gamma}_2), & l=2,\\
\hat{R}_{J}^{(1)}=\text{log}_2(1+\hat{\gamma}_{J}^{(1)})=\log_2(1+\hat{\gamma}_3), & l=3,
\end{cases}
\end{equation}
%\begin{equation}\label{hatR_I(1)}
%\hat{R}_{I(1)}=\log_2(1+\hat{\gamma}_{I(1)}),
%\end{equation}
%\begin{equation}\label{hatR_I1}
%\hat{R}_{I(2)}^{(1)}=\log_2(1+\hat{\gamma}_{I(2)}^{(1)}),
%\end{equation}
%\begin{equation}\label{hatR_I0}
%\hat{R}_{I(2)}^{(0)}=\text{log}_2(1+\hat{\gamma}_{I(2)}^{(0)}), 
%\end{equation}
%and
%\begin{equation}\label{hatR_J1}
%\hat{R}_{J}^{(1)}=\text{log}_2(1+\hat{\gamma}_{J}^{(1)}),
%\end{equation}
where the signal-to-interference-and-noise ratio (SINR) $ \gamma_l,\forall l $ is obtained as
\begin{equation}\label{hat_SNR}
\hat{\gamma}_l\!=\!\begin{cases}\!
%\hat{R}_0=
\hat{\gamma}_{I(1)}\!=\!\left| \mathbf{g}_{\textit{RI}}^{\mathrm{H}}\bm{\Lambda}\mathbf{G}\mathbf{w}_0+\hat{\mathbf{h}}_I^{\mathrm{H}}\mathbf{w}_0\right|^2\!/\sigma_I^2
\\\!=\!\!(\left| \mathbf{g}_{\textit{RI}}^{\mathrm{H}}\bm{\Lambda}\!\mathbf{G}\mathbf{w}\!_0\!\right|\!^2\!\!+\!\!L_I^2(\!1\!-\!\sigma_e^2)\!\left|\!\mathbf{1}\!_K^{\mathrm{H}}\mathbf{w}\!_0\!\right|\!^2)/\sigma_I^2, &  l=0,\\
%\hat{R}_1=
\hat{\gamma}_{I(2)}^{(1)}\!=\!\left| \mathbf{g}_{\textit{RI}}^{\mathrm{H}}\bm{\Theta}\mathbf{G}\mathbf{w}_I\right| ^2\!/\!(|\hat{\mathbf{h}}_I^{\mathrm{H}}\mathbf{w}_J|^2\!+\!\sigma_I^2)
\\\!=\!\!\left|\mathbf{g}_{\textit{RI}}^{\mathrm{H}}\bm{\Theta}\!\mathbf{G}\mathbf{w}\!_I\right|\! ^2\!/\!(L_I^2(\!1\!-\!\sigma_e^2)\!\left|\!\mathbf{1}\!_K^{\mathrm{H}}\mathbf{w}\!_J\!\right|\!^2\!\!+\!\sigma_I^2), 
%\frac{\left| \mathbf{g}_{\textit{RI}}^{\mathrm{H}}\bm{\Theta}\mathbf{G}\mathbf{w}_I\right| ^2}{|\hat{\mathbf{h}}_I^{\mathrm{H}}\mathbf{w}_J|^2+\sigma_I^2}, 
& l=1,\\
\hat{\gamma}_{I(2)}^{(0)}\!=\!\left| \mathbf{g}_{\textit{RI}}^{\mathrm{H}}\bm{\Theta}\mathbf{G}\mathbf{w}_I+\hat{\mathbf{h}}_I^{\mathrm{H}}\mathbf{w}_I\right| ^2\!/\sigma_I^2
\\\!=\!\!(\left| \mathbf{g}_{\textit{RI}}^{\mathrm{H}}\bm{\Theta}\!\mathbf{G}\mathbf{w}\!_I\!\right| \!^2\!+\!\!L_I^2(\!1\!-\!\sigma_e^2)\!\left|\!\mathbf{1}\!_K^{\mathrm{H}}\mathbf{w}\!_I\!\right|\!^2)/\sigma_I^2, & l=2,\\
\hat{\gamma}_J^{(1)}\!=\!|\hat{\mathbf{h}}_J^{\mathrm{H}}\mathbf{w}_J|^2\!/\!\left( |\mathbf{g}_{\textit{RJ}}^{\mathrm{H}}\bm{\Theta}\mathbf{G}\mathbf{w}_I|^2\!+\!\sigma_J^2\right)
\\\!=\!\!L_I^2(\!1\!-\!\sigma_e^2)\!\left|\!\mathbf{1}\!_K^{\mathrm{H}}\mathbf{w}\!_I\!\right|\!\!^2\!/\!\left( |\mathbf{g}_{\textit{RJ}}^{\mathrm{H}}\bm{\Theta}\!\mathbf{G}\mathbf{w}\!_I\!|^2\!\!+\!\sigma_J^2\right).
%\frac{|\hat{\mathbf{h}}_J^{\mathrm{H}}\mathbf{w}_J|^2}{|\mathbf{g}_{\textit{RJ}}^{\mathrm{H}}\bm{\Theta}\mathbf{G}\mathbf{w}_I|^2+\sigma_J^2}, 
& l=3,
\end{cases}
\end{equation}
where $ \hat{\mathbf{h}}_I=L_I\hat{\mathbf{f}}_I $ and $ \hat{\mathbf{h}}_J=L_J\hat{\mathbf{f}}_J $. 
%\begin{equation}\label{hatSNRI(1)}
%\hat{\gamma}_{I(1)}=\left| \mathbf{g}_{\textit{RI}}^{\mathrm{H}}\bm{\Lambda}\mathbf{G}\mathbf{w}_0+\hat{\mathbf{h}}_I^{\mathrm{H}}\mathbf{w}_0\right|^2\!/\sigma_I^2,
%\end{equation}
%\begin{equation}\label{hatSNRI1}
%\hat{\gamma}_{I(2)}^{(1)}=\frac{\left| \mathbf{g}_{\textit{RI}}^{\mathrm{H}}\bm{\Theta}\mathbf{G}\mathbf{w}_I\right| ^2}{|\hat{\mathbf{h}}_I^{\mathrm{H}}\mathbf{w}_J|^2+\sigma_I^2},
%\end{equation}
%\begin{equation}\label{hatSNRI0}
%\hat{\gamma}_{I(2)}^{(0)}=\left| \mathbf{g}_{\textit{RI}}^{\mathrm{H}}\bm{\Theta}\mathbf{G}\mathbf{w}_I+\hat{\mathbf{h}}_I^{\mathrm{H}}\mathbf{w}_I\right| ^2/\sigma_I^2,
%\end{equation}
%\begin{equation}\label{hatSNRJ}
%\hat{\gamma}_J^{(1)}=\frac{|\hat{\mathbf{h}}_J^{\mathrm{H}}\mathbf{w}_J|^2}{|\mathbf{g}_{\textit{RJ}}^{\mathrm{H}}\bm{\Theta}\mathbf{G}\mathbf{w}_I|^2+\sigma_J^2}.
%\end{equation}
In particular, $ l\!=\!0 $ represents the communication scenario during ISAC period where BS communicates with CU $ I $; $ l\!=\!1 $ represents the communication scenario during PC period where BS communicates with CU $ I $ when it detects TC $ J $ being present; $ l\!=\!2 $ represents the communication scenario during PC period where BS communicates with CU $ I $ when it detects TC $ J $ being absent ; $ l\!=\!3 $ represents the communication scenario during PC period where BS communicates with the TC $ J$  who is actually present and is correctly detected.
%	about $ \hat{\mathbf{h}}_I $ and $ \hat{\mathbf{h}}_J $ 
We observe that the effects of CSI estimation errors $ \sigma_e^2 $
on communication performances vary from users and transmission periods. 
Especially for $ \hat{\gamma}_{I(2)}^{(1)} $, the estimation error $ \sigma_e^2 $ affects the interference link rather than the signal links as in $ \hat{\gamma}_{I(1)} $, $ \hat{\gamma}_{I(2)}^{(0)} $, and $ \hat{\gamma}_{J}^{(1)} $. The reason is that when BS detects TC $ J $ being present and then transmits $ s_2 $ to TC $ J $, the CU $ I $ inevitably experiences interference from $ s_2 $ through the direct BS-CU $ I $ link. 
Therefore, the CSI estimation error $ \sigma_e^2 $ of $ \mathbf{h}_I $ effects the interference experienced by CU $ I $ when $ d=1 $ during the PC period.
Following Eq. \eqref{hat_SNR}, the probability of  $ \hat{R}_l,  l\in\left\lbrace1,2,3 \right\rbrace  $ is calculated as 
%\begin{equation}
$ P_l=\!\begin{cases}
P_{I(2)}^{(1)}=P_{11}+P_{01} , &l=1,\\
P_{I(2)}^{(0)}=P_{10}+P_{00} & l=2,\\
P_J^{(1)}=P_{11} , & l=3.
\end{cases} $
%\end{equation}
%$ P_{I(2)}^{(1)}=P_{11}+P_{01} $, $ P_{I(2)}^{(0)}=P_{10}+P_{00} $, and $ P_J^{(1)}=P_{11} $, based on the probabilities of four cases given in sensing model. 
%The probabilities of  $ \hat{R}_{I(2)}^{(1)} $, $ \hat{R}_{I(2)}^{(0)} $, and $ \hat{R}_{J}^{(1)} $ can be given by $ P_{I(2)}^{(1)}=P_{11}+P_{01} $, $ P_{I(2)}^{(0)}=P_{10}+P_{00} $, and $ P_J^{(1)}=P_{11} $, based on the probabilities of four cases given in sensing model.  

%detection and false alarm probabilities of TC $ J $ obtained at the ISAC period, we can get the probabilities of $ \hat{R}_I^{(1)} $, $ \hat{R}_I^{(0)} $, and $ \hat{R}_J $ as
%$ 
%\begin{equation}\label{PI1}
%\begin{split}
%P_I^{(1)}&=\Pr(a=1,b=1)+\Pr(a=0,b=1)\\
%&=P_{H_1}P_d+P_{H_0}P_f,
%\end{split}
%\end{equation}
%%$ 
%%$
%\begin{equation}\label{PI0}
%\begin{split}
% P_I^{(0)}&=\Pr(a=1,b=0)+\Pr(a=0,b=0)\\
% &=P_{H_1}(1-P_d)+P_{H_0}(1-P_f),
%\end{split}
%\end{equation}
%% $
%%and
%\begin{equation}\label{PJ}
%P_J=\Pr(a=1,b=1)=P_{H_1}P_d,
%\end{equation}
%respectively. $ P_{H_1} $ and $ P_{H_0} $ represent the probabilities of TC $ J $ being truly present and absent, respectively. 
%Ultimately, 
%Then, the average systemic throughput $ \hat{R}_T $ of the proposed R-SACE mechanism obtained with imperfect CSI during the entire timeline can be given by
%% of the R-SACE mechanism with imperfect CSI can be given by
%\begin{equation}\label{hat_R_SIGMA}
%\hat{R}_T=\frac{\tau\!-\!\tau_{1}}{\tau}\left( P_I^{(1)}\hat{R}_I^{(1)}+P_I^{(0)}\hat{R}_I^{(0)}+P_J\hat{R}_J\right) .
%\end{equation}

It's worth mentioning that data rates obtained with imperfect CSI may exceed achievable data rates obtained with perfect CSI due to estimation errors. 
Therefore, to address this issue,  as referred to \cite{estimation-error-model3},
we use average systemic throughput (AST) to evaluate the systemic communication performance of the proposed R-SACE mechanism obtained with imperfect CSI during the entire timeline.
%Accordingly, 
The AST can be calculated as 
%this paper utilizes the average outage sum capacity (AOSC) $ O_\Sigma $ to evaluate the systemic communication performance of the proposed R-SACE mechanism during the PC period. 
%Mathematically, we have 
\begin{equation}\label{O_T2}
\begin{split}
\!\!\!\!O_T\!=\!\!\frac{\tau_{1}}{\tau}\!\hat{R}_0\!\Pr\!\left[\!\hat{R}_0\!\!\le\! \!R_0|\hat{\mathbf{f}}_0 \right]\!\!+\!\!\frac{\tau\!\!-\!\tau_{1}}{\tau}\!\sum_{l=1}^{3}\!\hat{R}_lP_l\Pr\!\left[ \hat{R}_l\!\le\! R_l|\hat{\mathbf{f}_l}\right]
%\\
%=&\frac{\tau_{1}}{\tau}\hat{R}_0\Pr\left[\hat{R}_0\le R_0|\hat{\mathbf{f}}_0 \right] \\
%&+\frac{\tau\!-\!\tau_{1}}{\tau}\hat{R}_1P_1\Pr\left[\hat{R}_1\le R_1|\hat{\mathbf{f}}_1 \right] \\
%&+\frac{\tau\!-\!\tau_{1}}{\tau}\hat{R}_2P_2\Pr\left[\hat{R}_2\le R_2|\hat{\mathbf{f}}_2 \right]\\
%&+\frac{\tau\!-\!\tau_{1}}{\tau}\hat{R}_3P_3\Pr\left[\hat{R}_3\le R_3|\hat{\mathbf{f}}_3 \right]\\
\end{split}
\end{equation}
where $ \Pr\left[\hat{R}_l\le R_l|\hat{\mathbf{f}}_l \right], \forall l $
%$ \Pr\left[\hat{R}_{I(1)}\le R_{I(1)}|\hat{\mathbf{f}}_I \right] $, $ \Pr\left[\hat{R}_{I(2)}^{(1)}\!\le\! R_{I(2)}^{(1)}|\hat{\mathbf{f}}_I \right] $, $ \Pr\left[\hat{R}_{I(2)}^{(0)}\!\le\! R_{I(2)}^{(0)}|\hat{\mathbf{f}}_I \right] $, and $\Pr\left[\hat{R}_J^{(1)}\!\le\! R_J^{(1)}|\hat{\mathbf{f}}_J \right]  $ 
represent the probability that the data rate obtained with imperfect CSI deso not exceed the achievable data rate obtained with perfect CSI in each communication case. $ \hat{\mathbf{f}}_0\!=\!\hat{\mathbf{f}}_1\!=\! \hat{\mathbf{f}}_2\!=\!\hat{\mathbf{f}}_I $,$ \hat{\mathbf{f}}_3\!=\!\hat{\mathbf{f}}_J $.  
$ R_l\!=\!\begin{cases}
 R_{I(1)} , & \!\!l=0,\\
 R_{I(2)}^{(1)}, &\!\! l=1,\\
 R_{I(2)}^{(0)}, &\!\! l=2,\\
 R_J^{(1)}, &\!\! l=3,
\end{cases} $ is the achievable data rate obtained with perfect CSI.
We notice that the sensing performance, i.e., $ P_l $, obtained at the ISAC period takes important impacts on the average systemic throughput obtained at the PC period, i.e., the second term in Eq. \eqref{O_T2}.
Particularly, the communication performance of TC $ J $ is guaranteed only when the BS correctly detects the present TC $ J $. Therefore, the detection accuracy should be improved through effective methods, such as the adaptive threshold method\cite{Adaptive Threshold} or increasing sensing time $ \tau_1 $, i.e., the duration of ISAC period. 
Nevertheless, the increased $ \tau_1 $ causes the decreased duration of PC period, which thus degrades the communication performance at the PC period and consequently degrades the AST during the entire timeline. As a result, the S\&C balance in time allocation is a crucial issue for maximizing AST of the proposed R-SACE mechanism, which will be addressed in detail in the following section.

%\textcolor{blue}{To simplify notations, we give a subscript  $ l\in\left\lbrace0,1,2,3 \right\rbrace  $ to denote parameters in the following four communication cases.
%%	in Eqs. \eqref{hatR_I(1)}-\eqref{hatR_J1}. 
%Specifically,
%\begin{itemize}
%	\item \textit{$l=0 $} represents the parameters in the case that the communication of UE $ I $ performs during ISAC period. 
%%	Accordingly, $ \hat{R}_{I(1)} $ and $ \hat{\gamma}_{I(1)}  $ can be  re-expressed as $ \hat{R}_0 $ and $ \hat{\gamma}_0 $.
%	\item \textit{$l=1 $} represents the parameters in the case that the communication of UE $ I $ performs during PC period when $ d=1 $.
%	\item \textit{$l=2 $} represents the parameters in the case that the communication of UE $ I $ performs during PC period when $ d=0 $. 
%	\item \textit{$ l=3 $} represents the parameters in the case that the communication of the present TC $ J $ performs during PC period when $ d=1 $. 
%\end{itemize}}

%\textcolor{red}{As a result, the AST in Eq. \eqref{O_T} can be rewritten as }

\section{Problem Statement and Transformation}
\subsection{Problem Statement}
The goal of this paper is to maximize the systemic communication performance with imperfect CIS in the proposed R-SACE mechanism with assistance of RIS and the performance-guaranteed sensing result obtained in the ISAC period.
To be specific, we aim to maximize the AST $ O_T $ by jointly optimizing the time allocation $ \tau_1 $, the reflecting precoding $ \bm{\Theta} $ of RIS, and the downlink beamforming vectors $ \mathbf{w}_I $ and $ \mathbf{w}_J $ of the BS, under the probabilistic constraint and the sensing performance, transmission power, and communication interference constraints.
In a word, the AST maximization problem can be formulated as
\begin{alignat}{2}
(\mathbf{P1\!:}) \max_{\tau_1,\bm{\Theta},\mathbf{w}_\textit{I},\mathbf{w}_\textit{J}}& O_T
%\left( \tau_1,\bm{\Theta},\mathbf{w}_\textit{I},\mathbf{w}_\textit{J}\right) 
%(\tau_1,\bm{\Theta},\mathbf{w}_I,\mathbf{w}_J)
\label{p00}\\\nonumber
%\mbox{s.t.}\quad &C1: \Pr\left[ R_I^{(1)}<\hat{R}_I^{(1)}|\hat{\mathbf{f}}_I\right]\le\epsilon_{\mathrm{\mathrm{out}}}, 
%%&\tag{\ref{p00}{a}}\label{p00a}
%\\\nonumber
%&C2: \Pr\left[ R_I^{(0)}<\hat{R}_I^{(0)}|\hat{\mathbf{f}}_I\right]\le\epsilon_{\mathrm{\mathrm{out}}}, 
%%&\tag{\ref{p00}{b}}\label{p00b}
%\\\nonumber
%&C3: \Pr\left[ R_J<\hat{R}_J|\hat{\mathbf{f}}_J\right]\le\epsilon_{\mathrm{out}}, 
%%&\tag{\ref{p00}{c}}\label{p00c}
%\\\nonumber
%&C4:  P_{d}(\tau_1)\ge \bar{P_d}
%%&\tag{\ref{p00}{d}}\label{p00d}
%\\\nonumber
%& C5: P_{f}(\tau_1)\le \bar{P_f}
%%&\tag{\ref{p00}{e}}\label{p00e}
%\\ \nonumber
%& C6: \left\| \mathbf{w}_I\right\|^2p_I\le p_{\textit{Imax}}
%% &\tag{\ref{p00}{f}}\label{p00f}
% \\\nonumber
%&C7:  \left\| \mathbf{w}_J\right\|^2p_J\le p_{\textit{Jmax}}
%%&\tag{\ref{p00}{g}}\label{p00g} 
%\\ \nonumber
%& C8: |\mathbf{g}_{RJ}^{\mathrm{H}}\bm{\Theta}\mathbf{G}\mathbf{w}_\textit{I}|^2p_I\le \bar{\epsilon}_J
%%&\tag{\ref{p00}{h}}\label{p00h}
%\\
%\nonumber
%&C9: 0<\tau_1\le \tau_\Sigma, 
%%&\tag{\ref{p00}{i}}\label{p00i}
\mbox{s.t.}\quad &C1: \Pr\left[ R_l<\hat{R}_l|\hat{\mathbf{f}}_l\right]\le\epsilon_{\mathrm{out}}, \forall l
\\\nonumber
&C2:  P_{d}(\tau_1)\ge \bar{P_d}
%&\tag{\ref{p00}{d}}\label{p00d}
\\\nonumber
& C3: P_{f}(\tau_1)\le \bar{P_f}
%&\tag{\ref{p00}{e}}\label{p00e}
\\ \nonumber
& C4: \left\| \mathbf{w}_I\right\|^2\le p_{\mathrm{Imax}}
% &\tag{\ref{p00}{f}}\label{p00f}
\\\nonumber
&C5:  \left\| \mathbf{w}_J\right\|^2\le p_{\mathrm{Jmax}}
%&\tag{\ref{p00}{g}}\label{p00g} 
\\ \nonumber
& C6: |\mathbf{g}_{\textit{RJ}}^{\mathrm{H}}\bm{\Theta}\mathbf{G}\mathbf{w}_\textit{I}|^2\le \bar{\epsilon}_J
%&\tag{\ref{p00}{h}}\label{p00h}
\\
\nonumber
&C7: 0<\tau_1\le \tau, 
%&\tag{\ref{p00}{i}}\label{p00i}
\end{alignat}
where $ \epsilon_{\mathrm{out}} $ is the probability threshold;  
$ \bar{\epsilon}_J $ is the threshold of the tolerable interference of TC $ J $. 
$ p_{\mathrm{Imax}} $ and $ p_{\mathrm{Jmax}} $ represent the maximum transmission power of the BS. 
$ \bar{P_d} $ and $ \bar{P_f} $ represent the thresholds of detection probability and false alarm probability, respectively. 
In $ \mathbf{P1} $, $ C1 $ restricts the probability that the data rates obtained with imperfect CSI exceed the achievable data rates to less than $ \epsilon_{\mathrm{out}} $. 
$ C2 $ and $ C3 $ guarantee the sensing performance. 
$ C4 $ and $ C5 $ are transmission power limit constraints. 
$ C6 $ is the interference constraint and $ C7 $ gives the available region of $ \tau_1 $. 

For solving the non-convex probabilistic mixed problem effectively, we first transform $ \mathbf{P1} $ into a non-probabilistic optimization problem by incorporating the probabilistic constraint $ C1 $ into the objective function.

\subsection{Problem Transformation}
We first rewrite $ \hat{R}_l=\log_2(1+\hat{\gamma}_l)=\log_2(1+\frac{\hat{c}_l}{\hat{d}_l}) $ and $ R_l=\log_2(1+\gamma_l)=\log_2(1+\frac{c_l}{d_l}) $. Then, according to the total probability theorem\cite{estimation-error-model3}, the constraint $ C1 $ in Eq. \eqref{p00} can be recast as
\begin{equation}\label{gamma-constraint}
\begin{split}
 &\quad\Pr\left[\gamma_l<\hat{\gamma}_l|\hat{\mathbf{f}}_l \right] 
% \le \epsilon_{out}
% \\
% &
% \Leftrightarrow
 =\Pr\left[\frac{c_l}{d_l} <\frac{\hat{c}_l}{\hat{d}_l}|\hat{\mathbf{f}}_l \right]
 \\
% \le \epsilon_{out}
% \\
% &\Leftrightarrow\Pr\left[\frac{c_l}{d_l}<\left( 2^{\hat{R}_l}-1\right) |\hat{\mathbf{f}}_l \right] 
%\le \epsilon_{out}, \forall l.
%\end{split}
%\end{equation}
%And thus, we have $ R_l=\log_2(1+\gamma_l)=\log_2(1+\frac{c_l}{d_l}) $ and $ \hat{R}_l=\log_2(1+\hat{\gamma}_l)=\log_2(1+\frac{\hat{c}_l}{\hat{d}_l}) $.  
%We further rewrite $ \hat{\gamma}_l $ as $ \hat{\gamma}_l=2^{\hat{R}_l}-1 $ and the constraint in Eq. \eqref{gamma-constraint} can be rewritten as
%\begin{equation}
% \Pr\left[\frac{c_l}{d_l}<\frac{\hat{c}_l}{\hat{d}_l}|\hat{\mathbf{f}}_l \right] =\Pr\left[\frac{c_l}{d_l}<2^{\hat{R}_l}-1|\hat{\mathbf{f}}_l \right] \le \epsilon_{out}.
%\end{equation}
%According to the total probability theorem\cite{estimation-error-model3}, we have
%$ \Pr\left[ R_l<\hat{R}_l|\hat{\mathbf{f}}_l\right] $ 
%as
%\begin{equation}
%\begin{split}
%&
%\Pr\left[\frac{c_l}{d_l}<\frac{\hat{c}_l}{\hat{d}_l}|\hat{\mathbf{f}}_l \right]=
%\Pr\left[\frac{c_l}{d_l}<\left( 2^{\hat{R}_l}-1\right) |\hat{\mathbf{f}}_l \right] \\
&=\Pr\left[\frac{c_l}{d_l}< 2^{\hat{R}_l}-1 |c_l\le\hat{c}_l,\hat{\mathbf{f}}_l \right]\cdot\Pr\left[ c_l\le\hat{c}_l|\hat{\mathbf{f}}_l\right] \\
&\quad+\Pr\left[\frac{c_l}{d_l}< 2^{\hat{R}_l}-1 |c_l>\hat{c}_l,\hat{\mathbf{f}}_l \right]\cdot\Pr\left[ c_l>\hat{c}_l|\hat{\mathbf{f}}_l\right]\\
&\le\epsilon_{\mathrm{out}}
%&=\Pr\left[\frac{c_l}{d_l}<\frac{\hat{c}_l}{\hat{d}_l} |c_l\le\hat{c}_l,\hat{\mathbf{f}}_l \right]\cdot\Pr\left[ c_l\le\hat{c}_l|\hat{\mathbf{f}}_l\right] \\
%&\quad+\Pr\left[\frac{c_l}{d_l}<\frac{\hat{c}_l}{\hat{d}_l} |c_l>\hat{c}_l,\hat{\mathbf{f}}_l \right]\cdot\Pr\left[ c_l>\hat{c}_l|\hat{\mathbf{f}}_l\right]\\
%&\le\epsilon_{out}
\end{split}
\end{equation}

\textbf{\textit{Proposition 1:}}  Following \cite{estimation-error-model1,estimation-error-model2,estimation-error-model3}, the constraint $ C1 $ can be further derived from the following two equations
%}
\begin{equation}\label{di>di}
\Pr\left[ d_l\ge \hat{d}_l|\hat{\mathbf{f}}_l\right] \le \epsilon_{\mathrm{out}}/2
\end{equation}
\begin{equation}\label{ci<ci}
\Pr\left[ c_l\le \hat{c}_l|\hat{\mathbf{f}}_l\right] =\epsilon_{\mathrm{out}}/2.
\end{equation}

\textbf{\textit{Proof:}} Please refer to Appendix A.\quad\quad\quad\quad\quad\quad\quad\quad\quad$ \blacksquare $

Then, we integrate the probabilistic constraints in Eqs. \eqref{di>di} and \eqref{ci<ci}  into the objective function $ O_T $ of the problem $ \mathbf{P1} $.
%, aiming at transforming the probability-constrained optimization problem into a non-probabilistic one. 

We first analyze the Eq. \eqref{di>di}. 
Based on Eq. \eqref{hat_SNR},
%the definitions and transformations of $ \gamma_l $ and $ \hat{\gamma}_l $, 
we know that $ d_0=\hat{d}_0=\sigma_I^2 $, $ d_1=|\mathbf{h}_I^{\mathrm{H}}\mathbf{w}_J|^2+\sigma_I^2 $, $  $$ \hat{d}_1=|\mathbf{\hat{h}}_I^{\mathrm{H}}\mathbf{w}_J|^2+\sigma_I^2 $, $ d_2=\hat{d}_2=\sigma_I^2 $, $ d_3=\hat{d}_3=|\mathbf{g}_{\textit{RJ}}^{\mathrm{H}}\bm{\Theta}\mathbf{G}\mathbf{w}_I|^2+\sigma_J^2 $. Therefore, we only need to analyze $\Pr\left[ d_1\ge \hat{d}_1|\hat{\mathbf{f}}_1\right] \le \epsilon_{\mathrm{out}}/2 $. 
%where $ \Pr[ d_l\ge \hat{d}_l|\hat{\mathbf{f}}_l,l=1]\le \epsilon_{\mathrm{out}}/2  $.  
By utilizing the Markov inequality theorem\cite{estimation-error-model2,estimation-error-model3}, we have 
\begin{equation}\label{d1>d1}
\begin{split}
\Pr\left[d_1\ge \hat{d}_1|\hat{\mathbf{f}}_1\right]&\!=\!\Pr\left[\left| \mathbf{h}_I^{\mathrm{H}}\mathbf{w}_J\right| ^2\ge \left( \hat{d}_1-\sigma_I^2\right) |\hat{\mathbf{f}}_1\right]\\
&\le\!\frac{\mathbb{E}\left[ \left| \mathbf{h}_I^{\mathrm{H}}\mathbf{w}_J\right| ^2\right] }{\hat{d}_1-\sigma_I^2}\le\! \frac{\mathbb{E}\left[ \left\| \mathbf{h}_I\right\| ^2\left\| \mathbf{w}_J\right\| ^2\right] }{\hat{d}_1-\sigma_I^2}\\
&=\frac{L_I^2\left\| \mathbf{f}_I\right\| ^2\left\| \mathbf{w}_J\right\| ^2}{\hat{d}_1-\sigma_I^2}.
\end{split}
\end{equation}
Based on  $ \hat{\gamma}_1=\frac{\hat{c}_1}{\hat{d}_1}  $, we substitute $ \hat{d}_1=\hat{c}_1/\hat{\gamma}_1 $ and Eq. \eqref{d1>d1} into Eq. \eqref{di>di} and then we have
\begin{equation}\label{derive_gamma1}
\begin{split}
\frac{L_I^2\left\| \mathbf{f}_I\right\| ^2\!\left\| \mathbf{w}_J\right\| ^2\!}{\hat{d}_1-\sigma_I^2}\!=\!\frac{L_I^2(\|\hat{\mathbf{f}}_I\| ^2\!+\!\left\| \mathbf{1}\right\| ^2\!\sigma_e^2)\!\left\| \mathbf{w}_J\right\| ^2\!}{\hat{c}_1/\hat{\gamma}_1-\sigma_I^2}\!=\!\frac{\epsilon_{\mathrm{out}}}{2}.
\end{split}
\end{equation}
According to Eq. \eqref{derive_gamma1}, we get the transformed expression $ \tilde{\gamma}_1 $ of $ \hat{\gamma}_1 $ as
\begin{equation}\label{tildegamma1}
\tilde{\gamma}_1\triangleq\hat{\gamma}_1=\frac{\epsilon_{\mathrm{out}}\hat{c}_1}{2L_I^2(\|\hat{\mathbf{f}}_I\| ^2\!+\!\left\| \mathbf{1}_K\right\| ^2\!\sigma_e^2)\left\| \mathbf{w}_J\right\| ^2\!+\!\epsilon_{\mathrm{out}}\sigma_I^2},
\end{equation}
where $ \hat{c}_1=c_1=|\mathbf{g}_{\textit{RI}}^{\mathrm{H}}\bm{\Theta}\mathbf{G}\mathbf{w}_I|^2$.

Next, we analyze Eq. \eqref{ci<ci}. 
Considering $ c_1=\hat{c}_1$, we only need to analyze $ \Pr[ c_l\le \hat{c}_l|\hat{\mathbf{f}}_l,l=0,2,3]=\epsilon_{\mathrm{\mathrm{out}}}/2$.  According to Eq. \eqref{hat_SNR}, we have
\begin{equation}\label{ci<ci-transform3}
\begin{split}
\Pr\!\left[ c_l\le \hat{c}_l|\hat{\mathbf{f}}_l,l=3\right]\!=\!\Pr\!\left[ |\mathbf{h}_J^{\mathrm{H}}\mathbf{w}_J|^2\le \hat{c}_3 |\hat{\mathbf{f}}_J\right]\!=\epsilon_{\mathrm{out}}/2,
\end{split}
\end{equation}
and
\begin{equation}\label{ci<ci-transform1}
\begin{split}
&\Pr\left[ c_l\le \hat{c}_l|\hat{\mathbf{f}}_l,l=0,2\right]\\
=&\Pr\left[ |\mathbf{h}_l^{\mathrm{H}}\mathbf{w}_l|^2\le \left( \hat{c}_l-A_l\right) |\hat{\mathbf{f}}_l,l=0,2\right]=\epsilon_{\mathrm{out}}/2,
\end{split}
\end{equation}
where $ |\mathbf{h}_0^{\mathrm{H}}\mathbf{w}_0|^2=|\mathbf{h}_I^{\mathrm{H}}\mathbf{w}_0|^2 $, $ |\mathbf{h}_2^{\mathrm{H}}\mathbf{w}_2|^2=|\mathbf{h}_I^{\mathrm{H}}\mathbf{w}_I|^2 $, $ A_0=\left|  \mathbf{g}_{\textit{RI}}^{\mathrm{H}}\bm{\Lambda}\mathbf{G}\mathbf{w}_0\right|^2 $, $ A_2=\left|  \mathbf{g}_{\textit{RI}}^{\mathrm{H}}\bm{\Theta}\mathbf{G}\mathbf{w}_I\right|^2 $.  
Since $ |\mathbf{h}_l^{\mathrm{H}}\mathbf{w}_l|^2\le \|\mathbf{h}_l\|^2\|\mathbf{w}_l\|^2 $ and $ |\mathbf{h}_J^{\mathrm{H}}\mathbf{w}_J|^2\le \|\mathbf{h}_J\|^2\|\mathbf{w}_J\|^2 $, Eq. \eqref{ci<ci-transform3} and Eq. \eqref{ci<ci-transform1} can be respectively refined as
\begin{equation}\label{PR-fJ<}
\begin{split}
\Pr\!\left[\|\mathbf{h}_J\|^2\|\mathbf{w}\!_J\|^2\!\le\!\hat{c}_3 |\hat{\mathbf{f}}\!_J\right]
%=&\Pr\left[ L_l^2\|\mathbf{f}_l\|^2\|\mathbf{w}_l\|^2\le \left( \hat{c}_l-A_l\right) |\hat{\mathbf{f}}_l,l=0,2\right] \\
\!\!=\!\!\Pr\!\left[ \|\mathbf{f}_J\|^2\!\le\frac{\hat{c}_3 }{L_J^2\|\mathbf{w}\!_J\|^2}|\hat{\mathbf{f}}_J\right],
\end{split}
\end{equation}
and
\begin{equation}\label{PR-fi<}
\begin{split}
&\Pr\left[ \|\mathbf{h}_l\|^2\|\mathbf{w}_l\|^2\le \left( \hat{c}_l-A_l\right) |\hat{\mathbf{f}}_l,l=0,2\right]\\
%=&\Pr\left[ L_l^2\|\mathbf{f}_l\|^2\|\mathbf{w}_l\|^2\le \left( \hat{c}_l-A_l\right) |\hat{\mathbf{f}}_l,l=0,2\right] \\
=&\Pr\left[  \|\mathbf{f}_l\|^2\le \frac{\left( \hat{c}_l-A_l\right) }{L_l^2\|\mathbf{w}_l\|^2}|\hat{\mathbf{f}}_l,l=0,2\right].
\end{split}
\end{equation}

According to Eqs. \eqref{f_I} and \eqref{f_j} and the distributions of $\hat{\mathbf{f}}_I $, $\hat{\mathbf{f}}_J$,
%$ \hat{\mathbf{f}}_0=\hat{\mathbf{f}}_2=\hat{\mathbf{f}}_I $, 
$ \mathbf{e}_I $, $ \mathbf{e}_J $, 
%and $ \mathbf{e}_J $, 
we derive that $ \|\mathbf{f}_l\|^2\!\sim\! \mathcal{CN}(\|\hat{\mathbf{f}}_l\|^2,\|\mathbf{1}_K\|^2\sigma_e^2), l\in\left\lbrace 0,2,3\right\rbrace   $ follows a non-central chi-square distribution with two degrees of freedom\cite{estimation-error-model3}. Then, based on the cumulative distribution function (CDF) $ \mathcal{F} $ of the non-central chi-squared distributed $ \|\mathbf{f}_l\|^2, l\in\left\lbrace 0,2,3\right\rbrace $, we respectively derive Eqs. \eqref{PR-fi<} and \eqref{PR-fJ<} as
\begin{equation}\label{cdf-fi}
\begin{split}
&\Pr\left[  \|\mathbf{f}_l\|^2\le \frac{\left( \hat{c}_l-A_l\right) }{L_l^2\|\mathbf{w}_l\|^2}|\hat{\mathbf{f}}_l,l=0,2\right]=\frac{\epsilon_{\mathrm{out}}}{2}\\
&\!=\!\mathcal{F}_{\|\mathbf{f}_l\|^2}\left( \frac{ \hat{c}_l-A_l}{L_l^2\|\mathbf{w}_l\|^2}\right), l\!=\!0,2 \\
&\!=\!1\!-\!Q_1\!\!\left(\! \sqrt{\!\frac{2\|\hat{\mathbf{f}}_l\|^2}{\|\mathbf{1}_K\|^2\sigma_e^2}},\sqrt{\!\frac{2\left( \hat{c}_l-A_l\right)}{L_l^2\|\mathbf{w}_l\|^2\|\mathbf{1}_K\|^2\sigma_e^2}}\right)\!,l\!=\!0,2
\end{split}
\end{equation}
and
\begin{equation}\label{cdf-fj}
\begin{split}
&\Pr\left[  \|\mathbf{f}_J\|^2\le \frac{\hat{c}_3 }{L_J^2\|\mathbf{w}_J\|^2}|\hat{\mathbf{f}}_J\right]=\frac{\epsilon_{\mathrm{out}}}{2}\\
&=\mathcal{F}_{\|\mathbf{f}_J\|^2}\left( \frac{ \hat{c}_3}{A_J^2\|\mathbf{w}_J\|^2}\right), \\
&=1-Q_1\left( \sqrt{\frac{2\|\hat{\mathbf{f}}_J\|^2}{\|\mathbf{1}_K\|^2\sigma_e^2}},\sqrt{\frac{2 \hat{c}_3}{L_J^2\|\mathbf{w}_J\|^2\|\mathbf{1}_K\|^2\sigma_e^2}}\right),
\end{split}
\end{equation}
where $ Q_M(x,y)=\exp(-\frac{x^2+y^2}{2})\sum_{k=1-M}^{\infty} (\frac{x}{y})^k\textit{I}_k(xy)$ is the Marcum Q-function. $ I_k$ is the first-kind $ k $-th order modified Bessel function. 
Then, based on Eqs. \eqref{cdf-fi} and \eqref{cdf-fj}, we have 
\begin{equation}
\hat{c}_l=\mathcal{F}_{\|\mathbf{f}_l\|^2}^{-1}\left(\frac{\epsilon_{\mathrm{out}}}{2}\right) L_l^2\|\mathbf{w}_l\|^2+A_l, l=0,2
\end{equation}
\begin{equation}
\hat{c}_3=\mathcal{F}_{\|\mathbf{f}_J\|^2}^{-1}\left(\frac{\epsilon_{\mathrm{out}}}{2}\right) L_J^2\|\mathbf{w}_J\|^2,
\end{equation}
where $ \mathcal{F}^{-1} $ is the inverse function of $ \mathcal{F}$. 
As a result, we can derive the transformed expressions $ \tilde{\gamma}_l$ of $ \hat{\gamma}_l, l\in\left\lbrace 0,2,3\right\rbrace  $ as
\begin{equation}\label{tildegamma0}
\tilde{\gamma}_0\triangleq\hat{\gamma}_0\!=\!\frac{\hat{c}_0}{\hat{d}_0}\!=\!\frac{\mathcal{F}_{\|\mathbf{f}_I\|^2}^{-1}\!\!\left(\frac{\epsilon_{\mathrm{out}}}{2}\right)\!L_I^2\|\mathbf{w}_0\|^2\!+\!\left|  \mathbf{g}_{\textit{RI}}^{\mathrm{H}}\bm{\Lambda}\mathbf{G}\mathbf{w}_0\right|^2}{\sigma_I^2},
\end{equation}
\begin{equation}\label{tildegamma2}
\tilde{\gamma}_2\triangleq\hat{\gamma}_2\!=\!\frac{\hat{c}_2}{\hat{d}_2}\!=\!\frac{\mathcal{F}_{\|\mathbf{f}_I\|^2}^{-1}\!\!\left(\frac{\epsilon_{\mathrm{out}}}{2}\right)\!L_I^2\|\mathbf{w}_I\|^2\!+\!\left|  \mathbf{g}_{\textit{RI}}^{\mathrm{H}}\bm{\Theta}\mathbf{G}\mathbf{w}_I\right|^2}{\sigma_I^2},
\end{equation}
\begin{equation}\label{tildegamma3}
\tilde{\gamma}_3\triangleq\hat{\gamma}_3\!=\!\frac{\hat{c}_3}{\hat{d}_3}\!=\!\frac{\mathcal{F}_{\|\mathbf{f}_J\|^2}^{-1}\!\!\left(\frac{\epsilon_{\mathrm{out}}}{2}\right)L_J^2\|\mathbf{w}_J\|^2}{|\mathbf{g}_{\textit{RJ}}^{\mathrm{H}}\bm{\Theta}\mathbf{G}\mathbf{w}_I|^2+\sigma_J^2}.
\end{equation}

%\begin{equation}\label{tildegamma3}
%\tilde{\gamma}_2\triangleq\hat{\gamma}_3=\frac{\hat{c}_3}{\hat{d}_3}=\frac{\mathcal{F}_{\|\mathbf{f}_J\|^2}^{-1}\left(\frac{\epsilon_{out}}{2}\right) A_J^2\|\mathbf{w}_J\|^2p_J}{|\mathbf{g}_{\textit{RJ}}^{\mathrm{H}}\bm{\Theta}\mathbf{G}\mathbf{w}_I|^2p_I+\sigma_J^2}.
%\end{equation}

Finally, substituting Eqs. \eqref{tildegamma1} and \eqref{tildegamma0}-\eqref{tildegamma3} into the objective function in Eq. \eqref{p00}, 
%the problem $ \mathbf{P1} $ is transformed as  $ \mathbf{P2} $ in Eq. \eqref{p002},
%\begin{alignat}{2}
%(\mathbf{P2\!:}) \max_{\tau_1,\bm{\Theta},\mathbf{w}_\textit{I},\mathbf{w}_\textit{J}} &\tilde{O}_\Sigma(\tau_1,\bm{\Theta},\mathbf{w}_I,\mathbf{w}_J)\label{p002}=\sum_{l=1}^{3}\rho_l\tilde{R}_l\ \\\nonumber
%\mbox{s.t.}\quad &C_2-C_7, 
%\end{alignat}
%where $ \tilde{R}_l=\log_2(1+\tilde{\gamma}_l) $.
%$ \rho_l=r_l\cdot(1-\epsilon_{\mathrm{out}})=\frac{\tau_\Sigma\!-\!\tau_{1}}{\tau_\Sigma}P_l\cdot(1-\epsilon_{\mathrm{out}}) $ and $ P_l $ is specified in Eq. \eqref{p001},
% which concludes the proof. $ \quad\quad\quad\quad\quad\quad\quad\quad\quad\blacksquare $ 
%\textbf{\textit{Proposition 1: }} 
the probabilistic mixed problem $ \mathbf{P1} $ can be recast as the following problem
\begin{alignat}{2}
(\mathbf{P2\!:}) \max_{\tau_1,\bm{\Theta},\mathbf{w}_\textit{I},\mathbf{w}_\textit{J}} &\tilde{O}_T
%(\tau_1,\bm{\Theta},\mathbf{w}_I,\mathbf{w}_J)
%&
=\sum_{l=0}^{3}\rho_l\tilde{R}_l=\sum_{l=0}^{3}\rho_l\log_2\left( 1+\tilde{\gamma}_l\right)  \label{p002}\\\nonumber
\mbox{s.t.}\quad &C_2-C_7, 
\end{alignat}
where 
$ \rho_0
%=r_l(1-\epsilon_{\mathrm{out}})
=\frac{\tau_{1}}{\tau}(1-\epsilon_{\mathrm{out}}) $ and 
$ \rho_l
%=r_l(1-\epsilon_{\mathrm{out}})
=\frac{\tau-\tau_{1}}{\tau}P_l(1-\epsilon_{\mathrm{out}}), l\in\left\lbrace 1,2,3\right\rbrace  $, given $ \Pr\left[\hat{R}_l\le R_l|\hat{\mathbf{f}}_l \right]=1-\epsilon_{\mathrm{out}},\forall l\in\left\lbrace0,1,2,3 \right\rbrace  $.

% $\tilde{\gamma}_1=\frac{\epsilon_{out}|\mathbf{g}_{RI}^{\mathrm{H}}\bm{\Theta}\mathbf{G}\mathbf{w}_I|^2p_I}{2A_I^2(\|\hat{\mathbf{f}}_I\| ^2\!+\!\left\| \mathbf{1}_K\right\| ^2\!\sigma_e^2)\left\| \mathbf{w}_J\right\| ^2\!p_J+\epsilon_{out}\sigma_I^2}$; $ \tilde{\gamma}_2=\frac{\mathcal{F}_{\|\mathbf{f}_I\|^2}^{-1}\left(\frac{\epsilon_{out}}{2}\right) A_I^2\|\mathbf{w}_I\|^2p_I}{\sigma_I^2}$, and $ \tilde{\gamma}_3=\frac{\mathcal{F}_{\|\mathbf{f}_J\|^2}^{-1}\left(\frac{\epsilon_{out}}{2}\right) A_J^2\|\mathbf{w}_J\|^2p_J}{|\mathbf{g}_{RJ}^{\mathrm{H}}\bm{\Theta}\mathbf{G}\mathbf{w}_I|^2p_I+\sigma_J^2} $.

\section{Solution to the Average Systemic Throughput Maximization Problem}
%The iterative optimizations of $ \tau_1 $, $ \bm{\Theta} $, $ \mathbf{w}_I $, and $ \mathbf{w}_J $ are respectively illustrated in detail below.
%  method to decouple the problem $ \mathbf{P0} $ and alternatively optimize $ \tau_m$, $ \bm{\Theta} $, $ \mathbf{w}_I $, and $ \mathbf{w}_J $.
\subsection{Problem Solution}
This papers	propose a fixed-point iterative  (FPI)  algorithm for solving the transformed $ \mathbf{P2} $ by utilizing the QT method\cite{FP-1,FP-2} and typical convex optimization methods.  
The iterative optimizations of $ \tau_1 $, $ \bm{\Theta} $, $ \mathbf{w}_I $, $ \mathbf{w}_J $ are detailed as follows.	
\subsubsection{Optimization of $\tau_1 $}
When $ \bm{\Theta} $, $ \mathbf{w}_I $, and $ \mathbf{w}_J $ are fixed, the optimization of $ \tau_1 $  in $ \mathbf{P2} $ can be expressed as
\begin{alignat}{2}
(\mathbf{P3:}) \max_{\tau_1} &\quad \tilde{O}_T=\sum_{l=0}^{3}\rho_l(\tau_1)\tilde{R}_l\label{p01}\\\nonumber
%&=\frac{\tau_\Sigma-\tau_1}{\tau_\Sigma}\left(\! P_I^{(1\!)}\!(\!\tau_m\!)R_I^{(\!1\!)}\!\!+\!\!P_I^{(\!0\!)}\!(\!\tau_m\!)R_I^{(\!0\!)}\!\!+\!\!P_J^{(\!1\!)}(\!\tau_m\!)R_J^{(\!1\!)}\!\right)\nonumber\\
\mbox{s.t.}
&\quad C2: P_{d}(\tau_1)\ge \bar{P_d}
%&\tag{\ref{p01}{a}}\label{p01a}
\\\nonumber
&\quad C3: P_{f}(\tau_1)\le \bar{P_f}
%&\tag{\ref{p01}{b}}\label{p01b}
\\ \nonumber
&\quad C7: 0<\tau_1\le \tau, 
%&\tag{\ref{p01}{c}}\label{p01c}
\end{alignat}
where $\tilde{R}_l $ is a constant term when $ \bm{\Theta} $, $ \mathbf{w}_I $, and $ \mathbf{w}_J $  are fixed.

\textit{\textbf{Proposition 2:}} There exists a unique optimal $ \tau_1^\ast$ to maximize $ \tilde{O}_T $ in problem $ \mathbf{P3} $.
	
\textit{\textbf{Proof:}} According to Eq. \eqref{p002}, we get that
\begin{equation}
\rho_l (\tau_1)= \frac{\tau_1}{\tau} (1-\epsilon_{\mathrm{out}}), l=0,
\end{equation}
\begin{equation}
	 \rho_l (\tau_1)=\frac{\tau\!-\!\tau_{1}}{\tau}P_l(\tau_1) (1-\epsilon_{\mathrm{out}}),  l\in \left\lbrace 1,2,3\right\rbrace.
\end{equation}
%where $ P_1(\tau_1)$, $ P_2(\tau_1)$, and $P_3(\tau_1) $ are given in Eq. \eqref{O_T2}.
%\begin{equation}
% P_1(\tau_1)=P_I^{(1)} (\tau_1)=P_{H_1}P_d(\tau_1)+P_{H_0}P_f(\tau_1),
%\end{equation} 
%\begin{equation}
% P_2(\tau_1)=P_I^{(0)}(\tau_1)=P_{H_1}(1\!-\!P_d(\tau_1))+P_{H_0}(1\!-\!P_f(\tau_1)), 
%\end{equation}
%\begin{equation}
% P_3(\tau_1)=P_J(\tau_1)=P_{H_1}P_d(\tau_1), 
%\end{equation}
%based on Eqs. \eqref{PI0}-\eqref{PJ}.
Based on the closed-form expressions of $ P_f $, $ P_d $, and $ P_l(\tau_1), \forall l\in\left\lbrace 1,2,3\right\rbrace  $ in Section II, the proof of Proposition 2 can be implemented by deriving the first-order and second-order derivations of $ \rho_l(\tau_1), \forall l\in\left\lbrace0,1,2,3 \right\rbrace $.  
%The constraints in Eq. \eqref{p01} specify the feasible region for $ \tau_1^\ast $.	
Then, considering the feasible region for $ \tau_1$ in Eq. \eqref{p01}, the optimal $ \tau_1^\ast $ can be numerically obtained by referring to the detailed derivations in Appendix of \cite{ourWCL}. \quad\quad\quad\quad\quad\quad\quad\quad\quad\quad\quad\quad\quad\quad\quad\quad\quad$ \blacksquare $

\textit{\textbf{Proposition 3:}} When $ \tau_1 $ is fixed, the joint optimization of  $\bm{\Theta}$, $ \mathbf{w}_I $, $ \mathbf{w}_J $ in $ \mathbf{P2} $ can be recast as a polynomial expression as follows
\begin{alignat}{2}\label{p3}
\left( \mathbf{P4:}\right)  \max_{\bm{\Theta},\mathbf{w}\!_I,\mathbf{w}\!_J,\mathbf{u},\mathbf{v},\mathbf{y}} \!&\tilde{O}^{Q}_T
%(\bm{\Theta},\mathbf{w}\!_I,\mathbf{w}\!_J,\mathbf{v},\mathbf{y})
%\\
\!=\!\sum_{m=1}^{4}\rho_m\log_2(1\!+\!v_m)\!-\!\rho_mv_m\\
&\quad\quad\quad+\sum_{m=1}^{4}2\sqrt{\rho_m(1+v_m)\xi_m}\Re_m\nonumber\\
&\quad\quad\quad-|u_1|^2\sigma_{1}^2-|u_4|^2\sigma_{4}^2
%(\bm{\Theta},\mathbf{w}\!_I,\mathbf{w}\!_J)
\nonumber\\
&\quad\quad\quad-\sum_{m=2}^{3}\|\mathbf{y}_m\|^2\sigma_{m}^2
%(\bm{\Theta},\mathbf{w}\!_I,\mathbf{w}\!_J)
\nonumber\\
%&+2\sqrt{\rho_1(1+v_1)p_I}\mathbb{R}\left\lbrace y_1^\dagger\mathbf{g}_{\textit{RI}}^{\mathrm{H}}\bm{\Theta}\mathbf{G}_{\textit{BR}}\mathbf{w}_I\right\rbrace\nonumber\\
%&+2\sqrt{\rho_2(1+v_2)p_I}\mathbb{R}\left\lbrace y_2^\dagger\mathbf{h}_{\textit{I}}^{\mathrm{H}}\mathbf{w}_I\right\rbrace\nonumber\\
%&+2\sqrt{\rho_3(1+v_3)p_I}\mathbb{R}\left\lbrace y_3^\dagger\mathbf{h}_{\textit{BJ}}^{\mathrm{H}}\mathbf{w}_J\right\rbrace\nonumber\\
\mbox{s.t.}\quad  
& C4: \left\| \mathbf{w}_I\right\|^2\le p_{\textit{Imax}} 
%&\tag{\ref{p3}{a}}\label{p3a}
\nonumber\\
& C5: \left\| \mathbf{w}_J\right\|^2\le p_{\textit{Jmax}}
%&\tag{\ref{p3}{b}}\label{p3b} 
\nonumber\\ \nonumber
&C6: |\mathbf{g}_{\textit{RJ}}^{\mathrm{H}}\bm{\Theta}\mathbf{G}\mathbf{w}_\textit{I}|^2\le \bar{\epsilon}_J,
\end{alignat}	
where $ \mathbf{u}=[u_1,u_4] $, $ u_1$ and $ u_4$ are
%Based on the QT and Lagrangian dual transform methods\cite{FP-1,FP-2}, we introduce 
two auxiliary complex variables.  $ \mathbf{y}=[ \mathbf{y}_2^{\mathrm{H}},\mathbf{y}_3^{\mathrm{H}}]^{\mathrm{H}} $ in which $  \mathbf{y}_2\in\mathbb{C}^{K\times 1}$ and $  \mathbf{y}_3\in\mathbb{C}^{K\times 1}$ are two complex vectors.  $ \mathbf{v}=\left[v_1,v_2,v_3,v_4 \right] $ is an auxiliary real vector.
$ \rho_1 $, $ \rho_2 $, $ \rho_3 $ are given in Eq. \eqref{p002} and $ \rho_4=\rho_2 $. 
$ \xi_1=\epsilon_{\mathrm{out}} $, $ \xi_2=\mathcal{F}_{\|\mathbf{f}_I\|^2}^{-1}\left(\frac{\epsilon_{\mathrm{out}}}{2}\right) L_I^2 $, $ \xi_3=\mathcal{F}_{\|\mathbf{f}_J\|^2}^{-1}\left(\frac{\epsilon_{\mathrm{out}}}{2}\right) L_J^2 $, and $ \xi_4=1 $. 
$ \sigma_{1}^2\!=\!2L_I^2(\|\hat{\mathbf{f}}_I\| ^2\!+\!\left\| \mathbf{1}_K\right\| ^2\!\sigma_e^2)\left\| \mathbf{w}_J\right\| ^2\!+\!\epsilon_{\mathrm{out}}\sigma_I^2$, 
$ \sigma_{4}^2\!=\!\sigma_{2}^2\!=\!\sigma_I^2$, 
$ \sigma_{3}^2\!=\!|\mathbf{g}_{\textit{RJ}}^{\mathrm{H}}\bm{\Theta}\mathbf{G}\mathbf{w}_\textit{I}|^2\!+\!\sigma_J^2$.
%$ p_1\!=\!p_2\!=\!p_I $, $ p_3\!=\!p_J $. 
%\begin{equation}
$ \Re_m=\begin{cases}
\mathrm{Re}\left\lbrace u_1^{\mathrm{H}}\mathbf{g}_{\textit{RI}}^{\mathrm{H}}\bm{\Theta}\mathbf{G}\mathbf{w}_I\right\rbrace, &  m=1\\
\mathrm{Re}\left\lbrace\mathbf{y}_2^{\mathrm{H}}\mathbf{w}_I\right\rbrace, & m=2\\
\mathrm{Re}\left\lbrace \mathbf{y}_3^{\mathrm{H}}\mathbf{w}_J\right\rbrace, & m=3\\
\mathrm{Re}\left\lbrace u_4^{\mathrm{H}}\mathbf{g}_{\textit{RI}}^{\mathrm{H}}\bm{\Theta}\mathbf{G}\mathbf{w}_I \right\rbrace, & m=4
\end{cases} $
%\end{equation}

%$ \Re\left\lbrace\right\rbrace_1\!=\!\mathrm{Re}\left\lbrace y_1^{\mathrm{H}}\mathbf{g}_{\textit{RI}}^{\mathrm{H}}\bm{\Theta}\mathbf{G}\mathbf{w}_I\right\rbrace $, 
%$\Re\left\lbrace\right\rbrace_2\!=\!\mathrm{Re}\left\lbrace\mathbf{y}_2^{\mathrm{H}}\mathbf{w}_I\right\rbrace $, $\Re\left\lbrace\right\rbrace_3\!=\!\mathrm{Re}\left\lbrace \mathbf{y}_3^{\mathrm{H}}\mathbf{w}_J\right\rbrace $. 
%$ \dagger $ denotes the conjugate transpose operation.

\textit{\textbf{Proof:}} The proof is mainly based on the QT method and Lagrangian dual transform method  in \cite{FP-1,FP-2}. Please refer to Appendix B. \quad\quad\quad\quad\quad\quad\quad\quad\quad\quad\quad\quad\quad\quad\quad\quad\quad\quad\quad $ \blacksquare $

Based on the Proposition 3, the iterative optimizations of $ \mathbf{v} $, $ \mathbf{u} $, $ \mathbf{y} $, $ \mathbf{w}\!_J  $,  $\mathbf{w}_I$, and $ \bm{\Theta}$ are detailed as follows. 
\subsubsection{Optimize $ \mathbf{v} $}
Given  $ \mathbf{u} $, $ \mathbf{y} $, $ \mathbf{w}\!_J  $,  $\mathbf{w}_I$, and $ \bm{\Theta}$, $ -|u_1|^2\sigma_{D,1}^2-|u_4|^2\sigma_{D,4}^2-\sum_{m=2}^{3}\|\mathbf{y}_m\|^2\sigma_{D,m}^2 $ is a constant term in $ \mathbf{P4} $. 
Therefore, the optimization of $ \mathbf{v} $  can be recast as 
\begin{alignat}{2}\label{optimize-v}
 \max_{\mathbf{v}}\quad \tilde{O}^{Q}_T=&\sum_{m=1}^{4}\rho_m\log_2(1+v_m)-\rho_mv_m\\\nonumber
 &+\sum_{m=1}^{4}2\sqrt{\rho_m(1+v_m)\xi_m}\Re_m.
% \!-\!\text{const}( y_1,\! \mathbf{y},\! \bm{\Theta}, \!\mathbf{w}_I,\! \mathbf{w}_J ),
\end{alignat}	
%where $ \text{const}( y_1, \mathbf{y} , \bm{\Theta}, \mathbf{w}_I , \mathbf{w}_J ) $ is a  constant  term when   $ y_1 $, $ \mathbf{y} $, $\bm{\Theta}$, $ \mathbf{w}_I $, and $ \mathbf{w}_J $ are fixed.

According to the concavities of $ \log_2(\cdot) $ function and square-root $ \sqrt{\cdot} $ function,  it's obvious that $ \tilde{O}^{Q}_T $ is concave over $\mathbf{v} $. Therefore, the optimal $ \mathbf{v}^\ast $ can be obtained by solving $ \frac{\partial \tilde{O}_T^{Q}(v_m)}{\partial v_m}=0 $ as
\begin{equation}\label{optimal-v}
v_m^\ast=\frac{\xi_m\Re _m^2+\Re_m\sqrt{\xi_m^2\Re_m^2+4\xi_m\rho_m}}{2\rho_m}, \forall m.
\end{equation}

\subsubsection{Optimize $ \mathbf{u} $ and $ \mathbf{y} $}
Given $ \mathbf{v} $,  $ \mathbf{w}_J $, $ \mathbf{w}_I $, $\bm{\Theta}$,  $ \sum_{m=1}^{4}\rho_m\log_2(1+v_m)-\rho_mv_m+2\sqrt{\rho_m(1+v_m)\xi_m}\Re_m $ is a constant term in problem $ \mathbf{P4} $. 
Therefore, the optimizations of $ \mathbf{u} $ and $ \mathbf{y} $  can be recast as
\begin{alignat}{2}\label{optimize-y}
 \max_{\mathbf{u},\mathbf{y}} \quad\tilde{O}^{Q}_T=&-|u_1|^2\sigma_1^2-|u_4|^2\sigma_4^2-\sum_{m=2}^{3}\|\mathbf{y}_m\|^2\sigma_{m}^2\\\nonumber
% &\\\nonumber
&+\!\sum_{m=1}^{4}2\sqrt{\!\rho_m(1+v_m)\xi_m}\Re_m.
%+\!\text{const}(\bm{\Theta}, \!\mathbf{w}_I, \!\mathbf{w}_J, \!\mathbf{v} ).
\end{alignat}	
%where $ \text{const}(\bm{\Theta},  \mathbf{w}_I, \mathbf{w}_J , \mathbf{v} ) $ is a constant term when $ \bm{\Theta}$, $ \mathbf{w}_I$, $\mathbf{w}_J $ , and $\mathbf{v}$ are fixed.

Based on the concavity of  $-x^2$ function and the linearity of $ \Re_m $ over $ \mathbf{u}, \mathbf{y} $,  we get that $\tilde{O}^{Q}_T $ is concave over $ \mathbf{u} $ and $\mathbf{y}$, respectively. 
Therefore, the optimal $ \mathbf{u}^\ast $ and $ \mathbf{y}^\ast $ can ba obtained by solving $ \frac{\partial \tilde{O}_T^{Q}(u_m)}{\partial u_m}\!\!=\!\!0, m=1,4 $, 
%$ \frac{\partial \tilde{O}_T^{Q}(u_4)}{\partial u_1}\!\!=\!\!0 $, 
and $ \frac{\partial \tilde{O}_T^{Q}(\mathbf{y}_m)}{\partial \mathbf{y}_m}\!\!=\!\!0,m=2,3 $, respectively. Consequently, we obtain
\begin{equation}\label{optimal-y1}
u_m^\ast=\frac{\zeta_m\sqrt{\rho_m(1+v_m)\xi_m}}{\sigma_{m}^2}, m=1,4
\end{equation}
\begin{equation}\label{optimal-y}
\mathbf{y}_m^\ast=\frac{\bm{\zeta}_m\sqrt{\rho_m(1+v_m)\xi_m}}{\sigma_{m}^2}, m=2,3
\end{equation}
where $ \zeta_1=\mathbf{g}_{\textit{RI}}^{\mathrm{H}}\bm{\Theta}\mathbf{G}\mathbf{w}_I $, 
$ \bm{\zeta}_2=\mathbf{w}_I$, $ \bm{\zeta}_3=\mathbf{w}_J $.

\subsubsection{Optimize $ \mathbf{w}_J $}
Given $ \mathbf{v} $, $ \mathbf{u} $, $ \mathbf{y} $, $ \mathbf{w}_I $ and  $ \bm{\Theta} $, $ \sum_{m=1}^{4}\rho_m\log_2(1\!+\!v_m)\!-\!\rho_mv_m-|u_1|^2\epsilon_{\mathrm{out}}\sigma_I^2-|u_4|^2\sigma_{4}^2-\sum_{m=2}^{3}\|\mathbf{y}_m\|^2\sigma_{m}^2+\sum_{m\in\left\lbrace 1,2,4\right\rbrace}^{}2\sqrt{\rho_m(1+v_m)\xi_m}\Re_m $ is a constant term in $ \mathbf{P4} $. 
Therefore, the optimization of $ \mathbf{w}_J $ can be recast as
\begin{alignat}{2}\label{optimize Wj}
%\left( \mathbf{P6:}\right) 
\max_{\mathbf{w}\!_J} \quad &\tilde{O}_T^{Q}
%(\mathbf{w}_J)
= -|u_1|^2c_J\left\| \mathbf{w}\!_J\right\|  ^2\!\\\nonumber
%&\\\nonumber
&\quad\quad\quad+2\sqrt{\rho_3(1+v_3)\xi_3}\text{Re}\left\lbrace \mathbf{y}_3^{\mathrm{H}}\mathbf{w}\!_J\right\rbrace\!
%+\!\text{const}(\bm{\Theta}, \! \mathbf{w}_I,\! y_1,\! \mathbf{y},\! \mathbf{v})
%\text{const}_J
\\\nonumber
\mbox{s.t.}\quad  &\left\| \mathbf{w}\!_J\right\|^2\le p_{\textit{Jmax}},
\end{alignat}	
where $ c_{J}=2L_I^2(\|\hat{\mathbf{f}}_I\| ^2\!+\!\left\| \mathbf{1}_K\right\| ^2\!\sigma_e^2) $. 
%$ \text{const}_J=\text{const}(\bm{\Theta},  \mathbf{w}_I , y_1, \mathbf{y} , \mathbf{v})-|y_1|^2\epsilon_{out}\sigma_I^2 $, and 
%$ \text{const}(\bm{\Theta},  \mathbf{w}_I , y_1, \mathbf{y} , \mathbf{v}) $ is a constant term when $ \bm{\Theta} $, $ \mathbf{w}_I  $, $y_1  $, $ \mathbf{y} $, and $ \mathbf{v} $ are fixed. 
The $ \tilde{O}_T^{Q} $ in Eq. \eqref{optimize Wj} is convex about $ \mathbf{w}_J $ with the convex constraint. 
%Therefore, the optimal $ \mathbf{w}_J^\ast $ can be efficiently found using the standard numerical method\cite{FP-1}.
Therefore, utilizing the Lagrange multiplier method\cite{ActiveRIS?}, the optimal $ \mathbf{w}_J^\ast $ can be obtained as
\begin{equation}\label{optimal-Wj}
\mathbf{w}_J^\ast=\sqrt{\rho_3(1+v_3)\xi_3}\left(|u_1|^2c_{J}+\eta_J\mathbf{I}_K\right)\!^{-1}\mathbf{y}_3,
\end{equation}
where $ \eta_J $ is the dual variable introduced for the maximum power constraint of $ \mathbf{w}_J $. The optimal $ \eta_J^\ast$ can be obtained as $\eta_J^\ast=\text{min}\left\lbrace\eta\!_J\ge\!0:\|\mathbf{w}\!_J(\eta_J)\|^2\le p_{\textit{Jmax}} \!\right\rbrace   $ through the bisection search method\cite{FP-1}.

\subsubsection{Optimize $ \mathbf{w}_I $}
Given $ \mathbf{v} $, $ \mathbf{u} $, $ \mathbf{y} $, $ \mathbf{w}_J $ and  $ \bm{\Theta} $, $ \sum_{m=1}^{4}\rho_m\log_2(1\!+\!v_m)\!-\!\rho_mv_m\!+\!2\sqrt{\rho_3(1+v_3)\xi_3}\Re_3\!-\!|u_1|^2\sigma_{1}^2\!-\!|u_4|^2\sigma_{4}^2\!-\!\sum_{m=1}^{2}\|\mathbf{y}_m\|^2\sigma_{m}^2 $ is a constant term in problem $ \mathbf{P4} $. 
Therefore, the optimization of $ \mathbf{w}_I $  is recast as
\begin{alignat}{2}\label{optimize WI}
\max_{\mathbf{w}\!_I} \!\quad&\tilde{O}_T^{Q}
%(\mathbf{w}_I)
=\!-\|\mathbf{y}_3\|^2\mathbf{w}_I^{\mathrm{H}}\bm{\Gamma}\bm{\Gamma^{\mathrm{H}}}\mathbf{w}_I\!-\!\|\mathbf{y}_3\|^2\sigma_J^2\!+\!2\text{Re}\!\left\lbrace \bm{\zeta}^{\mathrm{H}}\mathbf{w}_I\!\right\rbrace
%\\\nonumber
%&\quad\quad\quad+\!2\text{Re}\left\lbrace \bm{\zeta}^{\mathrm{H}}\mathbf{w}_I\right\rbrace
%+\text{const}(\bm{\Theta},\!  \mathbf{w}_J,\! y_1,\! \mathbf{y},\! \mathbf{v}) 
\\\nonumber
	\mbox{s.t.}\quad  & \left\| \mathbf{w}_I\right\|^2\le p_{\textit{Imax}}\\\nonumber
	&|\mathbf{g}_{\textit{RJ}}^{\mathrm{H}}\bm{\Theta}\mathbf{G}\mathbf{w}_\textit{I}|^2\le \bar{\epsilon}_J,
\end{alignat}	
	where $ \bm{\Gamma^{\mathrm{H}}}=\mathbf{g}_{\textit{RJ}}^{\mathrm{H}}\bm{\Theta}\mathbf{G}$, $ \bm{\zeta}^{\mathrm{H}}=\sum_{m\in \left\lbrace1,2,4 \right\rbrace } \sqrt{\!\rho_m(\!1\!+\!v_m\!)\xi_m}\bm{\lambda}_m^{\mathrm{H}}$, $ \bm{\lambda}_1^{\mathrm{H}}=u_1^{\mathrm{H}}\mathbf{g}_{\textit{RI}}^{\mathrm{H}}\bm{\Theta}\mathbf{G} $, $ \bm{\lambda}_2^{\mathrm{H}}=\mathbf{y}_2^{\mathrm{H}} $, $ \bm{\lambda}_4^{\mathrm{H}}=u_4^{\mathrm{H}}\mathbf{g}_{\textit{RI}}^{\mathrm{H}}\bm{\Theta}\mathbf{G} $.

%	$ \text{const}(\bm{\Theta},  \mathbf{w}_J , y_1, \mathbf{y} , \mathbf{v}) $ is a constant term when $ \bm{\Theta} $, $ \mathbf{w}_J $, $y_1  $, $ \mathbf{y} $, and $ \mathbf{v} $ are fixed. 
%	As we can observe, 

The problem in \eqref{optimize WI} is a quadratically constrained quadratic problem (QCQP)\cite{ActiveRIS?} about $ \mathbf{w}_I $. Therefore, utilizing Lagrange multiplier method, the optimal $ \mathbf{w}_I^\ast $ can be obtained as
	\begin{equation}\label{optimal-Wi}
	\mathbf{w}_I^\ast=\left(\|\mathbf{y}_3\|^2\bm{\Gamma}\bm{\Gamma}^{\mathrm{H}}+\eta_I\mathbf{I}_K\right)\!^{-1}\bm{\zeta},
	\end{equation}
	where $ \eta_I $ is the dual variable introduced for the constraints of $ \mathbf{w}_I $. Similarly, the optimal $ \eta_I^\ast$ can be obtained as $\eta_I^\ast\!=\!\text{min} \left\lbrace\!\eta_I\!\ge\!0\!:\!\left\lbrace\!\|\mathbf{w}\!_I(\eta_I)\|^2\!\le\!p_{\textit{Imax}} \!\right\rbrace\!\bigcap\!\left\lbrace\!\mathbf{w}_I^{\mathrm{H}}(\eta_I)\bm{\Gamma}\bm{\Gamma}^{\mathrm{H}}\mathbf{w}_I(\eta_I)\!\le\!\bar{\epsilon}_J \!\right\rbrace\! \right\rbrace $ by utilizing the bisection search method.
	
\subsubsection{Optimize $ \bm{\Theta} $}
Given $ \mathbf{v} $, $ \mathbf{u} $, $ \mathbf{y} $, $ \mathbf{w}_J $ and  $ \mathbf{w}_I $, $ \sum_{m=1}^{4}\rho_m\log_2(1\!+\!v_m)\!-\!\rho_mv_m\!-\!|u_1|^2\sigma_{1}^2\!-\!|u_4|^2\sigma_{4}^2\!-\!\|\mathbf{y}_2\|^2\sigma_{2}^2\!+\!\sum_{m=2}^{3}2\sqrt{\rho_m(1+v_m)\xi_m}\Re_m\! $ is a constant term  in problem $ \mathbf{P4} $. 
Therefore, the optimization of $ \bm{\Theta}=\mathrm{diag}( \bm{\psi}=[e^{j\theta_1},...,e^{j\theta_{N\!_\mathrm{R}}}]^\mathrm{T}) $ can be recast as\footnote{
	Note that the 
%	optimal RIS phase shift obtained in this paper is continuous. 
%	The 
	practical discrete phase shift design involves additional constraint $ \theta_{n\!_R}\!\in\!\!\mathcal{F} $, 
	$ \forall n_{\mathrm{R}}\!\in\!\left\lbrace 1, \cdots, N\!_{\mathrm{R}} \right\rbrace  $, where $ \mathcal{F}=\left\lbrace 0, 2\pi/L,\cdots, 2\pi(L\!-\!1)/L \right\rbrace $ is the set of discrete phase shift values. $ L = 2b $ is the number of phase shift levels, and $ b $ represents the phase resolution in bits. 
%	The sub-optimal discrete phase shift $ \tilde{\bm{\psi}} $ can be obtained  
%	we can directly map each phase shift in $ \tilde{\bm{\psi}} $ to its nearest discrete value in $ \mathcal{F} $. 
%			Typically, RIS phase shift in practice is discrete. 
%	Under this situation, an extra constraint $ C8:\theta_{n\!_R}\in\mathcal{F} $, 
%	$ \forall n_R\in\left\lbrace 1, \cdots, N_R \right\rbrace  $ is introduced to $ \mathbf{P1} $, 
%	Nonetheless, the problem with $ C8 $ is NP-hard and needs to be further relaxed to the original problem $ \mathbf{P1} $. Then, after a series of transformations proposed in this paper, $ C8 $ is added to the problem in Eq. \eqref{optimize-theta}. Next, the near-optimal passive beamforming $ \tilde{\bm{\psi}} $ can be obtained 
%	through the semidefinite relaxation (SDR) method\cite{cascaded channel estimation2}. 
%	We can directly map each phase shift in $ \bm{\psi}^\ast $
%%	$ \tilde{\bm{\psi}} $ 
%	to its nearest discrete value in $ \mathcal{F} $, resulting in 
	Combining with the semi-definite relaxation (SDR) method, we can heuristically obtain the sub-optimal discrete phase shift $ \tilde{\bm{\psi}} $ by directly mapping each phase shift in $ \bm{\psi}^\ast $ to its nearest discrete value in $ \mathcal{F} $. 
	The larger $ b $ comes to the larger number of phase shift levels $ L $, leading to $ \tilde{\bm{\psi}} $ being closer to the optimal continuous phase shift $ \bm{\psi}^\ast $.  Otherwise, a smaller $ b $ degrades the optimality of RIS phase shift design, which inevitably degrades the RIS-assisted communication performance obtained at the PC period and further degrades the AST, as expressed in Eqs. \eqref{hat_SNR} and \eqref{O_T2}. 
	In a word, the optimal continuous RIS phase shift obtained in this research gives an examination of the upper-bound systemic communication performance of the proposed R-SACE mechanism.
}
\begin{alignat}{2}\label{optimize-theta}
 \max_{\bm{\Theta}}\quad&\tilde{O}_T^{Q}=-\|\mathbf{y}_3\|^2\bm{\psi}^{\mathrm{H}}\bm{\Pi}\bm{\psi}\!-\!\|\mathbf{y}_3\|^2\sigma_J^2\!+\!2\text{Re}\left\lbrace\! \bm{\psi}^{\mathrm{H}}\bm{\delta}\!\right\rbrace
% \\\nonumber
% &\quad\quad\quad+\!2\text{Re}\left\lbrace\! \bm{\psi}^{\mathrm{H}}\bm{\delta}\!\right\rbrace\!
% +\!\text{const}(\mathbf{w}_I,\!\mathbf{w}_J,\! y_1,\! \mathbf{y},\! \mathbf{v})
\\\nonumber
	\mbox{s.t.}\quad  & \bm{\psi}^{\mathrm{H}}\bm{\Pi}\bm{\psi}\le \bar{\epsilon}_J, 
	\end{alignat}	
	where $ \bm{\Pi}\!=\!\!\left(\text{diag}(\mathbf{G}\mathbf{w}_I) \right)^{\mathrm{H}}\!\mathbf{g}_{\textit{RJ}}\mathbf{g}_{\textit{RJ}}^{\mathrm{H}}\left(\text{diag}(\mathbf{G}\mathbf{w}_I)\right)$, 
	$ \bm{\delta}\!=\sum_{m\in\left\lbrace1,4 \right\rbrace }\sqrt{\!\!\rho_m(1\!+\!v_m)\xi_m}\text{diag }\!(u_m^{\mathrm{H}}\mathbf{g}_{\textit{RI}}^{\mathrm{H}})\mathbf{G}\mathbf{w}_I$. 
%	$ \text{const}(\mathbf{w}_I,  \mathbf{w}_J , y_1, \mathbf{y} , \mathbf{v}) $ is a constant term when $ \mathbf{w}_I $, $ \mathbf{w}_J $, $y_1  $, $ \mathbf{y} $, and $ \mathbf{v} $ are fixed.

The problem in \eqref{optimize-theta} is also a QCQP over $ \bm{\Theta} $. Therefore, based on the Lagrange multiplier method, the optimal $ \bm{\psi}^\ast$ can be obtained as
	\begin{equation}\label{optimal-theta}
	\bm{\psi}^\ast=\|\mathbf{y}_3\|^2\left( \bm{\Pi}+\eta_{\Theta}\mathbf{I}_{N\!_\textit{R}}\right)^{-1}\!\bm{\delta},
	\end{equation}
where $ \eta_{\Theta}$ is the dual variable introduced for the constraint of $ \bm{\psi} $.  Similarly, utilizing the  bisection search method, the optimal $ \eta_{\Theta}^\ast$ can be obtained as $\eta_{\Theta}^\ast=\text{min} \left\lbrace\eta_{\Theta}\ge0:\bm{\psi}^{\mathrm{H}}(\eta_{\Theta})\bm{\Pi}\bm{\psi}(\eta_{\Theta})\le\bar{\epsilon}_J/p_I \right\rbrace $.

Ultimately, the proposed FPI algorithm for solving the AST maximization problem $ \mathbf{P2} $ is summarized in Algorithm 1.

\begin{algorithm}[tb]
	\caption{FPI Algorithm for Solving AOSC Maximization Problem}\label{alg:alg1}
	\begin{algorithmic}
		%		\STATE 
%		\STATE {\textsc{\textbf{Input:}}} Initial $  \tau_1$, $ \bm{\Theta} $, $ \mathbf{w}_I $, and $ \mathbf{w}_J $ within feasible regions 
		\STATE \hspace{-0.3cm}{\textsc{\textbf{Initialize:}}} 
%		 Initialize 
		 $  \tau_1(0)$, $ \bm{\Theta}(0) $, $ \mathbf{w}_I(0) $, $ \mathbf{w}_J(0) $, $ \mathbf{u} $, $ \mathbf{y} $, and $ \mathbf{v} $ 
%		 to feasible values
		 , $ t \gets 1$
		\STATE \hspace{0cm}$ \textbf{Repeat}$
		%		\quad W \subset \mathbf{X}  $
		\STATE \hspace{0.4cm}$ \textbf{Part 1:}$
		\STATE \hspace{0.6cm} 
%		\begin{minipage}[t]{200pt}
			Update $\tau_1^\ast(t)$ by numerically solving the convex problem $ \mathbf{P3} $, given $ \bm{\Theta}(t-1) $, $ \mathbf{w}_I(t-1) $, and $ \mathbf{w}_J(t-1) $,
%		\end{minipage}
%		\STATE \hspace{0.6cm} 
%		\begin{minipage}[t]{208pt}
%		Transform $\mathbf{P2}$ to $ \mathbf{P5} $ according to Proposition 3,
%		\end{minipage}
		\STATE \hspace{0.4cm}$ \textbf{Part 2:}$
		\STATE \hspace{0.6cm} 
%		\begin{minipage}[t]{208pt}
Update $\mathbf{v}^\ast (t)$ by Eq. \eqref{optimal-v}, given $ \tau_1^\ast(t) $, $ \mathbf{u}(t-1) $, $  \mathbf{y}(t-1)$, $ \mathbf{w}_I(t-1) $, $ \mathbf{w}_J(t-1) $, and $ \bm{\Theta}(t-1) $,
%		\end{minipage}
\\
		\STATE \hspace{0.6cm} 
%		\begin{minipage}[t]{208pt}
Update $\mathbf{u}^\ast(t) $ and $ \mathbf{y}^\ast(t)$ by Eqs. \eqref{optimal-y1} and \eqref{optimal-y} respectively, given $ \tau_1^\ast(t) $, $  \mathbf{v}^\ast (t)$, $ \mathbf{w}_I(t-1) $, $ \mathbf{w}_J(t-1) $, and $ \bm{\Theta}(t-1) $,
%		\end{minipage}
\\
		\STATE \hspace{0.6cm} 
%		\begin{minipage}[t]{208pt}
Update $\mathbf{w}_J^\ast(t) $ by Eq. \eqref{optimal-Wj}, given $ \tau_1^\ast(t) $, $  \mathbf{v}^\ast(t)$, $\mathbf{u}^\ast(t) $, $ \mathbf{y}^\ast(t) $, $ \mathbf{w}_I(t-1) $, and $ \bm{\Theta}(t-1) $,
%		\end{minipage}
\\
		\STATE \hspace{0.6cm} 
%		\begin{minipage}[t]{208pt}
Update $\mathbf{w}_I^\ast(t) $ by Eq. \eqref{optimal-Wi}, given $ \tau_1^\ast(t) $, $  \mathbf{v}^\ast(t)$, $\mathbf{u}^\ast(t)$, $ \mathbf{y}^\ast(t) $, $ \mathbf{w}_J^\ast(t) $, and $ \bm{\Theta}(t-1) $,
%		\end{minipage}
		\\
		\STATE \hspace{0.6cm} 
		%		\begin{minipage}[t]{208pt}
Update $\bm{\Theta}^\ast(t)$ by Eq. \eqref{optimal-theta}, given $ \tau_1^\ast(t) $, $  \mathbf{v}^\ast(t) $, $\mathbf{u}^\ast(t) $, $ \mathbf{y}^\ast(t) $, $ \mathbf{w}_J^\ast(t) $, and $ \mathbf{w}_I^\ast(t) $,
		%		\end{minipage}
		\\
		\STATE \hspace{0.6cm} Calculate $ \tilde{O}_T(\tau_1^\ast(t),\bm{\Theta}^\ast(t) ,\mathbf{w}_I^\ast(t),  \mathbf{w}_J^\ast(t) )$, $t=t+1$,
		\STATE \hspace{0cm}\textbf{Until} 
		\STATE \hspace{0.6cm} $ \tilde{O}_T(\tau_1^\ast(t),\bm{\Theta}^\ast(t) ,\mathbf{w}_I^\ast(t),  \mathbf{w}_J^\ast(t) )-\tilde{O}_T(\tau_1^\ast(t\!-\!1),\bm{\Theta}^\ast(t\!-\!1) ,\mathbf{w}_I^\ast(t\!-\!1), \mathbf{w}_J^\ast(t\!-\!1) )\!\le\! \Delta$, $ \Delta $ is the convergence threshold.
		\STATE\hspace{-0.3cm} {\textsc{\textbf{Return:}}} The optimal solution $  \tau_1^\ast$, $ \bm{\Theta}^\ast$, $ \mathbf{w}_I^\ast$, $ \mathbf{w}_J^\ast $ of $ \mathbf{P2} $.
	\end{algorithmic}
	\label{alg1}
\end{algorithm}

\subsection{Convergence and Complexity Analysis }
We first analyze the convergence of the proposed FPI algorithm. 
As presented in Algorithm 1, there are Part 1 and Part 2 to solve the optimization of $ \tau_1 $ and the joint optimization of $ \bm{\Theta}$, $ \mathbf{w}_I$, $ \mathbf{w}_J $, respectively.
%First, we analyze the convergence of Part 1 in Algorithm 1. 
As presented in Proposition 2, the optimal $ \tau_1^\ast(t) $ can be numerically obtained by calculating the first- and second-order derivations of $ \rho_l(\tau_1),\forall l$. Therefore, we only need to analyze the convergence of Part 2 in Algorithm 1. 
Given $ \tau_1(t) $, the closed-form optimal $ \mathbf{w}_J^\ast(t)$, $ \mathbf{w}_I^\ast(t) $, and $ \bm{\Theta}^\ast(t)$ can be obtained through the fixed-point QT method, according to Eqs. \eqref{optimal-Wj}, \eqref{optimal-Wi}, and \eqref{optimal-theta}, respectively. 
The convergence of the fixed-point QT method to iteratively optimize variables has been  established in detail in Appendix A of \cite{FP-2}. Therefore, the part 2 in Algorithm 1 is convergent. 
Finally, the $ \tilde{O}_T $ in Eq. \eqref{p002} increases with each optimal iterations of $ \tau_1^\ast (t)$, $ \bm{\Theta}^\ast(t) $, $ \mathbf{w}_I^\ast(t) $, and $ \mathbf{w}_J^\ast(t)  $  in Algorithm 1.  Therefore, the overall convergence of the proposed FPI algorithm is guaranteed. 

% in which $ \epsilon_{\mathrm{out}}=0.1 $, $ N_{\mathrm{R}}=16$, and $ P_{\mathrm{J_1}}=0.5 $.

%\begin{figure}[tb]
%	\centering
%	\includegraphics[width=2.65in]{iteration_ts.eps}
%	\caption{\textcolor{red}{A comparison of the convergence iterations of different algorithms.} }
%	\label{convergence_iteration}
%\end{figure}

\begin{figure}[tb]
		\centering
\subfloat[]{\includegraphics[width=2.5in]{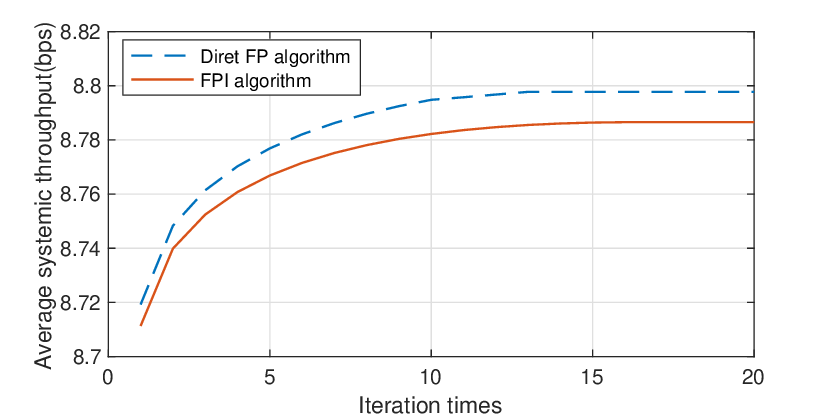}%
\label{convergence_iteration}
}
\hfil
\subfloat[]{\includegraphics[width=2.5in]{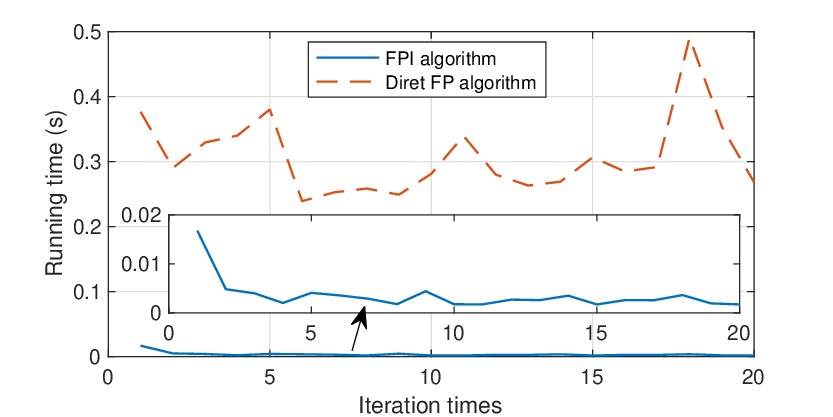}
	\label{runningtime}
}
\caption{A comparison of the convergence iterations and running time of different algorithms. (a) Convergence iteration. (b) Running time.}
\label{convergence_runningtime}
\end{figure}

Next, we analyze the complexity of the proposed FPI algorithm. 
According to Proposition 2, the optimal $ \tau_1^\ast $ is obtained by numerically calculating the first- and second-order derivatives of $ \tilde{O}_T$ over $ \tau_1 $.
Therefore, the complexity of optimizing $ \tau_1 $ in Part 1 in Algorithm 1 is $ \mathcal{O}(1) $, regardless of the number of RIS elements and communication users. This demonstrates the computational efficiency of FPI algorithm in optimizing the temporal resource allocation in ISAC systems.
Then, according to Proposition 3, the fixed-point QT method recasts the joint optimization of $ \bm{\Theta}$, $ \mathbf{w}_I$, and $ \mathbf{w}_J $ as a series of convex problems in Eqs. \eqref{optimize-v}-\eqref{optimal-theta}, where the closed-form optimal solutions are obtained through the bisection search method. 
Let $ I_{\Theta} $, $ I_{w1} $, and $ I_{w2} $ denote the maximal iterations of obtaining optimal $ \bm{\Theta}^\ast$, $ \mathbf{w}_I^\ast$, and $ \mathbf{w}_J^\ast$ through the bisection search method.
The complexity of jointly optimizing $ \bm{\Theta}^\ast(t) $, $ \mathbf{w}_J^\ast(t) $, and $ \mathbf{w}_I^\ast(t) $ in Part 2 in Algorithm 1 can be obtained as $ \mathcal{O}\left( \max \left\lbrace I_{\Theta}, I_{w1}, I_{w2} \right\rbrace \right)  $.
It's worth mentioning that with a larger $ N_{\mathrm{R}} $, the optimal $ \bm{\Theta}^\ast$ can also be obtained through the closed-form solution in Eq. \eqref{optimal-theta}. The iteration number of the bisection search to find the dual variable \( \eta_{\theta} \) remains unaffected. Therefore, the complexity of optimizing $ \bm{\Theta}^\ast$ does not increase with the number of RIS elements. 
Similarly, with larger number of TC $ J $ or CU $ I $, the complexities of optimizing $ \mathbf{w}_J^\ast(t) $ for each TC $ J $ and optimizing $ \mathbf{w}_I^\ast(t) $ for each CU $ I $ also remain unchanged.
%Therefore, the complexity of jointly optimizing $ \mathbf{w}_J^\ast(t) $, $ \mathbf{w}_I^\ast(t) $, and $ \bm{\Theta}^\ast(t) $ is independent of the number of communication users.
Then, let $ I_O $ denote the maximal iterations of the repeat operation in Algorithm 1. 
The complexity order of Algorithm 1 can be finally obtained as $ \mathcal{O}\left( I_O*\max \left\lbrace 1, \max \left\lbrace I_{\Theta}, I_{w1}, I_{w2} \right\rbrace \right\rbrace \right)=\mathcal{O}\left( I_O\!*\!\max \left\lbrace I_{\Theta}, I_{w1}, I_{w2} \right\rbrace \right) $. 
Note that, although $ I_O $ increases with more TC $ J $ and CU $ I $, it is still a finite number. 
Therefore, the complexity of solving AST maximization problem through the FPI algorithm is polynomial regardless of larger RIS configurations or more communication users.

We further analyze the convergence and running time of the adopted FPI algorithm compared with the direct fractional programming (FP) algorithm \cite{FP-1, FP-2} in Fig. \ref{convergence_runningtime}. 
In the direct FP algorithm, each iteration is performed by numerically solving a convex optimization problem, rather than in closed form as in the FPI algorithm. 
As shown in this figure, the adopted FPI algorithm takes more iterations to converge, but its running time per iteration is the lowest, thanks to the closed-form updates in each iteration. 
In summary, the adopted FPI algorithm is the fastest, resulting in greater practicality.

\begin{figure}[tb]
	\centering
	\includegraphics[width=2.5in]{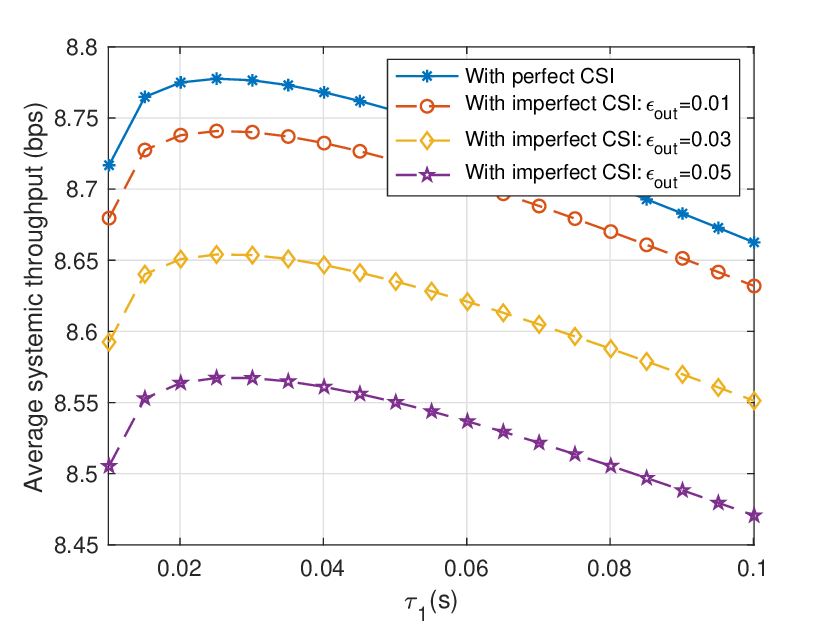}
	%	{RSACE_ts_opt_rand.eps}
	\caption{The AST obtained with perfect and imperfect CSIs changing with $ \tau_1 $. 
		%		In this simulation, $ \epsilon_{out}=0.1 $, $ \epsilon=2.4484\times10^{-11} $, $ P_{H_1}=P_{H_0}=0.5 $. 
	}
	\label{perfect_imperfect_CSI}
\end{figure}
%\begin{figure}[h!]
%	\centering
%	\includegraphics[width=2.5in]{sensing_performance_noise.eps}
%	\caption{\textcolor{red}{The sensing performance obtained at ISAC period changing with $ \sigma^2 $.}}
%	\label{Sensing_performance_ts}
%\end{figure}

\section{Simulation Results}
This section evaluates the effectivenesses of the proposed FPI algorithm and the R-SACE mechanism in comparison to baseline algorithms and mechanisms.
%that rely solely on RIS (indicated as the R-ACE mechanism) or sensing results (indicated as the S-ACE mechanism) for support.
The specific simulation parameters are set as follows.
The location coordinates for BS, CU $ I $, TC $ J $, and RIS are set as $ (-20,20) $, $ (20,0) $, $ (0,10)$, and $ (-5,15) $, respectively.
Referring to \cite{ActiveRIS?}, we set the large-scale path loss in this paper as $A_0 (\frac{d_X}{d_0})^{-\alpha}$, where $A_0=-30$dB denotes the path loss at reference distance $ d_0=1$m. $ d_X $, $ X\in\left\lbrace \textit{BR}, \textit{RI}, \textit{RJ}, I, J \right\rbrace $, denotes the distance of BS-RIS, RIS-CU $ I $, RIS-TC $ J $, BS-CU $ I $, and BS-TC $ J $.  $ \alpha $ denotes the path-loss exponent.
In particular,  $ \alpha_{\textit{BR}}=2.5 $, $ \alpha_{\textit{RI}}=2.5 $, $ \alpha_{\textit{RJ}}=2.5 $, $ \alpha_J=3 $, and $ \alpha_I=3.5 $ for $ \mathbf{g}_{\textit{RI}}$, $ \mathbf{g}_{\textit{RJ}} $,  $\hat{\mathbf{h}}_J $, and $  \hat{\mathbf{h}}_I $, respectively. 
Generally, $\mathbf{G}\!=\beta_G(\!\sqrt{g}\mathbf{G}^{\mathrm{LoS}}\!+\!\sqrt{1\!-g}\mathbf{G}^{\mathrm{NLoS}})$ follows Rician distribution\cite{RIS-ISAC2+1,STAR-RIS-Rician}, where $ \beta_G=\!\sqrt{\!\frac{A_0}{d_{\textit{BR}}^{\alpha_{\textit{BR}}}}} $ is distance dependent path loss, $ g=0.8 $ is Rician coefficient. 
$\mathbf{G}^{\mathrm{LoS}}=\mathbf{a}_{N\!_{\mathrm{R}}}(\theta_{\textit{RB}})\mathbf{a}_{K}^{\mathrm{H}}(\theta_{\textit{BR}})$ is the deterministic LoS component\cite{STAR-RIS-Rician,RIS-ISAC2+1}. $ \mathbf{a}_{N\!_R} $ and $ \mathbf{a}_{K} $ represent the steering vector and 
$ \theta_{\textit{RB}} $ and $ \theta_{\textit{BR}} $ represent the DoA and DoD. Each element in the non-LoS component $\mathbf{G}^{\mathrm{NLos}}$ follows $ \mathcal{CN}(0,1) $. 
Referring to \cite{RIS-ISAC10}, the duration of the entire timeline is $ \tau=1 $s. Additionally, $f_s=10$MHz, $K=4$, $N_{\mathrm{R}}=16$, $\sigma^2= \sigma_0^2=\sigma_I^2=\sigma_J^2=-110 $dBm, $ P_{\mathrm{J_1}}=P_{\mathrm{J_0}}=0.5 $, and $ \sigma_T^2=1 $ as referred to \cite{RIS-ISAC2+1}.

%	\begin{figure}[tb]
%	\centering
%	\includegraphics[width=2.5in]{Sensing_performance_ts.eps}
%	\caption{\textcolor{red}{The sensing performance obtained at ISAC period changing with $ \tau_1 $.}}
%	\label{Sensing_performance_ts}
%\end{figure}

\subsection{Performance Evaluation}

	Fig. \ref{perfect_imperfect_CSI} shows the difference between AST obtained with perfect CSI and imperfect CSI.
	This figure demonstrates that the AST obtained with imperfect CSI is lower than the upper-bound AST obtained with perfect CSI.
	Whereas, the gap in AST decreases with the probability threshold $ \epsilon_{\mathrm{out}} $. This is because lower $ \epsilon_{\mathrm{out}} $ helps achieve more desirable systemic communication performance by limiting the error probability of channel estimation to a lower level.
	Additionally, this figure shows that for every curve, there is an optimal $ \tau_1^\ast $ to maximize the AST.
	This is because increasing $ \tau_1 $ helps obtain more reliable sensing results, which facilitate communication during the PC period.
	However, increasing $\tau_1$ also reduces the time $\tau-\tau_1$ for communication in the PC period.
	Therefore, there exists an optimal time allocation $ \tau_1^\ast $ between ISAC and PC periods for achieving satisfied sensing performance and eventually contributing to the maximal communication performance AST.
	This simulation result further verifies the correctness of the theoretical analysis in Proposition 2.

%	\textcolor{red}{Fig. \ref{Sensing_performance_ts} illustrates the sensing performance changing with the time duration $ \tau_1 $ of the ISAC period. 
%		In this simulation, we introduce two additional sensing performance metrics: the error probability $ P_E\!=\!P_{\mathrm{J1}}(1\!-\!P\!_d)+(1\!-\!P_{\mathrm{J1}})P\!_f $ and the correct probability $ P\!_C\!=\!1\!-\!P_E $, for comprehensively evaluating the systemic sensing performance. These metrics are helpful to evaluate the robustness of the sensing performance in the proposed R-SACE mechanism.
%		As shown in this figure, the detection probability $ P_d $ increases with $ \tau_1 $, and the false alarm probability $ P_f $ decreases with $ \tau_1 $.
%		This outcome results from an increased $ \tau_1 $, indicating a longer sensing time and thus contributing to higher detection accuracy.
%		This result can be further verified through the theoretical expressions in Eqs. \eqref{P_d} and \eqref{P_f}. 
%		In the end, when $ \tau_1 $ rises, the total sensing performance gets better.
%		Specifically, both the correct probability $ P\!_C $ and the error probability $ P\!_E $ show favorable trends: $ P\!_C $ increases while $ P\!_E $ decreases with the longer time duration of the ISAC period $ \tau_1 $.
%	}

	\begin{figure}[h]
	\centering
	\includegraphics[width=2.5in]{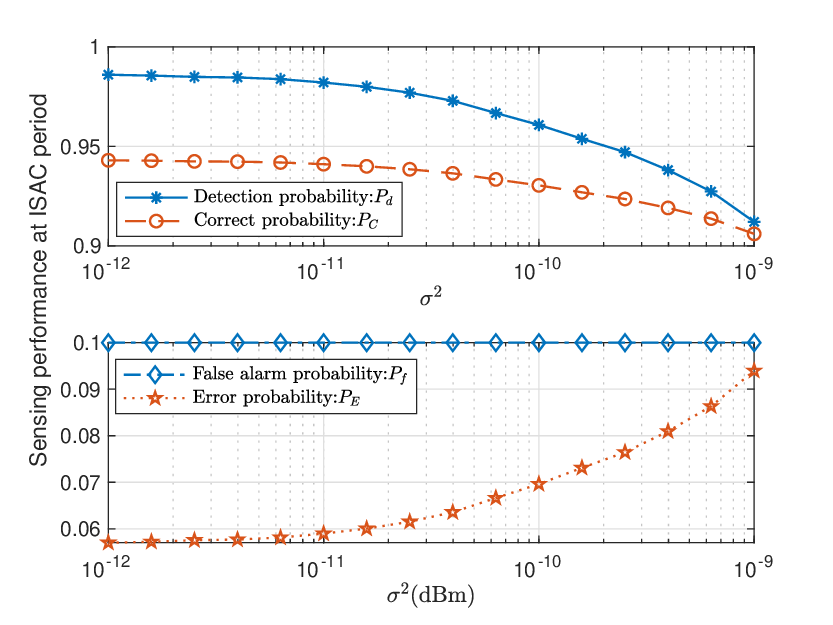}
	\caption{The sensing performance obtained at ISAC period changing with $ \sigma^2 $.}
	\label{Sensing_performance_ts}
\end{figure}

Fig. \ref{Sensing_performance_ts} illustrates the sensing performance obtained at the ISAC period changing with the noise power $ \sigma^2 $.
		The false alarm probability $ \bar{P}_f $ obtained at the BS is fixed as $ 0.1 $ in this simulation and thus remains unchangeable with the noise power. 
		In this simulation, we introduce two additional sensing performance metrics: the error probability $ P_E\!=\!P_{\mathrm{J1}}(1\!-\!P\!_d)+(1\!-\!P_{\mathrm{J1}})P\!_f $ and the correct probability $ P\!_C\!=\!1\!-\!P_E $, for comprehensively evaluating the systemic sensing performance. 
		As illustrated in this figure, with a fixed false alarm probability, the detection probability $ P_d $ decreases with the noise power $ \sigma^2 $.
		Furthermore, the degraded detection probability leads to a lower correct probability probability $ P_C $ and inevitably an increased error probability as noise power rises.
		Fortunately, the correct probability remains greater than 0.9, and the error probability does not exceed 0.1, even when the noise power is as large as $ 10^{-9} $dBm.
		This simulation result demonstrates the robustness of the proposed R-SACE mechanism in maintaining satisfactory sensing performance.
	%	This outcome results from a larger noise power causing a degraded sensing SNR longer sensing time and thus contributing to higher detection accuracy.
	%	This result can be further verified through the theoretical expressions in Eqs. \eqref{P_d} and \eqref{P_f}. 
	%	In the end, when $ \tau_1 $ rises, the total sensing performance gets better.
	%	Specifically, both the correct probability $ P\!_C $ and the error probability $ P\!_E $ show favorable trends: $ P\!_C $ increases while $ P\!_E $ decreases with the longer time duration of the ISAC period $ \tau_1 $.

\begin{figure}[tb]
	\centering
	\includegraphics[width=2.7in]{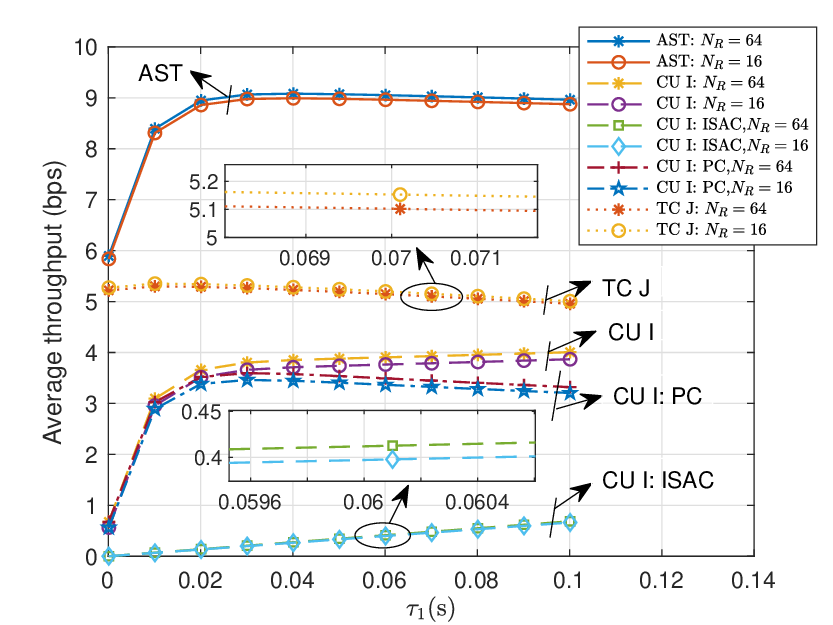}
	\caption{The average throughput changing with $ \tau_1 $ and $ N_{\mathrm{R}} $.}
	\label{Communication_performance_ts}
\end{figure}
	
Fig. \ref{Communication_performance_ts} presents the contributions of AST from CU $ I $ and TC $ J $ changing with $ \tau_1 $ and the number of RIS elements $ N_{\mathrm{R}} $. 
	It is clear that the average throughput of CU $ I $ during the ISAC period increases with $ \tau_1 $, owing to the longer communication time.
	During the PC period, the average throughput of CU $ I $ initially increases and then decreases with $ \tau_1 $.
	This phenomenon can be explained as follows.
	First, the sensing performance, i.e., the correct probability about TC $ J $, is enhanced with the increased sensing time $ \tau_1 $, leading to a higher average throughput for CU $ I $ during the PC period.
	Whereas as $ \tau_1 $ increases, the communication time $ \tau-\tau_1 $ during the PC period decreases, resulting in a lower average throughput of CU $ I $ during this period.
	The overall average throughput of CU $ I $ throughout the entire timeline increases with $ \tau_1 $ because the communication performance contribution from the ISAC period is greater than that from the PC period.
	Regarding TC $ J $, its average throughput during the PC period first increases and then decreases with $ \tau_1 $. 
	Specifically, the improved average throughput results from enhanced sensing accuracy, which increases with $ \tau_1 $ and allows for better transmission coordination between CU $ I $  and TC $ J $.
	Whereas, the lower average throughput is a result of the decreased communication time.
	Finally, the AST first increases and then decreases with $ \tau_1 $. 
	This phenomenon indicates that TC $ J $ has a greater contribution to AST than CU $ I $.
	Moreover, this result demonstrates that an optimal balance between S\&C in time resource allocation exists and must be achieved to maximize the AST of the proposed R-SACE mechanism. 
%	 This phenomenon further proves that an optimal trade-off exists between the ISAC period and the PC period for maximizing the AST.
%	More specifically, this result emphasizes the need to optimize the S\&C time allocation in the proposed R-SACE mechanism.
%		This result is primarily due to the stronger direct BS-TC $ J $ link than the reflected BS-RIS-CU $ I $ link and the RIS-assisted suppressed interference from CU $ I $ during the much longer PC period. 

Fig. \ref{Communication_performance_ts} further shows that the average throughput of CU $ I $ increases with the number of RIS elements during the ISAC period, PC period, and the entire period.
This result is because the larger $ N_{\mathrm{R}} $ enhances the received signal strength from the BS-RIS-CU $ I $ link, thereby contributing to a higher communication SINR for CU $ I $.
As for the TC $ J $, its average throughput decreases with $ N_{\mathrm{R}} $.
This is due to the increased number of RIS elements resulting in severe interference from the BS-RIS-CU $ I $ link. 
Finally, the AST increases with $ N_{\mathrm{R}} $, since the RIS-assisted increment in the average throughput of CU $ I $ is greater than the decrement in the average throughput of TC $ J $.
This result demonstrates the effectiveness of utilizing RIS in improving systemic communication performance of the proposed R-SACE mechanism by coordinating transmission and suppressing mutual interference.

	\begin{figure}[tb]
	\centering
	\includegraphics[width=2.7in]{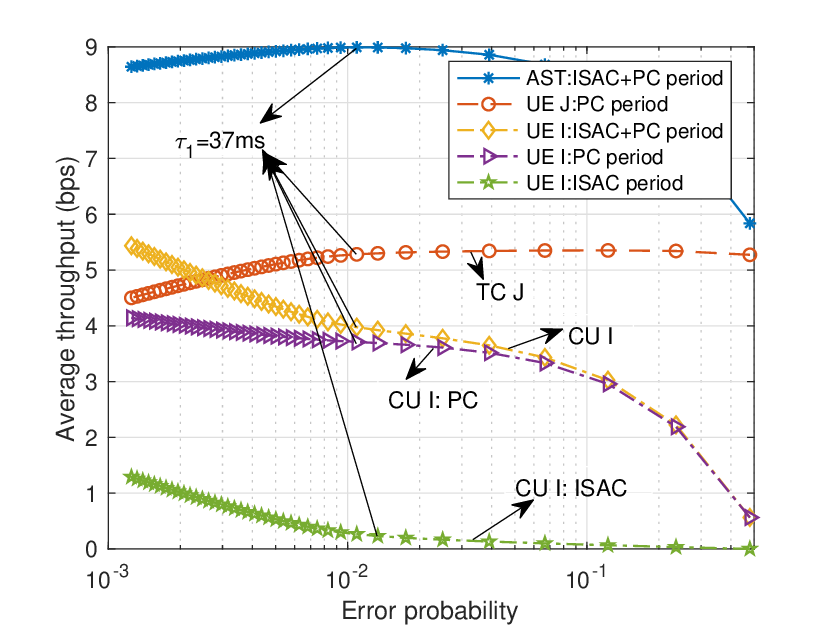}
	\caption{The average throughput vs. error probability.}
	\label{Tradeoff_C_S_ts}
\end{figure}

Fig. \ref{Tradeoff_C_S_ts} presents the trade-off between S\&C  performances affected by time resource allocation in the proposed R-SACE mechanism. 
The average throughputs and the error probability are presented in this figure.
	In this figure, the average throughput of CU $ I $ during the ISAC period increases as the error probability $ P_E $ decreases.  
	This is due to the longer duration $ \tau_1 $ leading to a lower $ P_E $ and a higher average throughput of CU $ I $ during the ISAC period.
	During the PC period, the average throughput of CU $ I $ increases as the error probability $ P_E $ decreases.
	This phenomenon can be explained as follows.
%	As $ \tau_1 $ increases, the communication performance enhancement of CU $ I$ benefiting from the continuously decreased $ P_E $ compensates the decrement in the average throughput of CU $ I $ incurred by the deduced time duration of PC period, as shown in Fig. \ref{Communication_performance_ts}. 
	As $ \tau_1 $ increases, the improvement in communication performance of CU $ I$, which benefits from the continuously decreasing $ P_E $, offsets the decrease in average throughput of CU $ I$ caused by the reduced duration of the PC period as shown in Fig. \ref{Communication_performance_ts}.  
%	Synthesizing the above results, 
	Therefore, the overall average throughput of CU $ I $ during the entire timeline increases as the error probability decreases.
	For TC $ J $, the average throughput first increases and then decreases as $ P_E $ decreases.
	This result demonstrates that there exists an optimal S\&C time allocation between ISAC and PC periods, which maximizes sensing-assisted communication performance for TC $ J $.
%	This occurs because the decreased $ P_E $ benefits from the increased $ \tau_1 $, meaning the reduced time duration of the PC period ultimately results in decreased average throughput for TC $ J $. 
	Finally, we can observe that a maximal AST exists when $ \tau_1=37 $ ms, as the error probability $ P_E $ varies.
	This result shows that an optimal S\&C time allocation exists for achieving the maximal systemic communication performance.
	More importantly, this result demonstrates that the proposed R-SACE mechanism optimally balances the resource competition and functional assistance between S\&C, leading to maximal systemic communication performance.
	Last but not least, after $ \tau_1=37 $ms, the reduction in AST becomes much more serious, even when the increase in error probability $ P_E $ is minimal.
	This highlights the need for improving detection accuracy using effective methods, such as the adaptive threshold method\cite{Adaptive Threshold}.

\subsection{Comparisons of optimization algorithms}
In this subsection, we evaluate the effectiveness of the proposed FPI algorithm in obtaining higher AST by jointly optimizing time allocation, transmission beamforming, and phase shift of RIS.
Three baseline algorithms are presented for comparison: the joint beamforming and phase shift optimization (J-BPO) algorithm\cite{joint beamforming and RIS 1}, 
%\cite{RIS-ISAC9},
the joint time allocation and phase shift optimization (J-TPO) algorithm\cite{ourWCL}, 
and the joint time allocation and beamforming optimization (J-TBO) algorithm. 
In the J-BPO, J-TPO, and J-TBO algorithms, the time allocation $ \tau_1 $, the beamforming $ \mathbf{w}_I $ and $ \mathbf{w}_J $, and the phase shift $ \bm{\Theta} $ are randomly designed, respectively.

\begin{figure}[tb]
	\centering
%	\subfloat[ FPI algorithm compares with J-BPO algorithm. ]{\includegraphics[width=2.8in]{RSACE_ts_opt_rand.eps}
%		\label{Opt_random_tau}}
%	\hfil
%	\subfloat[ FPI algorithm compares with J-TPO algorithm.]{\includegraphics[width=2.8in]{RSACE_wiwj_opt_rand_20points.eps}
%		\label{Opt_random_beamforming}}
%	\hfil
%	\subfloat[ FPI algorithm compares with J-TBO algorithm. ]{\includegraphics[width=2.8in]{RSACE_theta_opt_rand_20points.eps}
%		\label{Opt_random_theta}}
	\includegraphics[width=2.7in]{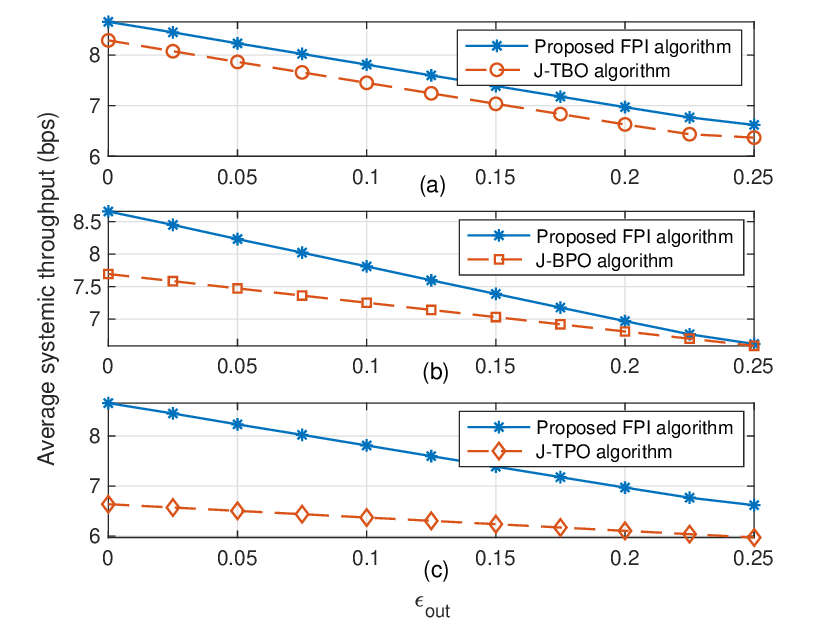}
	\caption{The AST v.s. $ \epsilon_{\mathrm{out}} $ obtained with different algorithms.
%		 In these simulations, $ \epsilon=2.4484\times10^{-11} $, $ P_{H_1}=P_{H_0}=0.5 $, and  $ \epsilon_{out}=0.1 $ in $ (a) $ and $ (b) $.
	 }
	\label{performance evaluation}
\end{figure}

Fig. \ref{performance evaluation}(a) shows the superiority of the proposed FPI algorithm in achieving higher AST than the J-TBO algorithm.
The reason is that the phase shift $ \bm{\Theta} $ of RIS  is randomly designed in the J-TBO algorithm. While in the proposed FPI algorithm, $ \bm{\Theta} $ is jointly optimized with time allocation $ \tau_1 $ and transmission beamforming, which thus helps further improve the systemic communication performance of the R-SACE mechanism. 
Fig. \ref{performance evaluation}(b) shows that the proposed FPI algorithm achieves higher AST than the J-BPO algorithm.
	%, where the obtained AOSC varies with the coverage probability $ 1-\epsilon_{out} $. 
This advantage is attributed to that optimizing the time allocation $ \tau_1 $ between ISAC and PC periods in the FPI algorithm contributes to better coordination of S\&C performances. In other words, compared to the randomly designed ISAC time protocol in the J-BPO algorithm, optimizing $ \tau_1 $ in the FPI algorithm guarantees the sensing performance that further helps improve the systemic communication performance of the R-SACE mechanism.

% Furthermore, according to the outcome shown in Figure \ref{performance evaluation}(b), the AOSC rises with $ \tau_1 $ before reaching the convergence point.

Fig. \ref{performance evaluation}(c) shows that the proposed FPI algorithm achieves higher AST than the J-TPO algorithm by further optimizing the transmission beamforming $ \mathbf{w}_I $ and $ \mathbf{w}_J $. 
More specifically, in contrast to the randomly designed beamforming in J-TPO algorithm, the joint optimization of $ \mathbf{w}_I $ and $ \mathbf{w}_J $ in the FPI algorithm helps coordinate the communication performances of two UEs and hence contributes to the improvement of the systemic communication performance. 
%We also find that the AOSC increases with $ \tau_1 $ and then converges to a stationary point in Fig. \ref{performance evaluation}(b). 
Moreover, as we can see in Fig. \ref{performance evaluation}, the AST obtained through differently algorithms decreases with $ \epsilon_{\mathrm{out}} $. This result is because that the lower $ \epsilon_{\mathrm{out}} $ means the lower probability that the data rates obtained with imperfect CSI exceed the achievable data rates obtained with perfect CSI. In other words, the lower $ \epsilon_{\mathrm{out}} $ restricts the outage probability of the data rate obtained with imperfect CSI, which thus contributes to a higher AST. This result can be further verified by the theoretical analysis in Eq. \eqref{p002}.

\begin{table*}[tbp]
	\renewcommand{\arraystretch}{1}
	\caption{Comparisons of Different Communication Enhance Mechanisms with Imperfect CSI}
	\label{Table_comparison}
	\centering
	\begin{tabular}{|c|c|c|c|c|}
		\hline
		\multirow{2}{*}{\bfseries Mechanisms} & \multicolumn{2}{c|}{\bfseries ISAC Period} & \multicolumn{2}{c|}{\bfseries PC Period} \\\cline{2-5}
		%		\hline
		& \bfseries Sensing   & \bfseries Communication & \bfseries Average data rate of CU $ I $  & \bfseries Average data rate of TC $ J $\\\cline{2-5}
		\hline
		R-SACE & $ \checkmark $ & $ \log_2(1+\hat{\gamma}_{I(1)}) $ & $ P_{I(2)}^{(1)}\!\log_2\left( 1\!+\!\hat{\gamma}_{I(2)}^{(1)}\right) \!+\!P_{I(2)}^{(0)}\!\log_2\left( 1\!+\!\hat{\gamma}_{I(2)}^{(0)}\right) $ & $ P_J^{(1)}\!\log_2\left( 1\!+\!\hat{\gamma}_J^{(1)}\right) $ \\
		\hline
		R-ACE & $ \times $ & \tabincell{c}{
			$ \log_2(1+\hat{\gamma}_{I(1)})  $
%			$ P_{\textit{AJ}}\log_2\left( 1\!+\!\hat{\gamma}_{I}^{R(1)}\right)$\\+$(1\!-\!P_{\textit{AJ}})\log_2\left( 1\!+\!\hat{\gamma}_{I}^{R(0)}\right) $\\+$P_J^{'}\log_2\left( 1\!+\!\hat{\gamma}_J^{(1)}\right) $  
		} & $ P_{\mathrm{AJ}}\log_2\left( 1\!+\!\hat{\gamma}_{I}^{R(1)}\right)\!+\!(1\!-\!P_{\mathrm{AJ}})\log_2\left( 1\!+\!\hat{\gamma}_{I}^{R(0)}\right) $ &\tabincell{c}{ $P_J^{R}\log_2\left( 1\!+\!\hat{\gamma}_J^{(1)}\right) $ } \\
		\hline
		S-ACE & $ \checkmark $ & $ \times $ & \tabincell{c}{ $P_{I(2)}^{(1)}\!\log_2\left( 1\!+\!\hat{\gamma}_I^{S(1)}\!\right) \!+\!P_{I(2)}^{(0)}\log_2\left( 1\!+\!\hat{\gamma}_{I}^{S(0)}\!\right) $} & \tabincell{c}{$P_J^{(1)}\log_2\left( 1\!+\!\hat{\gamma}_J^{S(1)}\right)  $}\\
		\hline
	\end{tabular}
\end{table*}

\subsection{Comparisons of communication enhancement mechanisms }
In this subsection, we evaluate the effectiveness of the proposed R-SACE mechanism by comparing with two baseline mechanisms, i.e., the RIS-assisted communication enhancement (R-ACE) mechanism\cite{R-ACE} and the sensing-assisted communication enhancement (S-ACE) mechanism\cite{S-ACE1}.
The main differences among these mechanisms lie in that in the ISAC period, the proposed R-SACE mechanism performs both sensing and communication, while the R-ACE mechanism only performs communication and the S-ACE mechanism only performs sensing. 
%The sensing or communication performed during the ISAC period has a significant impact on the communication performance obtained during the PC period. 
Table I summarizes different mechanisms. 
Note that the communication performances of R-ACE and S-ACE mechanism are evaluated with imperfect CSI also.
Then, we first detail the communication performances of the R-ACE and the S-ACE mechanisms, respectively.

In the R-ACE mechanism, the BS does not sense but communicates with CU $ I $ during the ISAC period, where the average data rate of CU $ I $ is the same as in the R-SACE mechanism.
	Without sensing, the BS assumes the probability of TC $ J $ being present as $ P_{\mathrm{AJ}}$ during the PC period.
	When the BS assumes that TC $ J $ is present, it communicates with CU $ I $ via the BS-RIS-CU $ I $ link while simultaneously communicating with TC $ J $ via the BS-TC $ J $ link.
	Otherwise, when the BS assumes TC $ J $ is not present, it communicates with CU $ I $ via both the BS-RIS-CU $ I $ link and the BS-CU $ I$ link.
	As a result, the average data rate of the present TC $ J $ in the PC period is given by $P_J^{R}\log_2\!\left( \!1\!+\!\hat{\gamma}_J^{(1)}\!\right)  $, where $P_J^{R}\!=\!P_{\mathrm{AJ}}P_{\mathrm{J_1}} $ and $\!\hat{\gamma}_J^{(1)}  $ is given in Eq. \eqref{hat_SNR}.
	The average data rate of CU $ I $ in the PC period is given by $ P_{\mathrm{AJ}}\log_2\!\left(\! 1\!+\!\hat{\gamma}_{I}^{R(1)}\!\right)\!+\!(1\!-\!P_{\mathrm{AJ}})\log_2\!\left(\! 1\!+\!\hat{\gamma}_{I}^{R(0)}\!\right) $, where $\hat{\gamma}_{I}^{R(1)}\!\!=\!\hat{\gamma}_{I(2)}^{(1)}  $ and $ \hat{\gamma}_{I}^{R(0)}\!\!=\!\!\hat{\gamma}_{I(2)}^{(0)} $ as shown in Eq. \eqref{hat_SNR}.
	The constraints in the R-ACE mechanism are the same as those in the R-SACE mechanism.

%Without RIS, the BS communicates with two UEs through direct links and utilizes the sensing result and power control to alleviate the mutual interference. 

In the S-ACE mechanism, the BS performs sensing during the ISAC period and then uses the sensing results to aid communication during the PC period.
When the BS detects the presence of TC $ J $, it communicates with CU $ I $ and TC $ J $ via the direct BS-CU $ I $ and BS-TC $ J $ links, respectively.
To alleviate mutual interference, the communication of CU $I$ is limited by the tolerable interference of TC $J$.  
Otherwise, when BS detects the absence of TC $ J $, it communicates only with CU $ I $ via the direct BS-CU $ I $ link. 
Therefore, the average outage rate of CU $ I $ during the PC period is given by $ P_{I(2)}^{(1)}\!\log_2\left( 1\!+\!\hat{\gamma}_I^{S(1)}\!\right) \!+\!P_{I(2)}^{(0)}\!\log_2\left( 1\!+\!\hat{\gamma}_I^{S(0)}\!\right) $, where $\hat{\gamma}_I^{S(1)}\!\!=\!|\hat{\mathbf{h}}_I^{\mathrm{H}}\mathbf{w}_I|^2/\!\left(|\hat{\mathbf{h}}_I^{\mathrm{H}}\mathbf{w}_J| ^2\!+\!\sigma^2\right) $ and $ \hat{\gamma}_I^{S(0)}\!\!=\!|\hat{\mathbf{h}}_I^{\mathrm{H}}\mathbf{w}_I|^2\!/\sigma^2 $. 
The average data rate of the present TC $ J $ during the PC period is given by $ P_J^{(1)}\log_2(1\!+\!\hat{\gamma}_J^{S(1)}) $, where
$\hat{\gamma}_J^{S(1)}\!=\!\frac{|\hat{\mathbf{h}}_J\mathbf{w}_J|^2\!}{ |\hat{\mathbf{h}}_J\mathbf{w}_I|^2\!+\!\sigma^2 }$. 
The interference constraint is $|\hat{\mathbf{h}}_J^{\mathrm{H}}\mathbf{w}_I|^2\!\le\!\bar{\epsilon}_J $, which is different from $ C6 $ in the R-SACE mechanism.

%It should be noted that the probabilistic AOSC maximization problems in R-ACE and S-ACE mechanisms can be transformed into non-probabilistic problems using the theoretical derivations in Proposition 3.

\begin{figure}[tb]
	\centering
	\includegraphics[width=2.5in]{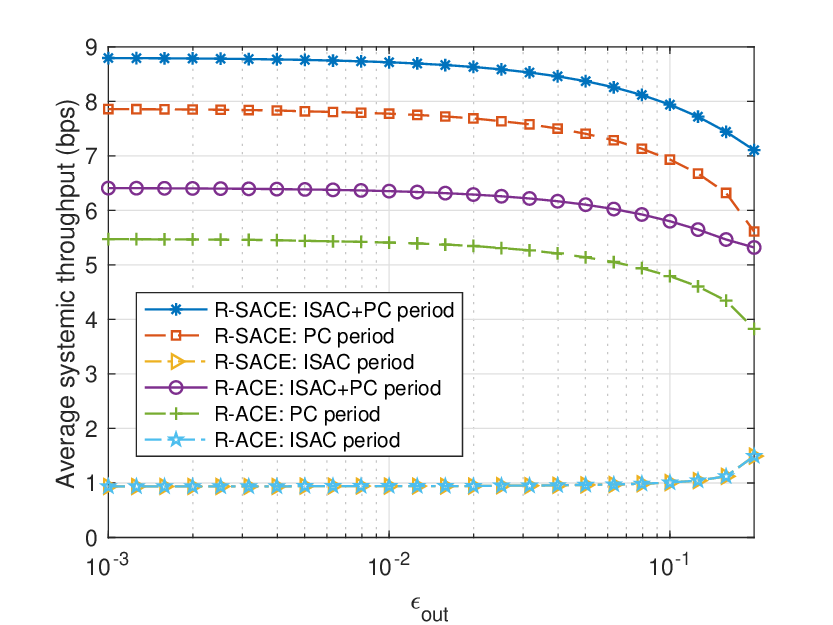}
	\caption{The AST  v.s. $ \epsilon_{\mathrm{out}} $ in R-SACE and R-ACE mechanisms. }
	\label{RSACE-RACE_epsilon}
\end{figure}

\subsubsection{Compare with R-ACE mechanism}
Fig. \ref{RSACE-RACE_epsilon} illustrates the AST varying with $ \epsilon_{\mathrm{out}} $ in the proposed R-SACE mechanism and the baseline R-ACE mechanism. 
	This figure shows that during the ISAC period, the throughput achieved by the R-SACE mechanism is the same as that achieved by the R-ACE mechanism, as shown in Table I.
	During the PC period, the proposed R-SACE mechanism achieves higher throughput than the R-ACE mechanism.
	This result is due to the performance-guaranteed sensing result obtained during the ISAC period in the R-SACE mechanism, which aids in better communication coordination between two UEs during the PC period.
	As a result, the proposed R-SACE mechanism achieves higher throughput in the PC period, which further benefits the higher systemic throughput compared to the R-ACE mechanism.

\begin{figure}[tb]
	\centering
	\includegraphics[width=2.5in]{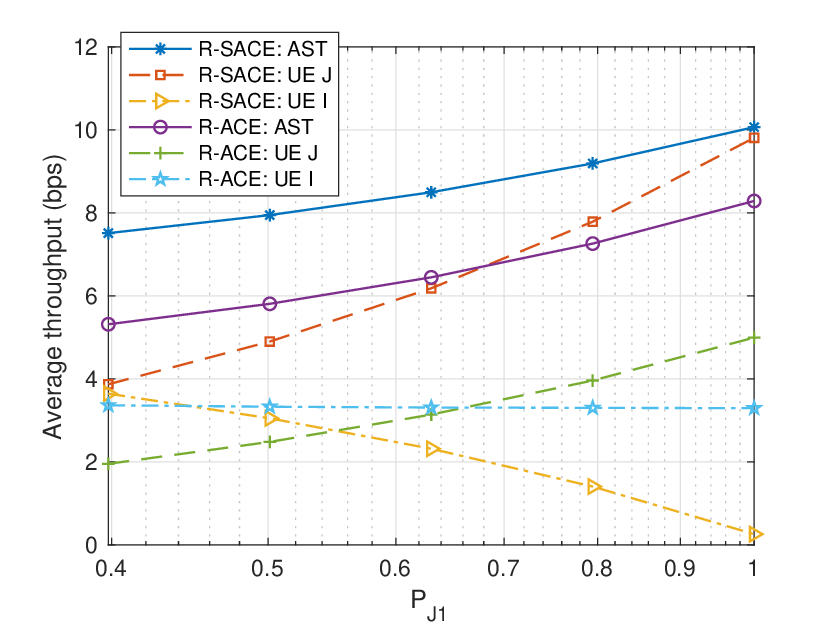}
	\caption{Average throughput v.s. $ P_{\mathrm{J_1}} $ in R-SACE and R-ACE mechanisms.  }
	\label{RSACE-RACE_PH1}
\end{figure}

%\begin{figure}[tb]
%	\centering
%	\includegraphics[width=2.7in]{RSACE_SACE_epsilon_capacity.eps}
%	\caption{The maximal outage capacities in R-SACE and S-ACE mechanisms changing with outage probability threshold. In this simulation, $ \epsilon=2.4484\times10^{-11} $, $ P_{H_1}=P_{H_0}=0.5 $. }
%	\label{RSACE_SACE_epsilon_capacity}
%\end{figure}

Fig. \ref{RSACE-RACE_PH1} illustrates the average throughputs of CU $ I $, TC $ J $, and the entire system changing with the probability $ P_{\mathrm{J_1}} $  in R-SACE and R-ACE mechanisms. 
%The average throughputs m are presented, respectively. 
As shown in this figure, the average throughput of CU $ I $ decreases with $ P_{\mathrm{J_1}}  $ in both R-SACE and R-ACE mechanisms. 
This phenomenon occurs because a higher $P_{\mathrm{J_1}} $ indicates a higher probability of TC $ J $ being present. 
As a result, the possibility of CU $ I $ encountering interference from the BS-TC $ J $ link increases, leading to a decrease in the average throughput of CU $ I $.
On the other hand, the average throughput of TC $ J $ in both R-SACE and R-ACE mechanisms increases with $ P_{\mathrm{J_1}}$.
This result is obvious, as evidenced by the theoretical expressions of the average data rate of TC $ J $ in Table II.
Finally, the average throughput of the entire system, i.e., the AST,  increases with $ P_{\mathrm{J_1}} $ because the improvement in the throughput of TC $ J $ is greater than the decrease in the throughput of CU $ I$.
Furthermore, the proposed R-SACE mechanism achieves a higher AST than the R-ACE mechanism due to the effective assistance of the performance-guaranteed sensing results obtained during the ISAC period.

%\begin{figure}[tb]
%	\centering
%	\includegraphics[width=3in]{RSACE_RACE_vsPH1_epi00501015_capacity.eps}
%	\caption{The maximal outage capacities in R-SACE and R-ACE mechanisms changing with outage probability threshold and the probability of UE $ J $ being present. In this simulation, $ \epsilon=2.4484\times10^{-11} $, $ P_{H_1}=P_{H_0}=0.5 $. }
%	\label{RSACE-RACE_PH1_epi}
%\end{figure}

\subsubsection{Comparison with S-ACE mechanism}

\begin{figure}[tb]
	\centering
	\includegraphics[width=2.5in]{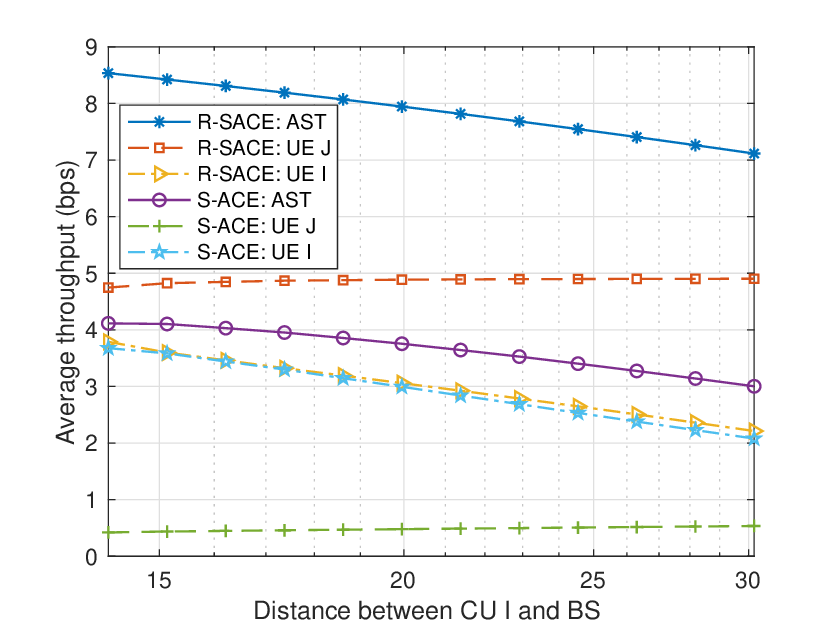}
	\caption{Average throughput v.s. the distance between CU $ I $ and BS. }
	\label{RSACE_SACE_UEIX_capacity}
\end{figure}

Fig. \ref{RSACE_SACE_UEIX_capacity} presents the average throughputs of CU $ I $, TC $ J $, and the entire system changing with the BS-CU $ I $ distance in R-SACE and S-ACE mechanisms.
As shown in this figure, the average throughput of CU $ I $ decreases with the distance between BS and CU $I $. This is because the increasing BS-CU $ I $ distance results in more serious large-scale fading on the BS-CU $ I $ and BS-RIS-CU $ I $ links.
On the other hand, the average throughput of TC $ J $ slightly increases with the distance between BS and CU $ I $. 
This is because TC $ J $ experiences less interference from the communication links between BS and CU $ I $, as the BS-CU $ I $ distance increases.
This phenomenon can be further explained by the theoretical expression of SINR of TC $ J $ in Eq. \eqref{hat_SNR}.

\begin{figure}[tb]
	\centering
	\includegraphics[width=2.5in]{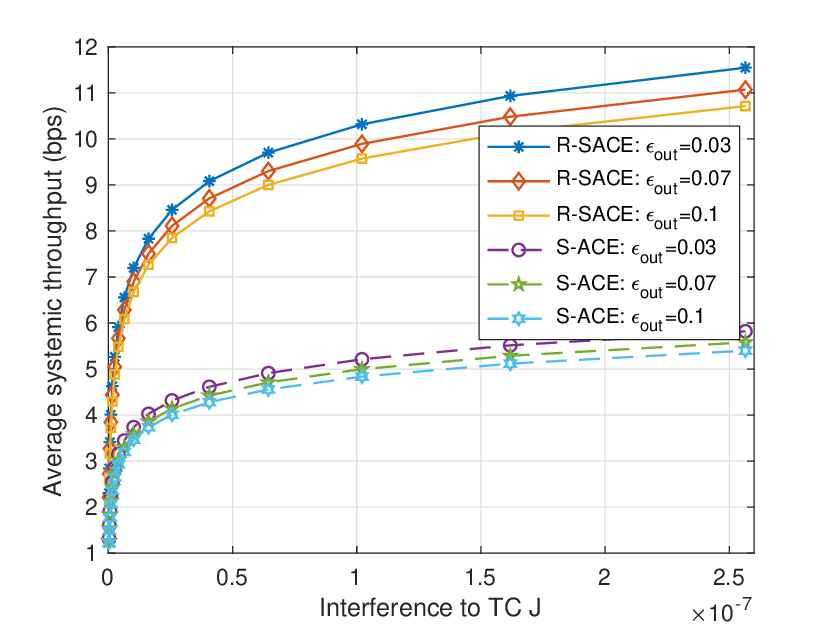}
	\caption{The AST and interference on TC $ J $ changing with $ \epsilon_{\mathrm{out}} $.}
	\label{RSACE_SACE_InfJ_capacity_epi00701}
\end{figure}

Fig. \ref{RSACE_SACE_UEIX_capacity} also shows that the average throughput of CU $ I $ in the proposed R-SACE mechanism and the S-ACE mechanism is nearly the same. 
	This result indicates that the RIS in the R-SACE mechanism offers negligible assistance in enhancing the communication performance for  CU $ \!I $ in comparison to the S-ACE mechanism.
	The reason can be explained as follows.
	First, when TC $ \!J $ is present during the PC period, the double fading of the RIS-assisted BS-RIS-CU $ I $ link degrades the SINR of CU $ I $ in the R-SACE mechanism. 
	Alternatively, when TC $ \!\!\!J $ is absent during the PC period, RIS contributes little to the increment of SINR of CU $I $ due to the stronger direct BS-CU $ I $ link. 
	Therefore, even if CU $ I $ achieves additional communication performance during the ISAC period, the RIS-assisted communication performance gain for CU $ \!\!I $ is insignificant in the R-SACE mechanism when compared to the S-ACE mechanism.
	%This simulation result can further be verified by comparing the theoretical expressions of the communication performances of CU $ I $ in R-SACE and S-ACE mechanisms, as given in Table \ref{Table_comparison}.
	For TC $\! \!J $, the RIS in the R-SACE mechanism significantly improves its communication performance when compared to the S-ACE mechanism.
	This advantage stems from the significantly reduced interference encountered by TC $ \!J $ from CU $ \!I $ with the assistance of RIS during the PC period. 
	The theoretical expressions of $ \hat{\gamma}_J^{(\!1\!)} $ in Eq. \eqref{hat_SNR} and $ \hat{\gamma}_J^{S(\!1\!)} $ in Table \ref{Table_comparison} further support this result.
	Finally, the R-SACE mechanism outperforms the S-ACE mechanism in terms of the systemic communication performance, by inheriting the RIS-assisted notable performance enhancement for TC $ J $.
%This result demonstrates the systemic effectiveness of RIS in coordinating downlink transmission by reshaping the communication environment.

Fig. \ref{RSACE_SACE_InfJ_capacity_epi00701} illustrates the AST and the interference to TC $ J $  considering different $ \epsilon_{\mathrm{out}}$ in R-SACE and S-ACE mechanisms. 
	As shown in this figure, the AST in either the R-SACE or the S-ACE mechanism decreases with $\epsilon_{\mathrm{out}}$. 
	This phenomenon is consistent with that in Fig. \ref{performance evaluation} and Fig. \ref{RSACE-RACE_epsilon}. 
	Furthermore, this figure shows that the proposed R-SACE mechanism achieves a higher AST than the S-ACE mechanism under the same interference to the TC $ J $.
	In other words, the TC $ J $ experiences more severe interference in the S-ACE mechanism than in the proposed R-SACE mechanism when the AST is the same. 
	On the other hand, from the CU $ I $ perspective, the presence of the RIS allows for a more relaxed constraint under the same interference threshold $ \bar{\epsilon}_J $.
	Specifically, the constraint condition $ |\mathbf{g}_{\textit{RJ}}^{\mathrm{H}}\bm{\Theta}\mathbf{G}\mathbf{w}_\textit{I}|^2\!\le\! \bar{\epsilon}_J $, to the beamforming $ \mathbf{w}_\textit{I} $ is less stringent compared to the interference constraint condition $|\hat{\mathbf{h}}_J^{\mathrm{H}}\mathbf{w}_I|^2\!\le\!\bar{\epsilon}_J $ in the S-ACE mechanism.
	This result further demonstrates the effectiveness of RIS in suppressing mutual interference and coordinating communication performance between CU $ I $ and TC $ J $.
%	This simulation result can be further verified by the theoretical expressions of interference constraints in $|\mathbf{g}_{\textit{RJ}}^{\mathrm{H}}\bm{\Theta}\mathbf{G}\mathbf{w}_\textit{I}|^2\le \bar{\epsilon}_J  $ of the R-SACE mechanism and in $ |\hat{\mathbf{h}}_J^{\mathrm{H}}\mathbf{w}_I|^2\!\le\!\bar{\epsilon}_J  $ of the S-ACE mechanism.

\section{Conclusion}
This paper investigated the RIS-assisted sensing-aided communication enhancement (R-SACE) with imperfect CSI.
We proposed a new R-SACE model where an ISAC BS provides downlink communication services to both a static communication user and an aperiodically present target that needs to be detected first.
The availabilities of either or both the direct link and the RIS-assisted reflected link are considered.
Average systemic throughput (AST) was maximized by jointly optimizing the time protocol design, the transmission beamforming of BS, and the phase shift of RIS.
%We considered constraints of sensing performance, transmission power, and communication interference, as well as the probabilistic constraint related to the data rate obtained with perfect CSI.
The non-convex probabilistic mixed optimization problem was transformed and then solved by the proposed fixed-point iterative (FPI) algorithm. 
Simulation results verified that the proposed R-SACE mechanism achieved an optimal balance between sensing and communication in functional assistance and competing resources, including both temporal and spatial resources. This optimal balance ultimately enhanced systemic communication performance by at least 21.53\% and 75\% through the proposed R-SACE mechanism when compared to the R-ACE and S-ACE mechanisms, respectively.
%the maximal systemic communication performance with assistance of satisfied sensing performance. 
%Simulation results also showed 
%Moreover, simulation results shows that the proposed R-SACE mechanism suppressed the interference by 
%In future work, we will investigate 
The multi-objective optimization problem that optimizes both S\&C performances will be further investigated in future work.
%, building on the contributions in this paper.

%\section{Acknowledgment}
%This work is supported by the project of the National Natural Science Foundation of China (61971239), the Jiangsu Provincial Key Research and Development Program (No. BE2022068-2), the Fundamental Research Program of Shanxi Province (202203021212290), and the University Science and Technology Innovation Project of Shanxi Province (2022L028). 

\appendices
\section{Proof of Proposition 1}
Given $ \hat{d}_l=\frac{\hat{c}_l}{2^{\hat{R}_l}-1} $, Eq. \eqref{di>di} can be rewritten as
\begin{equation}
\begin{split}
\quad\Pr\left[ d_l\ge \hat{d}_l|\hat{\mathbf{f}}_l\right]&=\Pr\left[d_l\ge \left( \hat{c}_l/( 2^{\hat{R}_l}-1)\right) |\hat{\mathbf{f}}_l \right]\\ 
&=\Pr\left[\frac{\hat{c}_l}{d_l}\le 2^{\hat{R}_l}-1|\hat{\mathbf{f}}_l \right] \le \epsilon_{\mathrm{out}}/2.
\end{split}
\end{equation}
When $ c_l\ge \hat{c}_l $, it's easy to obtain that 
\begin{equation}\label{Pr E2}
\Pr\left[\frac{c_l}{d_l}\le 2^{\hat{R}_l}-1 |c_l>\hat{c}_l,\hat{\mathbf{f}}_l \right]\le \epsilon_{\mathrm{out}}/2.
\end{equation}
Then, according to Eq. \eqref{ci<ci}, we can get 
\begin{equation}\label{cl>hat_cl}
 \Pr\left[ c_l>\hat{c}_l|\hat{\mathbf{f}}_l\right]=1-\epsilon_{\mathrm{out}}/2. 
\end{equation}
Since $ \Pr\left[\frac{c_l}{d_l}< 2^{\hat{R}_l}-1 |c_l\le\hat{c}_l,\hat{\mathbf{f}}_l \right]\le1 $, by substituting Eqs. \eqref{ci<ci}, \eqref{Pr E2}, and \eqref{cl>hat_cl} into Eq. \eqref{gamma-constraint}, we have
\begin{equation}
\begin{split}
&\quad\Pr\left[\frac{c_l}{d_l}< 2^{\hat{R}_l}-1 |c_l\le\hat{c}_l,\hat{\mathbf{f}}_l \right]\cdot\Pr\left[ c_l\le\hat{c}_l|\hat{\mathbf{f}}_l\right] \\
&+\Pr\left[\frac{c_l}{d_l}< 2^{\hat{R}_l}-1 |c_l>\hat{c}_l,\hat{\mathbf{f}}_l \right]\cdot\Pr\left[ c_l>\hat{c}_l|\hat{\mathbf{f}}_l\right]\\
&\le \epsilon_{\mathrm{out}}/2+\epsilon_{\mathrm{out}}/2(1-\epsilon_{\mathrm{out}}/2)\\
&=\epsilon_{\mathrm{out}}-\epsilon_{\mathrm{out}}^2/4\approx \epsilon_{\mathrm{out}},
\end{split}
\end{equation}
for $ \epsilon_{\mathrm{out}}\ll 1 $. As a conclusion, the constraint $ C_1 $ can be rewritten as Eqs. \eqref{di>di} and \eqref{ci<ci}.

\section{Proof of Proposition 3}
When $ \tau_1 $ is fixed, $ \rho_0\tilde{R}_0 $ is a constant term in problem $ \mathbf{P2} $. 
Thus, the joint optimization of $ \bm{\Theta} $, $ \mathbf{w}\!_I $, $ \mathbf{w}\!_J $ can be recast as 
\begin{alignat}{2}\label{appendix1}
\max_{\bm{\Theta},\mathbf{w}\!_I,\mathbf{w}\!_J}  &\tilde{O}_T
=\sum_{l=1}^{3}\rho_l\tilde{R}_l=\sum_{l=1}^{3}\rho_l\log_2(1+\tilde{\gamma}_l) \\\nonumber
&\quad=\rho_1\log_2\left( 1+\tilde{\gamma}_1(\bm{\Theta},\mathbf{w}\!_I,\mathbf{w}\!_J)\right)\\
&\quad+\rho_2\log_2\left( 1+\tilde{\gamma}_2(\bm{\Theta},\mathbf{w}\!_I)\right)\nonumber\\
&\quad+\rho_3\log_2\left( 1+\tilde{\gamma}_3(\bm{\Theta},\mathbf{w}\!_I,\mathbf{w}\!_J)\right)\nonumber\\\nonumber
\mbox{s.t.}\quad  
 &C4-C6.
\end{alignat}
Let $ \tilde{\gamma}_{21}=\frac{\mathcal{F}_{\|\mathbf{f}_I\|^2}^{-1}\!\left(\frac{\epsilon_{\mathrm{out}}}{2}\right)L_I^2\|\mathbf{w}_I\|^2\!}{\sigma_I^2}$, $ \tilde{\gamma}_{22}=\frac{\left|  \mathbf{g}_{\textit{RI}}^{\mathrm{H}}\bm{\Theta}\mathbf{G}\mathbf{w}_I\right|^2}{\sigma_I^2} $.  We define $\tilde{R}_2^{'}=\rho_2\log_2\left( 1+\tilde{\gamma}_{21}(\mathbf{w}\!_I)\right)+\rho_2\log_2\left( 1+\tilde{\gamma}_{22}(\bm{\Theta},\mathbf{w}\!_I)\right) $.
Then, by solving the functions of first-order derivatives $ \partial \tilde{R}_2^{'}/\partial \bm{\Theta}=0 $ and  $ \partial \tilde{R}_{2}/\partial \bm{\Theta}=0 $, we get that $ \tilde{R}_2^{'} $ and $ \tilde{R}_{2} $ keep the same tendency changing with $ \bm{\Theta} $ and meanwhile have the same extreme point within the feasible region of $ \bm{\Theta} $. 
As a result, the optimizations of $ \bm{\Theta} $, $ \mathbf{w}\!_I $, $ \mathbf{w}\!_J $ in Eq. \eqref{appendix1} is equivalent to
\begin{alignat}{2}\label{appendix2}
\max_{\bm{\Theta},\mathbf{w}\!_I,\mathbf{w}\!_J}  \tilde{O}_T
%=&\sum_{m=1}^{4}\rho_m\tilde{R}_m
=&\sum_{m=1}^{4}\rho_m\log_2(1+\tilde{\gamma}_m^{'})\\\nonumber
=&\rho_1\log_2\left( 1+\tilde{\gamma}_1^{'}(\bm{\Theta},\mathbf{w}\!_I,\mathbf{w}\!_J)\right)\\
&+\rho_2\log_2\left( 1+\tilde{\gamma}_{2}^{'}(\mathbf{w}\!_I)\right)\nonumber\\
&+\rho_3\log_2\left( 1+\tilde{\gamma}_3^{'}(\bm{\Theta},\mathbf{w}\!_I,\mathbf{w}\!_J)\right)\nonumber\\\nonumber
&+\rho_4\log_2\left( 1+\tilde{\gamma}_{4}^{'}(\bm{\Theta},\mathbf{w}\!_I)\right)\nonumber\\\nonumber
\mbox{s.t.}\quad  
&C4-C6,
\end{alignat}
where $ \rho_4=\rho_2 $ and 
%\begin{equation}
$ \tilde{\gamma}_m^{'}=\begin{cases}
\tilde{\gamma}_1, &  m=1\\\nonumber
\tilde{\gamma}_{21}, & m=2\\\nonumber
\tilde{\gamma}_3, & m=3\\\nonumber
\tilde{\gamma}_{22}, & m=4.
\end{cases} $
%\end{equation}

Next, we utilize the Lagrangian dual transform method to recast the sum-of-logarithm problem in Eq. \eqref{appendix2} as a sum-of-ratio problem. 
%As shown in Eq. \eqref{p2}, the objective function of problem $ \mathbf{P4} $ is the sum of  three weighted $ \log $ functions, i.e., $ \rho_l\log_2( 1+\tilde{\gamma}_l), l\in\left\lbrace1,2,3 \right\rbrace $.
According to Theorems $ 3 $ and $ 4 $ in \cite{FP-2}, we introduce an auxiliary vector $ \mathbf{v}=[v_1,v_2,v_3,v_4] $ and thus the sum-of-logarithm objective function can be recast as
\begin{alignat}{2}\label{appendix3}
%\tilde{O}_\Sigma^{LT}(\mathbf{v},\!\bm{\Theta},\!\mathbf{w}\!_I,\!\mathbf{w}\!_J)\!&=\!
\sum_{m=1}^{4}\rho_m\log_2(1\!+\!v_m)\!-\rho_mv_m\!+\!\sum_{m=1}^{4}\rho_m(1\!+\!v_m)\tilde{\gamma}_m^{'}.
\end{alignat}
We rewrite $ \sum_{m=1}^{4}\rho_m(1\!+\!v_m)\tilde{\gamma}_m^{'}$ as $  \sum_{m=1}^{4}\rho_m(1\!+\!v_m)\frac{\tilde{\gamma}_{m}^{n'}}{\tilde{\gamma}_{m}^{d'}}$.

Then,  we utilize the QT method\cite{FP-1} to recast the sum-of-ratio fractional programming problem.
Specifically, we introduce two auxiliary complex vectors $ \mathbf{u}=[u_1,u_4] $ and $ \mathbf{y}=[\mathbf{y}_2,\mathbf{y}_3] $, where $ u_1 $ and $ u_4 $ are two complex variables and $ \mathbf{y}_2\in\mathbb{C}^{K\times 1} $ and $ \mathbf{y}_3\in\mathbb{C}^{K\times 1} $ are two complex vectors.
Following that, the numerator functions in $\rho_m(1+v_m)\frac{\tilde{\gamma}_{m}^{n'}}{\tilde{\gamma}_{m}^{d'}},\forall m$ can be recast as 
\begin{equation}\label{numer-1}
\rho_m\!(\!1\!+v\!_m\!)\tilde{\gamma}_{m}^{n'}\!\!\Rightarrow\!\!\begin{cases}
\!2\sqrt{\!\!\rho_1(\!1\!\!+\!\!v_1)\epsilon_{\mathrm{out}}}\text{Re}\!\left\lbrace\! u_1^{\mathrm{H}}\mathbf{g}_{\textit{RI}}^{\mathrm{H}}\bm{\Theta}\mathbf{G}\mathbf{w}\!_I\!\right\rbrace\!, \!\!&m\!=\!1,\\
%\end{equation}
%\begin{equation}\label{numer-2}
\!2\sqrt{\!\!\rho_2(\!1\!\!+\!\!v_2)\mathcal{F}\!_{\|\mathbf{f}_I\!\|\!^2}^{-1}\!\!\left(\!\frac{\epsilon_{\mathrm{out}}}{2}\!\right)\!\!L_I^2}\text{Re}\!\left\lbrace\!\mathbf{y}_2^{\mathrm{H}}\!\mathbf{w}\!_I\!\right\rbrace\!, \!\!&m\!=\!2,\\
%\end{equation}
%\begin{equation}\label{numer-3}
\!2\sqrt{\!\!\rho_3(\!1\!\!+\!\!v_3)\mathcal{F}\!_{\|\mathbf{f}_J\!\|\!^2}^{-1}\!\!\left(\!\frac{\epsilon_{\mathrm{out}}}{2}\!\right)\!\!L_J^2}\text{Re}\!\left\lbrace\!\mathbf{y}_3^{\mathrm{H}}\!\mathbf{w}\!_J\!\right\rbrace\!, \!\!&m\!=\!3,\\
%\end{equation}
%\begin{equation}\label{numer-4}
\!2\sqrt{\!\!\rho_4(\!1\!\!+\!\!v_4)}\text{Re}\!\left\lbrace\! u_4^{\mathrm{H}}\mathbf{g}_{\textit{RI}}^{\mathrm{H}}\bm{\Theta}\mathbf{G}\mathbf{w}\!_I\!\right\rbrace, \!\!&m\!=\!4.
\end{cases}
\end{equation}
The denominator functions in $\rho_m(1+v_m)\frac{\tilde{\gamma}_{m}^{n'}}{\tilde{\gamma}_{m}^{d'}},\forall m$ can be recast as 
\begin{equation}\label{denom-1}
\tilde{\gamma}_{m}^{d'}\!\!\Rightarrow\!\!\begin{cases}
\!\!-|u_1|^2\!\left(\!2L_I^2(\|\hat{\mathbf{f}}_I\|^2\!\!+\!\!\left\|\mathbf{1}\!_K\right\|^2\!\!\!\sigma_e^2)\!\left\|\mathbf{w}\!_J\right\| ^2\!\!+\!\epsilon_{\mathrm{\mathrm{out}}}\sigma_I^2\!\right), &m=1,\\
%\end{equation}
%\begin{equation}\label{denom-2}
\!\!-\|\mathbf{y}_2\|^2\sigma_I^2, &m=2,\\
%\end{equation}
%\begin{equation}\label{denom-3}
\!\!-\|\mathbf{y}_3\|^2\!\left(  |\mathbf{g}_{\textit{RJ}}^{\mathrm{H}}\bm{\Theta}\mathbf{G}\mathbf{w}_\textit{I}|^2\!+\!\sigma_J^2\right), &m=3,\\
%\end{equation}
%\begin{equation}\label{denom-4}
\!\!-|u_4|^2\sigma_I^2, &m=4.
\end{cases}
\end{equation}
Finally, by substituting \eqref{numer-1} and \eqref{denom-1} into \eqref{appendix3}, the objective function in Eq. \eqref{appendix2} can be recast as the transformed objective function in Eq. \eqref{p3}. 
Last but not the least, the consistencies between the original problem and the QT-based transformed problem in both the optimal solutions and the optimal value of the objective function are demonstrated in \cite{FP-1} and \cite{FP-2}.

\vspace{-25pt}
\begin{IEEEbiographynophoto}
	%	[{\includegraphics[width=1in,height=1.25in,clip,keepaspectratio]{Xiaohui_Li.eps}}]
	{Xiaohui Li} received the Ph.D. degree in communication and information system from the Key Wireless Laboratory of Jiangsu Province, the School of Telecommunication and Information Engineering, Nanjing University of Posts and Telecommunications (NUPT), Nanjing, China.
	She was a Visiting Student at the Department of Electrical and Computer Engineering, Western University, London, Canada, in 2018. 
	She is currently a lecturer with the School of Telecommunication and Information Engineering, NUPT, Nanjing, China. 
	Her current research interests include Internet of Things, crowd sensing, incentive mechanism, and dynamic spectrum access and sharing.
\end{IEEEbiographynophoto}
%\end{IEEEbiography}
\vspace{-25pt}
%\begin{IEEEbiography}
\begin{IEEEbiographynophoto}
	%	[{\includegraphics[width=1in,height=1.25in,clip,keepaspectratio]{Qi_Zhu.eps}}]
	{Qi Zhu} received the bachelor’s and master’s degrees in radio engineering from the Nanjing University of Posts and Telecommunications (NUPT), Nanjing, China, in 1986 and 1989, respectively. 
	She is currently a Professor with the School of Telecommunication and Information Engineering, NUPT, Nanjing, China. Her research interests include technology of next-generation communication, broadband wireless access, orthogonal frequency division multiplexing, and dynamic allocation of radio resources.
	%\end{IEEEbiography}
\end{IEEEbiographynophoto}
\vspace{-25pt}
\begin{IEEEbiographynophoto}
	%	%	[{\includegraphics[width=1in,height=1.25in,clip,keepaspectratio]{Yunpei_Chen.eps}}]
	{Yunpei Chen} received the Ph.D. degree in communication and information systems from Nanjing University of Posts and Telecommunications, Nanjing, China, in 2023. His research interests are in the areas of next-generation broadband networks and wireless communication systems, with an emphasis on analysis of Non-Orthogonal Multiple Access (NOMA) and stochastic geometry applied to wireless systems. 
	%	%\end{IEEEbiography}
\end{IEEEbiographynophoto}
\vspace{-25pt}
\begin{IEEEbiographynophoto}
	{Chadi Assi} received the M.A.Sc. and Ph.D. degrees from the Graduate Center, City
	University of New York. He is a Professor with the Concordia Institute for Information Systems
	Engineering, Concordia University, Montreal, Canada, where he currently holds a Tier I University Research Chair. Before joining Concordia University in 2003, he was a Visiting Scientist with Nokia Research Center, Boston, from 2002 to 2003, working on quality-of-service in optical access networks. His current research interests are in the general areas of Networks (wired and cellular), network design and modeling, network optimization, and cyber security. He received the prestigious Mina Rees Dissertation Award from the City University of New York in August 2002 for his research on wavelength-division-multiplexing optical networks and lightpath provisioning. He served on the editorial board of several IEEE journals and is currently serving as an Associate Editor for the IEEE TRANSACTIONS ON VEHICULAR TECHNOLOGY, IEEE TRANSACTIONS ON MOBILE COMPUTING, and IEEE TRANSACTIONS ON NETWORK AND
	SERVICE MANAGEMENT.
\end{IEEEbiographynophoto}
\vspace{-20pt}
\begin{IEEEbiographynophoto}
	%	%	[{\includegraphics[width=1in,height=1.25in,clip,keepaspectratio]{Yunpei_Chen.eps}}]
	{Yifei Yuan} received Bachelor degree and Master degree from Tsinghua University in China and Ph. D degree from Carnegie Mellon University in USA. He was with Alcatel-Lucent from 2000 to 2008. From 2008 to 2020, he was with ZTE Corporation as a technical director and chief engineer, responsible for standards \& research of 4G LTE-Advanced and 5G technologies. He joined the China Mobile Research Institute in 2020 as a Chief Expert, responsible for 6G wireless research. He has extensive publications, including 10 books on LTE-Advanced, 5G and 6G. He has over 60 granted US patents.
	%	%\end{IEEEbiography}
\end{IEEEbiographynophoto}

\end{document}